\documentclass[a4paper,12pt,twoside,openright]{book}
\usepackage{custom_packages}

\begin{document}
\thispagestyle{empty}
%
%
%
%
%

\begin{center}
    {
    \setlength\intextsep{0pt}
    \begin{figure}[h!]
        \centering
        \includegraphics[width=0.9\textwidth]{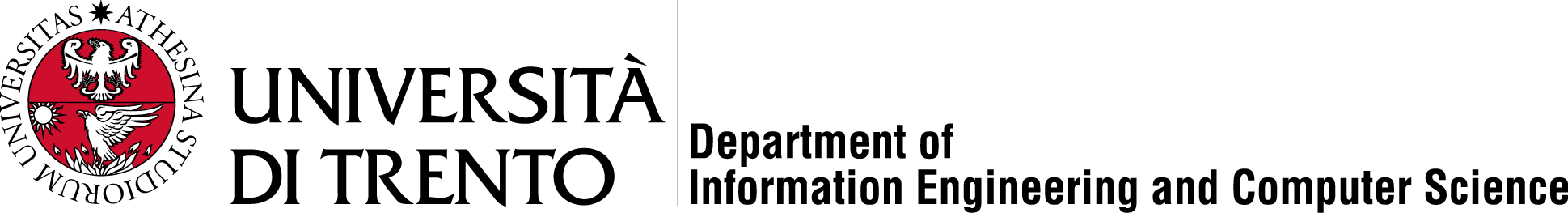}
        \label{fig:university-logo}
    \end{figure}
    }

    \par\noindent\rule{\textwidth}{0.1pt}

    \vspace{0.5 cm}  

    \large\textbf{Doctoral Programme in} \\
    \large\textbf{Information Engineering and Computer Science}

\bigskip
\bigskip

    \vspace{1 cm}
    
    \Large\textsc{\textbf{Language and Knowledge Representation}}\\\textbf{A Stratified Approach}

    \vspace{1 cm}
    
\bigskip
\bigskip

\begin{tabular}{ll}
\multicolumn{2}{l}{\large Author:}\\
 & \large MAYUKH BAGCHI, \\
  & \large PhD Candidate,\\
 & \large University of Trento, Italy.\\
\end{tabular}

\vspace{1 cm}

\bigskip
\bigskip


\begin{tabular}{ll}
\multicolumn{2}{l}{\large \ \ \ \ Advisor:}\\
 & \large \ \ \ \ FAUSTO GIUNCHIGLIA,\\
  & \large \ \ \ \ Professor of Computer Science,\\
 & \large \ \ \ \ University of Trento, Italy.\\
\end{tabular}

\vspace{1cm}

\bigskip


\begin{tabular}{ll}
\multicolumn{2}{l} 
\large A THESIS SUBMITTED FOR THE DEGREE OF\\
  & \large \ \ \ \ \ \ \ \ \ \ \ \ \ \ \ DOCTOR OF PHILOSOPHY (PhD)
\end{tabular}

\vspace{1cm}

\end{center}

\begin{center}
    \vspace{1cm}
    \hrule
    \vspace{4pt}
    2025
\end{center}

\clearemptydoublepage

\frontmatter
\singlespacing
    \chapter{Acknowledgement}

\begin{itshape}
At the very outset, I would like to convey my deepest and unfeigned gratitude to my advisor Professor Fausto Giunchiglia for heartily enabling me access to his immense depth as well as width of knowledge in my research area and allied research areas. His unmatched devotion, vision, expertise, strategy and discipline to advance research and pedagogy is worthy of emulation. I am blessed to have learnt a lot from him.

I would also like to acknowledge my sincere regards for Professor Giancarlo Guizzardi who hosted me at the University of Twente, Enschede, The Netherlands, for my PhD research internship.

Further, I would like to thank each and every member of the KnowDive research group and collaborators from research projects, especially Dr. Feroz Farazi, for the mutually fruitful research collaboration(s) achieved during my doctoral period.

Next, I would like to acknowledge the early encouragement and continued inspiration I draw from excellent teachers who taught me during my postgraduate scholarship at the Indian Statistical Institute (ISI), namely, Dr. M.P. Satija, Dr. Biswanath Dutta, Dr. A.R.D. Prasad, Dr. B.S. Daya Sagar, Dr. Devika P. Madalli and Dr. M. Krishnamurthy. They equipped us well with knowledge, discipline and practical sense to get ahead in whatever interested us.

I am also grateful to my seniors, friends and public figures from whom I have learned new perspectives on academia, research and the world beyond it, including, Sourav Bhattacharyya, Mayukh Sarkar, Dr. Subhashis Das, Abhishek Hemarajan, Indranil Naskar, Dr. Abhishek Singh and Alexandra Elbakyan.

Finally, I would like to acknowledge the unconditional affection, inspiration, encouragement and support I draw from my parents (my mother and my late father), without whom, the doctoral journey would have been impossible.

The research reported in this thesis was financially enabled by the DELPhi (DiscovEring Life Patterns) project funded by the MIUR (PRIN) 2017 as well as the JIDEP project funded by Horizon Europe Research and Innovation under grant agreement number 101058732.
\end{itshape}

    \chapter*{Abstract}
\begin{itshape}
Semantic heterogeneity is conventionally understood as the existence of variance in the representation of the same target reality when computationally modelled by independent parties. It can have serious implications in critical application scenarios like that of Knowledge Graph-based multilingual data integration. In view of the above, the thesis argues that the current understanding of the problem of semantic heterogeneity as the `existence of variance', while being crucially necessary, is not sufficient and under-characterized. There can be no variance without a prior notion of a unifying reference taken as the basis for computing the variance itself. To that end, the thesis proposes the problem of representation heterogeneity to emphasize the fact that heterogeneity is an intrinsic property of any representation, wherein, different observers encode different representations of the same target reality in a stratified manner using different concepts, language and knowledge (as well as data). The thesis then advances a top-down solution approach to the above stratified problem of representation heterogeneity in terms of several solution components, namely: (i) a representation formalism stratified into concept level, language level, knowledge level and data level to accommodate representation heterogeneity, (ii) a top-down language representation using Universal Knowledge Core (UKC), UKC namespaces and domain languages to tackle the conceptual and language level heterogeneity, (iii) a top-down knowledge representation using the notions of language teleontology and knowledge teleontology to tackle the knowledge level heterogeneity, (iv) the usage and further development of the existing LiveKnowledge catalog for enforcing iterative reuse and sharing of language and knowledge representations, and, (v) the kTelos methodology integrating the solution components above to iteratively generate the language and knowledge representations absolving representation heterogeneity. The thesis also includes proof-of-concepts of the language and knowledge representations developed for two international research projects - DataScientia (data catalogs) and JIDEP (materials modelling). Finally, the thesis concludes with future lines of research.
\end{itshape}

\section*{Keywords}
\textit{Stratification of Representation, Language Representation, Knowledge Representation.}

    \tableofcontents
    \listoftables
    \listoffigures

\mainmatter
\onehalfspacing
    \chapter{INTRODUCTION}

\section{Context}
The phenomenon of \emph{Semantic Heterogeneity} is conventionally understood as the \emph{existence of variance} in the representation of the same target reality when computationally modelled by independent parties in database schemas or data sets \cite{2005-SH}. 
The principal ramifications of semantic heterogeneity in data management include \emph{representational incompleteness} and
\emph{inconsistency} with the consequent \emph{loss of semantic interoperability} \cite{ETR}. This is a widely studied problem in increasingly emergent application scenarios like Knowledge Graph-based multilingual data integration (see, for instance, \cite{US3,2021-KGCW,KGSWC}) which serve as the principal motivation for the problem and the cumulative solution advanced by this thesis. Note that many \emph{partial solutions} at the schema and data level have been proposed for the aforementioned problem (see, for instance, \cite{2012-KARMA,2013-HDI}). However, so far, there is no unifying model of \textit{why} and \textit{how}  semantic heterogeneity manifests itself and, even less, a general solution.

In this thesis, let us consider \textit{Representation Heterogeneity} \cite{FMKD}, rather than \textit{Semantic Heterogeneity}, to emphasize the fact that heterogeneity is an intrinsic property of any representation \cite{2019-JOWO}, wherein, different observers encode different representations of the same target reality depending on the local context \cite{T1,T2}. Representation heterogeneity is, in turn, rooted in the more general omnipresent phenomenon of \emph{World Heterogeneity}. 
Thus, for example, there is a need of determining whether two different (occurrences of) musical instruments are actually the same instrument. In this perspective, the problem of representation heterogeneity is as follows. Given that (i) there are no two identical occurrences of reality, not even of the same reality, and that (ii) there are no two identical representations of even the same occurrence of reality, then (iii) how can we establish whether \textit{two heterogeneous representations actually represent the same reality?}. 

\section{Problem}
The thesis argues that the current understanding of the problem of semantic heterogeneity as the \emph{`existence of variance'}, while being crucially \emph{necessary}, is not \emph{sufficient} and under-characterized.  \emph{There can be no variance without a prior notion of a unifying reference taken as the basis for computing the variance itself}. To that end, the thesis proposes the notion of representation heterogeneity as briefed above. It grounds the notion of representation heterogeneity in that of world heterogeneity which, in turn, is modelled as the \emph{co-occurrence} of \emph{(World) Unity} and \emph{(World) Diversity}. \emph{Unity}\footnote{The notion of \emph{Unity} is unrelated to its namesake in OntoClean \cite{OC}.} models the ability of recognizing two different real world phenomena as \textit{different occurrences} of the same target reality, and, given \emph{Unity}, \emph{Diversity} models the ability of recognizing the existence of differences among them. In turn, this allows to disambiguate the (also omnipresent) phenomenon of representation heterogeneity into the two distinct phenomena of \textit{Representation Unity} which models the fact that two representations represent the same target reality, and given \emph{Representation Unity}, the notion of \emph{Representation Diversity} models the differences between the representations. Finally, these two notions are modelled in terms of the co-occurrence of \emph{Unity} and \emph{Diversity} into two distinct ordered layers, i.e., \emph{Language} and \emph{Knowledge}. The \emph{Language} layer comprises the heterogeneity arising in the conceptual as well as the (lexical) language level. Instead, the \emph{Knowledge} layer comprises of the heterogeneity arising in the (schematic) knowledge and data level. Notice that the principal focus of this thesis is on accommodating and thereby resolving the representation heterogeneity underlying the conceptual, language and knowledge level, with a focus towards resolving data heterogeneity. The resolution of data heterogeneity, while being a motivation of the problem and solutions advanced by this thesis, is out of scope for this thesis. 

\subsubsection{World Heterogeneity}
\begin{figure}[htp]
\centering
\includegraphics[width=14cm,height=5cm]{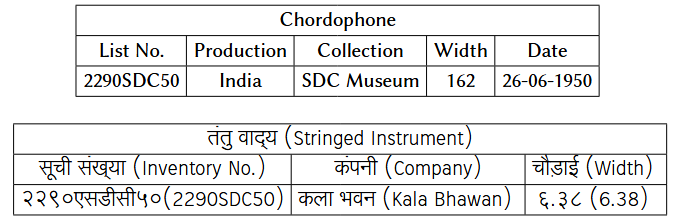}
\caption{Two heterogeneous datasets encoding information about musical instruments.}
\label{T10}
\end{figure}

Consider the motivating example (see figure \ref{T10}) of two representationally heterogeneous datasets encoding information about the same target reality, e.g., a musical instrument identified as \texttt{`2290SDC50'}. The first dataset is a record in a musical instruments catalog from Europe capturing some geophysical details of the aforementioned instrument such as production, collection, width etc. The second dataset, instead, is a record from the instrument's host museum in India encoding details such as its company and width. There are at least four levels which complicate the \emph{representation} of the entity (with label) \texttt{`2290SDC50'}. First, the fact that the same musical instrument is conceptualized differently in the two datasets, \emph{viz.} chordophone and stringed instrument respectively. Second, but non-trivially, the fact that the first dataset is in English whereas the second dataset is in Hindi. Third, the observation that for modelling the same entity, each dataset employ a different set of properties thus, essentially, leading to two different descriptions. Finally, the fact that even for a common property such as the width, the values recorded are different due to the adherence to different units of measurement.

This example is a direct instantiation of the phenomenon of \textit{Representation Heterogeneity}, which is conventionally understood as the \emph{existence of variance} in the representation, interpretation and resulting meaning. Representation heterogeneity is pervasive, wherein, even for the same target reality, different observers encode different representations depending on the local context, purpose, focus or other factors. In turn, representation heterogeneity is rooted in the \emph{unavoidable} phenomenon of \emph{World Heterogeneity}. Genetic diversity allows species to adapt to changes in the environment, production diversity allows economies to adapt to changes in market dynamics, and social and cultural diversity fuel progress in the society. Heterogeneity is the key distinguishing feature of life as there will never be two identical places or two identical individuals. Still, despite this, we are able to determine whether two heterogeneous occurrences of reality are actually occurrences of the same reality. Thus for instance, we can determine whether (or not) two different (occurrences of) objects are two instances of a musical instrument or whether (or not) two different (occurrences of) musical instruments are two occurrences of the same instrument. We formalize the intuitions above as follows.

\vspace{0.1cm}
\noindent
\textit{World Heterogeneity:} there are no two identical occurrences of the same or different realities;

\vspace{0.1cm}
\noindent
\textit{Representation Heterogeneity:} there are no two identical representations of the same or different occurrences of reality.

\vspace{0.1cm}
\noindent
Notice how the latter is an instance of the former and, as such, it is also unavoidable, as it is the  consequent \emph{(non-)generality of any representation} \cite{KD-1995-bouquet}. It is \emph{impossible to construct a  representation capable of capturing the infinite richness of the real world} and also the infinite ways, provided by language, to describe the world itself. Thus, on one hand, for any chosen representation, there will always be an aspect of the world which is not captured and, on the other hand, there will always be an alternative way to represent the same aspect of the world. 

Based on these premises, it should be evident that the understanding of semantic heterogeneity as the \emph{`existence of variance'}, while being crucially \emph{necessary}, is not \emph{sufficient} to characterize the heterogeneity of representations, and even less to suggest a way to handle it. 
The crucial observation is that, being everything (and every representation) different from everything else (and any other representation), \emph{a proper notion of variance can only be given based on a prior notion of a unifying reference taken as the basis for computing the variance itself}. And, once defined the unifying reference, we need also, as a second step, dependent on the previous, to make precise the basis on which we compute what is different. 

\subsubsection{Representation Heterogeneity}
We model representation heterogeneity in terms of the \emph{co-occurrence} of the two notions of \emph{Representation Unity} and \emph{Representation Diversity}. \emph{Representation Unity} models the fact that two different representations actually \textit{represent} the same target reality, e.g., two encounters with the same chordophone \texttt{`2290SDC50'}, or two encounters with a \textit{musical instrument}. Given a \emph{Representation Unity}, namely for any two representations for which it has been identified the reason why there is Unity, \textit{Representation Diversity} models the ability of recognizing their mutual differences. Thus, for instance, we can recognize that two musical instruments (unity) are a \emph{chordophone} and an \emph{aerophone} (diversity), or the fact that two musical instruments have different width. Thus, while establishing \emph{Representation Unity} allows to \emph{determine the space of what exists} across multiple perceptions of the same target reality, establishing \emph{Representation Diversity} allows to \emph{determine the space of variations}, e.g., the properties, of any target reality which was decided to exist. 
Based on this, we propose the following \textit{solution to the problem of representation heterogeneity}:

 \vspace{0.1cm}
\noindent
\ \ \ \ \ For any two representations, given that:
 
\ \ \ \ \  \textit{(representation heterogeneity):} there are no two identical representations of reality,
 
  \noindent
\ \ \ \ \  decide whether:
 
\ \ \ \ \ \  \textit{the two representations represent the same reality}.

\vspace{0.1cm}
\noindent
If this is the case then we say that we have the unity of representations, otherwise we say that we have representation diversity. But how to establish this fact? We do this by recursively reducing the problem of representation heterogeneity to the problems of language and knowledge heterogeneity and, in turn, to the co-occurrence of \emph{Unity} and \emph{Diversity} in the \emph{Language} and the \emph{Knowledge} layers. Let us briefly highlight how this works in practice.

In the \emph{Language} layer, the \emph{need} for co-occurrence of \emph{Language Unity} and \emph{Language Diversity} are primarily due to the different lexico-conceptual hierarchies into which (the same) target realities are hierarchically modelled (e.g., in terms of \textit{Genus} and \textit{Differentia} \cite{UKC1,2023-iconf}). Thus, for instance, \emph{Koto}, \emph{Dulcimer} and \emph{Guitar} are different while being \emph{string instruments}.
\emph{Language Unity} models the ability of recognizing two different concepts as \textit{different occurrences} of the same common concept. Thus for instance, in the above example, we establish language unity by establishing that we have musical instruments.
\emph{Language Diversity} models the ability of recognizing the differences between them, in the example above, that we have two different musical experiments. 
So, we always have both language unity and language diversity. The one which is selected depends on the level of abstraction at which we are thinking. Are we looking for two musical instruments, and we do not care which one, or are we looking for a specific one, e.g., a Guitar?

A similar situation happens in the \emph{Knowledge} layer. If in the language layer we need to define the objects we are looking for, in the knowledge layer we need to clarify the specific properties (of such objects) we are interested in. \emph{Knowledge Unity} models the ability of recognizing two different entities as \textit{different occurrences} of the same common entity, while, for any occurrence of \emph{Knowledge Unity}, \emph{Knowledge Diversity} models the ability of recognizing the differences between the entities. Thus, for instance, once we have decided at the language level that we are interested in Guitars, we may establish that we are interested in guitars of a certain form, or of a certain color. The selected set of properties determines which unity and diversity we are looking for. 

The process by which we decide on representation heterogeneity will lead to different results depending on the specifics of what we are looking for, as it is the case in our everyday life. 
We formalize this in the following notion

\vspace{0.1cm}
\noindent
\textit{Language Heterogeneity:} whether there is language unity or language diversity depends on the selected level of abstraction in the lexical-semantic conceptual hierarchy;

\vspace{0.1cm}
\noindent
\textit{Knowledge Heterogeneity:} given a certain language unity, whether there is knowledge unity or knowledge diversity depends on the properties selected as reference.

\vspace{0.1cm}
\noindent
This leads to the following refined \textit{solution to the problem of representation heterogeneity}:

 \vspace{0.1cm}
\noindent
\ \ \ \ \ For any two representations, given that:
 
\ \ \ \ \  \textit{(representation heterogeneity):} there are no two identical representations of reality,
 
  \noindent
\ \ \ \ \  decide whether:

\ \ \ \ \ \  \textit{the two representations represent the same reality},

  \noindent
\ \ \ \ \  based on:

\ \ \ \ \ \  \textit{the selected language layer of abstraction and follow-up selected knowledge layer of abstraction}.

\vspace{0.1cm}
\noindent
In other words, a general top-down solution to the problem of representation heterogeneity must be characterized in terms of two precise (top-down) choices at the language level and knowledge level.


\section{Solution}
The thesis advances a top-down solution approach to the problem of representation heterogeneity as discussed and exemplified above in terms of several solution components which are characteristically independent as each tackle a specific aspect of the problem but functionally interrelated to provide an overarching solution general enough to tackle the representation heterogeneity problem in its entirety as well. The solution components are as follows:
\begin{itemize}
    \item \textit{Stratification of Representation}: First, the thesis advances a representation formalism which is stratified into concept level, language level, knowledge level and data level to representationally and methodologically accommodate the unity and diversity existing within each level in line with the stratified character of representation heterogeneity.
    \item \textit{Language Representation}: Second, given the stratification of representation, the thesis advocates a top-down language representation using Universal Knowledge Core (UKC), UKC namespaces and domain languages to tackle the conceptual and language level unity and diversity existent within the overall problem of representation heterogeneity.
    \item \textit{Knowledge Representation}: Third, given the stratification of representation and language representation, the thesis advocates a top-down knowledge representation using the notions of language teleontology\footnote{teleontology is a portmanteau based on the Greek words \textit{telos}, \textit{ont} and \textit{logia}.} and knowledge teleontology to tackle the knowledge level unity and diversity  existent within the overall problem of representation heterogeneity.
    \item \textit{LiveKnowledge Catalog}: Fourth, the thesis emphasizes the usage and further development of the existing LiveKnowledge Catalog for enforcing iterative reuse and sharing of language and knowledge resources with an aim to increasingly minimize representation heterogeneity with each successive iteration of reuse and sharing.
    \item \textit{kTelos\footnote{a portmanteau based on the word \textit{knowledge} and the Greek word \textit{telos}.} Methodology}: Finally, the top-down \textit{kTelos} methodology elucidates the steps which methodologically integrates the solution components above to iteratively generate the language and knowledge representations absolving representation heterogeneity.
\end{itemize}
The thesis also includes proof-of-concepts relevant to two international research projects - DataScientia (data catalogs) and JIDEP (materials modelling) - wherein highlights of the language and knowledge representations adopted for the projects are briefly described and exemplified. Last but not the least, the thesis also includes a distributed review of related work comparing, as and when relevant, the solution components listed above to state-of-the-art scientific literature in their research sub-areas.

\section{Structure of the thesis}
The remainder of the thesis is organized as follows. The stratification of representation is described and exemplified in Chapter 2. The language representation and knowledge representation is described and illustrated in Chapter 3 and Chapter 4, respectively. The LiveKnowledge Catalog is described and exemplified in Chapter 5. The \textit{kTelos} methodology is described and exemplified in Chapter 6. The language and knowledge representation highlights for data catalogs and materials modelling are described and exemplified in Chapter 7 and Chapter 8, respectively. Finally, the thesis concludes with open issues and future work in Chapter 9. It is important to note that while the ideas in chapters 2, 3, 4, 5 and 6 are exemplified by employing suitable motivating and illustrative examples of different types, the proof-of-concepts in chapters 7 and 8 are illustrated using a continuing running example, especially in relation to the language and knowledge representations.

\section{Research Publications}
The following research publications and pre-prints are directly related to the ideas expressed in the thesis.
\begin{itemize}
    \item Giunchiglia, Fausto, and Mayukh Bagchi. "Representation Heterogeneity." In Joint Ontology WOrkshop (JOWO), CEUR, vol. 3249, 2022 (ideas from this paper have been reused and adapted in chapter 1).
    \item Giunchiglia, Fausto, Alessio Zamboni, Mayukh Bagchi, and Simone Bocca. "Stratified Data Integration." In the Second International Workshop on Knowledge Graph Construction (KGCW), ESWC, 2021 (ideas from this paper have been reused and adapted in chapter 2).
    \item Giunchiglia, Fausto, Simone Bocca, Mattia Fumagalli, Mayukh Bagchi, and Alessio Zamboni. "Popularity driven data integration." In Iberoamerican Knowledge Graphs and Semantic Web Conference, pp. 277-284. Cham: Springer International Publishing, 2022 (ideas from this paper have been reused and adapted in chapter 2).
    \item Bocca, Simone, Alessio Zamboni, Gábor Bella, Yamini Chandrashekar, Mayukh Bagchi, Gabriel Kuper, Paolo Bouquet, and Fausto Giunchiglia. "Building Interoperable Electronic Health Records as Purpose Driven Knowledge Graphs." In Data Science and Artificial Intelligence for Digital Healthcare: Communications Technologies for Epidemic Models, pp. 237-254. Cham: Springer International Publishing, 2024 (this paper is an example case study of the motivation expressed in chapter 2).
    \item Bagchi, Mayukh. "A diversity-aware domain development methodology." Doctoral Consortium - ER Conference (2022) (ideas from this paper have been reused and adapted in chapter 3 and 4).
    \item Fumagalli, Mattia, Marco Boffo, Daqian Shi, Mayukh Bagchi, and Fausto Giunchiglia. "Towards a gateway for knowledge graph schemas collection, analysis, and embedding." FOIS-Joint Ontology WOrkshop (JOWO), (2023) (ideas from this paper have been reused and adapted in chapter 5).
    \item Giunchiglia, Fausto, and Mayukh Bagchi. "From Knowledge Representation to Knowledge Organization and Back." International Conference on Information. Cham: Springer Nature Switzerland, 2024 (ideas from this paper have been reused and adapted in chapter 6).
    \item Giunchiglia, Fausto, Mayukh Bagchi, and Subhashis Das. "From Knowledge Organization to Knowledge Representation and Back." Annals of Library and Information Studies 71, no. 1 (2024): 137-149 (ideas from this paper have been reused and adapted in chapter 6).
    \item Seddiqui, Md Hanif, (...), Mayukh Bagchi, (...), et al. "A Material Passport Ontology for a Circular Economy." Available at SSRN 4993406 (this pre-print is linked to ideas expressed in chapter 8).     
\end{itemize}

\noindent The following research publications were published during the doctoral period but are not directly related to the ideas expressed in the thesis. 
\begin{itemize}
    \item Giunchiglia, Fausto, Mayukh Bagchi, and Xiaolei Diao. "A semantics-driven methodology for high-quality image annotation." European Conference on Artificial Intelligence (ECAI), (2023).
    \item Giunchiglia, Fausto, Mayukh Bagchi, and Xiaolei Diao. "Aligning visual and lexical semantics." In International Conference on Information, pp. 294-302. Cham: Springer Nature Switzerland, 2023.
    \item Giunchiglia, Fausto, Xiaolei Diao, and Mayukh Bagchi. "Incremental Image Labeling Via Iterative Refinement." In 2023 IEEE International Conference on Acoustics, Speech, and Signal Processing Workshops (ICASSPW), pp. 1-5. IEEE, 2023.
    \item Giunchiglia, Fausto, and Mayukh Bagchi. "Object Recognition as Classification via Visual Properties." In Knowledge Organization across Disciplines, Domains, Services and Technologies, pp. 87-100. Ergon-Verlag, 2022.
    \item Giunchiglia, Fausto, and Mayukh Bagchi. "Millikan+ Ranganathan–from perception to classification." In CAOS-Joint Ontology WOrkshop (JOWO), vol. 2969. CEUR, 2021.
\end{itemize}

    \chapter{STRATIFICATION OF REPRESENTATION}

\section{Motivation}
\begin{figure}[htp]
    \centering
    \includegraphics[width=16cm, height=8cm]{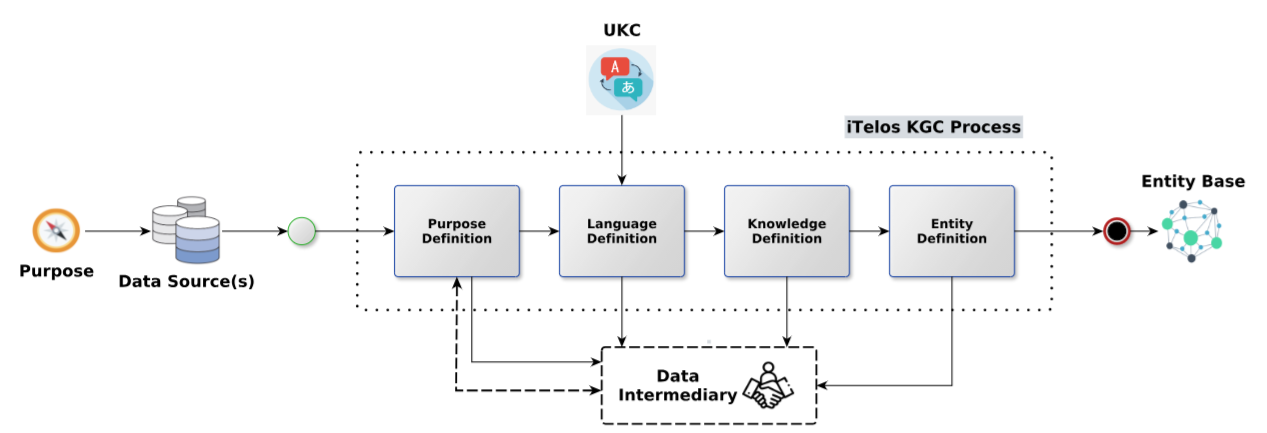}
    \caption{Motivation - \textit{iTelos} methodology.}
    \label{KGC}
\end{figure}

\noindent As briefly noted in the introductory chapter, the principal motivation behind the stratification of representation heterogeneity is Knowledge Graph Construction (KGC) based integration of representationally heterogeneous data, e.g., as developed within the context of the \textit{iTelos} methodology (please refer to \cite{sbp} for details regrading the (phases of the) methodology which, in itself, is only a motivation for this thesis and extrinsic to the main scope of this thesis). The figure \ref{KGC} illustrates the high-level view of the iTelos KGC process including the inputs and outputs. Specifically, the figure outlines the four phases of the KGC process: Purpose Definition, Language Definition, Knowledge Definition and Entity Definition. Each phase is characterized by its own internal (sub)process, requires phase-specific inputs and generates outputs that serve as inputs for the subsequent phase. The figure also depicts the Unified Knowledge Core (UKC) at the top (to be introduced in greater detail in chapter (3)) which provides the lexical-semantic background knowledge required for the iTelos representation process. Additionally, the Data Intermediary (DI) at the bottom collects intermediate outputs from each phase and facilitates their reuse. A high-level overview of the primary objectives for each phase is as follows.

The first phase of the methodology is that of Purpose Definition, wherein, the first objective is to formalize the informal purpose provided by the user. The notion of formalization here refers to the explicit definition of requirements that the final KG must fulfill. The second objective involves gathering data necessary for constructing the final KG. This data is sourced from the inputs specified by the user and from the DI, which serves as a repository of diversity-aware data. In this phase, the iTelos KGC process addresses representation heterogeneity at the data source level (notice that this level of heterogeneity is a motivation to the thesis and as such is extrinsic to the scope of the thesis).

The next phase of Language Definition takes as input the outputs of the previous phase, including, the formalized user purpose and the collected data. Additionally, the iTelos KGC process leverages the UKC during this phase to model the purpose-specific language representation which will compose the conceptual and language layer of the final KG to be produced following the iTelos data representation framework. In this phase, the iTelos KGC process addresses representation heterogeneity at the conceptual level and the language level.
 
The third phase of Knowledge Definition involves the formal user purpose, the language representations produced in the preceding phase and reference knowledge representations (to be introduced in detail in chapter (4)) to construct the final KG. The objective of this phase is to develop the purpose-specific teleology \cite{T1,T2} which represents the Entity Type Graph (ETG) of the final KG \cite{KGSWC} based on the aforementioned knowledge representation. In this phase, the iTelos KGC process addresses representation heterogeneity at the knowledge level.

Finally, the Entity Definition phase of iTelos KGC concentrates on constructing the final KG, referred to as the Entity Graph (EG), by integrating the teleology with the dataset values representing real-world entities included in the final KG. In this phase, the iTelos KGC process addresses representation heterogeneity at the data value level (notice, again, that this level of heterogeneity is a motivation to the thesis and as such is out of the scope of the thesis).

It is important to note that the iTelos methodology is founded on a \textit{middle-out} approach \cite{KGSWC,giunchiglia2021itelos}, whereby, it concretely reuses and harmonizes both \textit{top-down} language and knowledge representations (of the kind described in the thesis later) and \textit{bottom-up} purpose-specific language and knowledge definitions (out of scope for this thesis) to compose the EG. Given the concrete motivation of integrating representationally heterogeneous data via the iTelos KGC process while addressing heterogeneity at each layer, the focus now shifts to the formalism of stratification of representation diversity which underlies the \textit{iTelos} data and EG representation. 

\section{Stratification of Representation}

\setlength{\tabcolsep}{0.3em} 
{\renewcommand{\arraystretch}{1.2}
\begin{table}
\caption{Three representationally heterogeneous schemas containing a heterogeneous description of the same entity.}
\centering
\begin{tabular}{|c|c|c|c|c|}
\hline
\multicolumn{5}{|c|}{\textbf{Car}} \\ \hline
\textbf{\begin{tabular}[c]{@{}c@{}}\\ Nameplate\end{tabular}} &
  \textbf{\begin{tabular}[c]{@{}c@{}}schema:\\ speed\end{tabular}} &
  \textbf{\begin{tabular}[c]{@{}c@{}}schema:\\ fuelCapacity\end{tabular}} &
  \textbf{\begin{tabular}[c]{@{}c@{}}schema:\\ fuelType\end{tabular}} &
  \textbf{\begin{tabular}[c]{@{}c@{}}schema:\\ modelDate\end{tabular}} \\ \hline
FP372MK &
  150 &
  62 &
  Petrol &
  2020-11-25 \\ \hline
\end{tabular}
\bigskip

\centering
\begin{tabular}{ |c|c|c| } 
\hline
  \multicolumn{3}{|c|}{\textbf{Vettura}}\\
\hline
    \textbf{Targa}  & \textbf{Velocità} & \textbf{Tipo di corpo}  \\
\hline
    FP372MK  & 158 & Coupé  \\
\hline
\end{tabular}
\bigskip

\centering
\begin{tabular}{ |c|c|c|c| } 
\hline
  \multicolumn{4}{|c|}{\textbf{Vehicle}}\\
\hline
    \textbf{vso:VIN}  & \textbf{vso:feature} & \textbf{vso:modelDate} & \textbf{vso:speed}  \\
\hline
    FP372MK  & Armrest & 2020-11-25 & 155.0 \\
\hline
\end{tabular}
\label{table:ME3}
\bigskip\vspace{-0.2cm}
\end{table}
}

In this section, let us examine how representation heterogeneity can be stratified and accommodated as four levels of characteristically independent but functionally interrelated unity and diversity via the following motivating example (see also Table \ref{table:ME3}).

\vspace{0.2cm}
\noindent \emph{There are three datasets \{Car, Vettura, Vehicle\} which refer to the same real world entity, namely a car, which we assume has plate ‘FP372MK’. The first dataset describes FP372MK as a car, having five attributes, four of which are expressed using the automotive extension of schema.org.\footnote{https://schema.org/docs/automotive.html} The second dataset also considers the entity as a car but its description is provided in Italian. The third dataset encodes the entity as a vehicle having four attributes expressed employing the Vehicle Sales Ontology\footnote{http://www.heppnetz.de/ontologies/vso/ns}  namespace ‘vso:’.}

\vspace{0.2cm}\noindent
Taking a close look at the above example, it is easy to notice unity and diversity implicit at four different levels which are briefly described as follows:

\begin{itemize}
    \item \emph{Concept}: the same real world object is mentioned in two datasets using the concept denoted by the word car while in the third data set is called vehicle, namely using a more general term (because, for instance, in this latter case, there is no need to distinguish among the various types of vehicles as the issue is that of counting the number of free parking lots).
    
    \item \emph{Language}: the same real object is described using three different lexicons, namely, that of a natural language, i.e., Italian, and two namespaces, i.e., from the automobile extension of schema.org and from the Vehicle Sales Ontology, both of which use a different natural language, i.e., English, as the base language. Notice that these are three different lexicons, independently developed where the meaning of the terms used are intuitively similar but formally unrelated. 
    
    \item \emph{Knowledge}: the same real world object is described using different properties, the motivation being most likely in the different focus of the three databases. Thus, for instance, the first could be the description used in an online car rental which codifies its data using schema.org, the second could be the description used by the Italian Automobile Club and the third could be the description used in an online sales portal.
    
    \item \emph{Data}: the same real world object is described in a way that, even when associated with the same properties, the corresponding values are different. There can be many reasons for this last source of heterogeneity, for instance, different approximations, different formats, different units of measure, different reference standards (e.g., date standards for dates) and so on.
\end{itemize}
\noindent Let us consider these four layers in detail.
 
\vspace{0.2cm}
\noindent \emph{Concepts.} The notion of concept is well known in the philosophy of language literature, see, e.g. \cite{millikan1987language}, and in computational linguistics, see, e.g. \cite{PWN}. Here we follow the work as described in \cite{UKC1,UKC2}, and take concepts to be \textit{unique alinguistic identifiers}. 
To that end, concepts are organized in multiple hierarchies wherein a child concept is taken to be more specialized than the father. For instance, in the example in Table \ref{table:ME3}, the concept of car is a direct specialization of the concept of vehicle. 

\vspace{0.2cm} \noindent
\emph{Language.} Languages are taken here in a very broad sense to include, e.g.,  natural languages, namespaces and formal languages. Language unity and diversity occurs both \emph{across and within languages}. Thus, there are multiple languages available for the purpose of representing the same concept, but also, even within the same language, linguistic phenomena like \emph{polysemy} and \emph{synonymy} allow for multiple diverse representations of entities. As a result there is a \textit{many-to-many mapping} between words and concepts, both within the same language and across languages \cite{UKC1,UKC2}.

\vspace{0.2cm} \noindent
\emph{Knowledge.} We model knowledge as a set of \textit{entity types}, also called \textit{etypes}, meaning by this, classes of entities with associated \textit{properties}. Knowledge unity and diversity
arises from the \emph{many-to-many mapping} between etypes and the properties employed to describe them \cite{ETR}, and can appear in one of two different forms. The first appears when there are ‘\(n\)’ representations of different etypes described in terms of the same set of properties. The second manifests itself when there are ‘\(n\)’ representations of the same etype with different sets of properties. As an example of the latter situation, in Table \ref{table:ME3}, two datasets describe the same etype car, but the two etypes are associated to three different groups of properties.

\vspace{0.2cm}\noindent
\emph{Data.} We model data, meaning by this the concrete, ground knowledge, that we have about objects in the world, as \textit{entities} each associated with \textit{property values}, where properties are inherited from the etype of the entity. Data unity and diversity \cite{2005-SH} exists because of the fact that the \textit{mapping} between entities and the property values used to describe them is \textit{many-to-many}. Data unity and diversity appears as well in two different forms, wherein the same real world entity, associated with the same properties, is described using different data values, while dually, two different real world entities, still associated with the same properties, can be described using the same data values. As an example of the latter situation, there can be two identical cars which are both described by a set of attributes which do not contain their plate or any other identifying attribute. The example in Table \ref{table:ME3} provides an example of the former situation. Here, the three datasets refer to the same entity, a car with plate ‘FP372MK’, which shares a common attribute which is the car speed, but this property has three different values.

Notice how the stratification of representation heterogeneity presented above has the following crucial characteristics. The first is that each layer models a different type of phenomenon and the corresponding type of unity and diversity. The second is that how the unity and diversity appearing in one layer is completely (characteristically) independent of how it appears in any other layer. The third is that the conceptual layer provides the grounding for unifying the various types of unity and diversity as they appear in the other layers. In fact, in all respects, the conceptual layer is a \textit{formal logical theory}, which can be codified in a \textit{logical language}, e.g., Description Logics \cite{DL}, where the semantics of all terms (the alinguistic identifiers) is unequivocally defined by the links of the hierarchy.  The fourth and last observation is that the unity and diversity mappings which appear in concept, language, knowledge and data layer are all \textit{many-to-many} and this generates the type of combinatorial complexity which makes it so difficult to handle the problem of representation heterogeneity. As already hinted to in the introduction, the stratification of representation heterogeneity provides a major advantage in that it allows to structure it in four independent and much smaller problems, where each problem can be treated uniformly by developing methods and techniques which are specialized just for that layer. This last observation is the basis for the work presented in the remainder of this thesis. Note that the data layer is not dealt within the scope of this thesis but it remains a major motivation behind the work. 

\begin{figure}[htp]
    \centering
    \includegraphics[width=15cm, height=10cm]{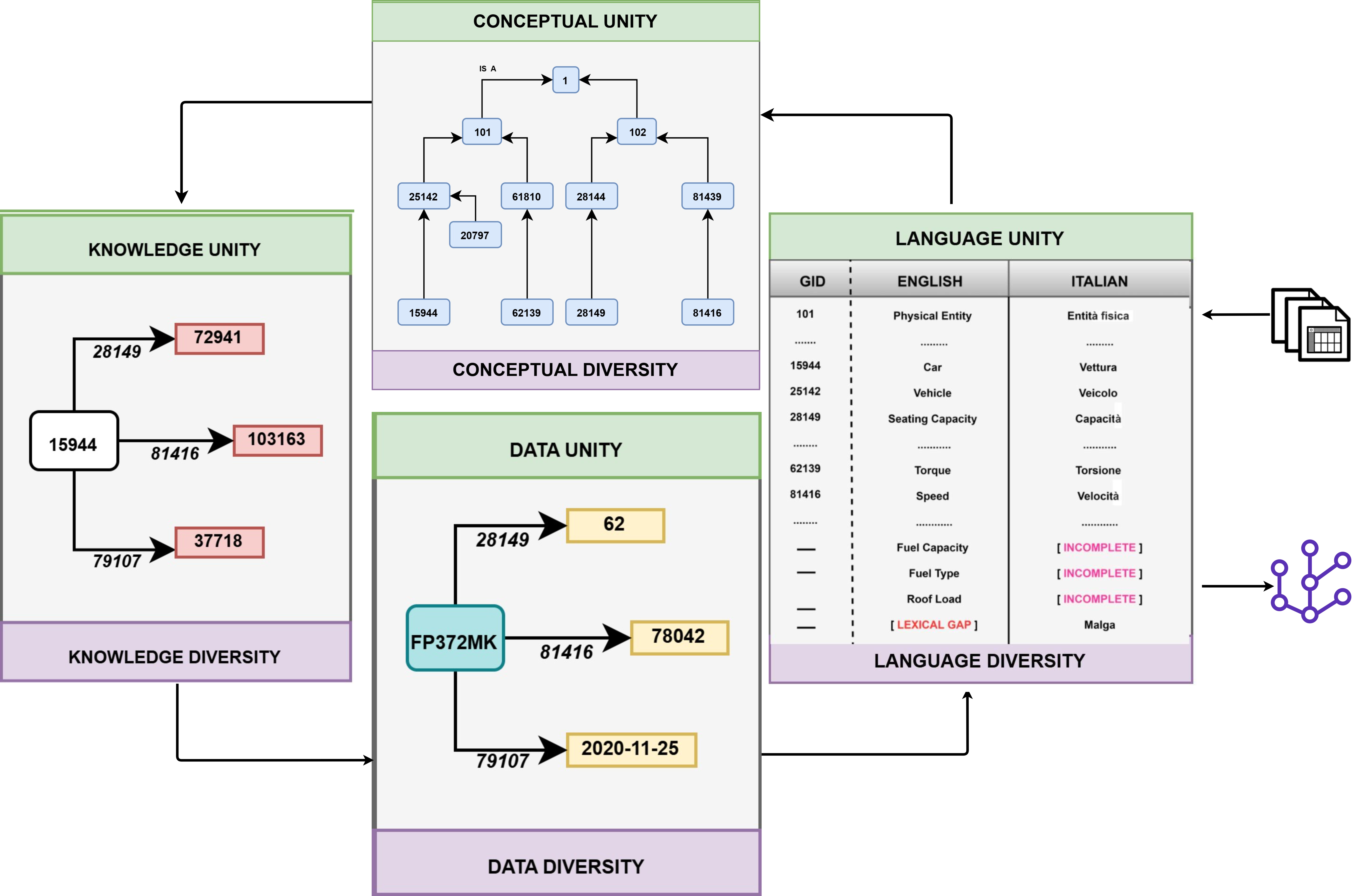}
    \caption{The Representation Heterogeneity Architecture.}
    \label{fig:RDN3}
\end{figure}

The figure \ref{fig:RDN3} depicts a \emph{representation heterogeneity architecture} illustrating the representation heterogeneity formalism to accommodate and address different layers of representation heterogeneity as proposed and advanced by this thesis. The illustration in Figure \ref{fig:RDN3} is instantiated (partially, for lack of space) to the example in Table \ref{table:ME3}. Here, the arrows represent the \emph{functional dependencies} which must be enforced among the different layers and, therefore, implicitly define the \emph{order of execution} which must be followed, e.g., during a KG-based multilingual data integration project (as in iTelos), starting from the user input and concluding with the fully integrated data as an EG. 

The language representation layer appears first and last in the architecture in Fig.\ref{fig:RDN3}. It enforces the input and the output dependence of the representation of data on the user language. In fact, language is the key enabler of the bidirectional interaction between users and the process. In the first phase, the input language is translated into the system internal conceptual language and the input language is only resumed during the last step, when the results of the data integration steps are presented back to the user. To this extent, notice how, in figure \ref{fig:RDN3} the language, knowledge and data used are just identifiers, while the conversion table of the first phase in the repository is the mapping from internal and external language(s). In this process, the language layer is key in keeping completely distinct the multilingual user-defined data representation and the alinguistic system-level data representation.
It is also important to notice how the proposed representation heterogeneity architecture is natively multilingual as a result of the alinguistic concepts being the convergence of semantically equivalent words in different languages. A very important case which can be dealt by this architecture is the \emph{heterogeneity of namespaces}, as also reflected in the running example. Any number of namespaces and natural languages can in fact be seamlessly integrated following the same uniform process. Further, notice how this layer helps accommodate and establish representation heterogeneity at the language level in terms of unity and diversity. For instance, in the figure \ref{fig:RDN3}, we can establish language unity by modelling \texttt{vehicle} and \texttt{veicolo} to be synonymous across the English and the Italian languages. On the other hand, given the language unity, we can establish language diversity by modelling different specializations of vehicles, e.g., \texttt{car} and \texttt{vettura} in the English and the Italian language, as per the level of abstraction required for a specific application scenario. This layer also tackles complications arising out of the occurrence of lexical gaps \cite{LG} and language incompleteness which can be corollary issues relating to language unity and diversity in specific cases of KG-based multilingual data integration.

The management of concepts, which functionally comes next in sequence, involves the organization of the alinguistic concepts, as identified in the first step, into a lexical-semantic graph which codifies the semantic relations across concepts (and, therefore, among, the corresponding input words). In order to achieve this goal we exploit, as \textit{a priori} knowledge, a multilingual lexical-semantic resource, called \emph{Universal Knowledge Core (UKC)} \cite{UKC1,UKC2} which represents words and synonyms quite similarly to the Princeton Wordnet \cite{PWN}, still with important differences \cite{UKC1} (the UKC is introduced in greater detail in chapter (3)). The net result of this phase is the graph with the following properties:
\begin{itemize}
\item the concepts identified during the first phase are the all and only the nodes in this graph;
\item these nodes are annotated with the input terms, across languages;
\item these nodes are organized into a hierarchy which preserves the ordering across the links of the UKC.
\end{itemize}
Note how this layer helps accommodate and establish representation heterogeneity at the conceptual level in terms of unity and diversity. For instance, in the figure \ref{fig:RDN3}, we can establish conceptual unity by modelling the concept \texttt{25142} (denoting \texttt{vehicle} and \texttt{veicolo}). On the other hand, given the conceptual unity, we can establish conceptual diversity by modelling different specializations of the concept \texttt{25142}, e.g., \texttt{15944} (denoting \texttt{car} and \texttt{vettura}), \texttt{20797}, etc., as per the level of abstraction required for a specific application scenario.

Given the concept layer and the language layer, the third representation layer is dedicated to the management of \emph{knowledge} and involves the construction of a knowledge-level schema encoded using \textit{only} concepts occurring in the language layer constructed during the previous two phases. In this phase, the first step is to distinguish concepts into \textit{etype} names and \textit{property} names (both object properties and datatype properties) while the second step is to organize them into a \emph{subsumption hierarchy} of etypes interrelated and described via object and data properties. The key observation here, which constitutes a major departure from previous work, is that etype subsumption, as encoded in the knowledge layer, is enforced to be coherent with the concept hierarchy encoded in the language layer. Thus for instance, as from the above example, the object with plate \texttt{FP372MK}, being a car, can be encoded to be an entity of etype vehicle, but \textit{not} of, e.g., etype organism. This fact, which is natively enforced by the lexical-semantic hierarchy of the UKC for what concerns natural languages (in the above example, Italian), is extended to cover also terms belonging to namespaces (in the above example, that of schema.org and that of the Vehicle Sales Ontology). This alignment of meanings across languages and namespaces, which absorbs a major source of representation heterogeneity present in the (Semantic) Web, is a natural consequence of the language and conceptual alignment performed during the first two phases.

Further, notice how this layer helps accommodate and establish representation heterogeneity at the knowledge level in terms of unity and diversity. For instance, in the figure \ref{fig:RDN3} (partially illustrated), we can establish knowledge unity by modelling the etype car \texttt{15944} described by data properties such as \texttt{81416}, etc., having data type, e.g., \texttt{103163} (integer). On the other hand, given the knowledge unity, we can establish knowledge diversity by modelling \texttt{15944} with different sets of (object and data) properties, as per the level of abstraction required for a specific application scenario.

In the fourth representation layer, we tackle the heterogeneity of data values by employing a \emph{data-level knowledge graph} populating the knowledge level schema with the entities extracted from the input datasets. This graph is constituted of a backbone of alinguistic identifiers, each annotated with the input terms where, for each term, the system implementing the architecture remembers the dataset it comes from. This information is crucial in case of iterated (multi-phased) data integration, as it is usually the case, as the system needs to remember which new terms and values substitute which old terms and values. This mechanism is implemented via a \textit{provenance} mechanism, not represented in figure \ref{fig:RDN3}, which applies to all the input dataset elements, both at the schema and at the data level. A last observation is that in figure \ref{fig:RDN3} the unique identifier (id, in short) identifying the car with plate FP72MK is \texttt{\#589625}, a new identifier which never appeared before. In fact, any time a new entity is identified, it is associated with a unique id which is managed internally by the system and that the user can see and also query, but not modify.

Finally, notice how this layer helps accommodate and establish representation heterogeneity at the data level in terms of unity and diversity. For instance, in the figure \ref{fig:RDN3} (partially illustrated), we can establish data unity by modelling the enity (here, the car FP72MK) \texttt{\#589625} described by instantiations of data properties such as \texttt{81416}, etc., having data values, e.g., \texttt{78042}. On the other hand, given the data unity, we can establish data diversity by modelling \texttt{\#589625} in terms of representations of data values using, e.g., different data types.

\section{Related Work}

\begin{figure}[htp]
\vspace{-0.3cm}
    \centering
    \includegraphics[width=13cm, height=3cm]{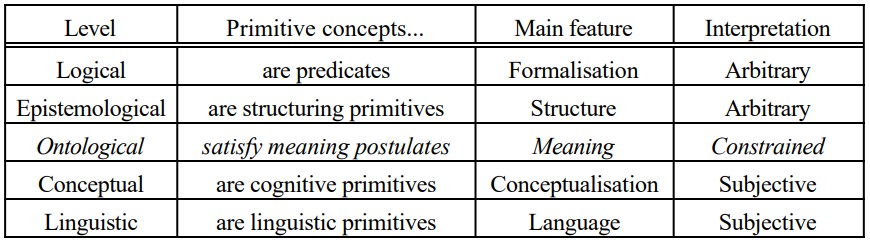}
    \caption{Knowledge Representation Levels - Guarino (taken from \cite{1994-OL}).}
    \label{KLG}
\end{figure}


\noindent 
The approach proposed in this thesis to tackle and address representation heterogeneity is based on a \textit{stratification} of representations where, starting from the  heterogeneity of the world, the thesis articulates the unity and diversity of representations in terms of a set of choices made at the conceptual, language and knowledge level. 

This is not the first time a stratification of representation has been provided. Most noticeable is the model proposed first by Brachman \cite{1979-Brachman} and later modified by Guarino \cite{1994-OL,2009-Guarino} (see Figure \ref{KLG}). Withe respect to Brachman's proposal, Guarino emphasized that the focus of the epistemological level was on structuring and \emph{formal reasoning} and not on \emph{formal representation} of concepts which remained arbitrary and \emph{neutral} as concerns ontological commitment. He argued against this very \emph{ontological neutrality} and advocated that a rigorous \emph{ontological analysis} can greatly \emph{``improve the quality of the knowledge engineering process"} \cite{1994-OL}. To that end, he proposed a new knowledge representation level termed as the \emph{ontological level} positioned between epistemological and conceptual level (see Figure \ref{KLG}). The \emph{ontological level} was intended as the \emph{``level of meaning"} and offered primitives which \emph{`` satisfy formal meaning postulates ... restrict[ing] the interpretation of a logical theory on the basis of formal ontology, intended as a theory of a priori distinctions"} \cite{1994-OL}. As part of the ontological level, several distinctions (in the form of \emph{metaproperties}, e.g., \emph{sortal}, \emph{rigidity}) were proposed (see \cite{2009-Guarino} for full details). 

The stratification of representation proposed here is orthogonal to the one proposed by Brachman and Guarino. Their work was focused on the knowledge engineering process by which one would generate a certain model of reality. The work proposed in this thesis is focused on how to solve the problem of representation heterogeneity (and, by implication, semantic heterogeneity). The solution proposed is a stratified model of how we represent the world and of how we deal with representation heterogeneity. Representation heterogeneity should therefore be seen, quoting from  \cite{2006-ECAI}, a \emph{``feature which must be maintained and exploited"} and not a \emph{``defect that must be absorbed"}, as it is the means by which we cope with the complexity of the world heterogeneity. The work so far, as partially cited in this thesis, make a step ahead in this direction.

\section{Summary}
To summarize, given the ubiquitous problem of addressing and accommodating the stratified character of representation heterogeneity and the complications it entail for a KG-based data integration process, this chapter proposed a formalism for stratifying representation unity and diversity in terms of concepts, language, knowledge and data. It also exemplified a formalism for the stratification of representation in terms of a concrete representation heterogeneity architecture instantiated on a motivating example of three datasets on vehicles. The next two chapters, namely, chapter 3 and chapter 4 will concentrate on describing and illustrating the concrete language representation and knowledge representation, respectively, which implement and are based on the formalism of conceptual unity/diversity, language unity/diversity (both modelled via language representation) and knowledge unity/diversity (modelled via knowledge representation) as outlined in this chapter.
    \chapter{LANGUAGE REPRESENTATION}

\section{Introduction}
In the context of the thesis, language representation can be defined as the informal or formal representation of \emph{``the meaning of words and how they relate to each other"} \cite{LR}. As highlighted in the previous chapter on the formalism developed for the stratification of representation, a language representation is key to model both conceptual unity/diversity and language unity/diversity (see figure \ref{fig:RDN3}), both of which, in turn, are core to address and accommodate the overarching issue of representation heterogeneity. To that end, this chapter elucidates and exemplifies the design of language representations in terms of:
\begin{itemize}
    \item Universal Knowledge Core (UKC) \cite{UKC1,UKC2}, a multilingual lexical-semantic resource forming the foundation of the proposed language representation,
    \item the notion of a \textit{Topic} which is employed to reuse existing informal language representations,
    \item the UKC annotation process to transform informal language representations into UKC-conforming formal language representations, and,
    \item the notion of UKC namespaces integrated as domain languages within the UKC.
\end{itemize}
Let us now explain and exemplify each of the above aspects in greater detail.

\section{Universal Knowledge Core}
The Universal Knowledge Core (UKC) \cite{UKC1,UKC2} is a multilingual lexical-semantic resource \cite{LSR} which forms the foundation of language representation adopted in this thesis. The UKC represents what exists in the world (via concepts) and what we can name and define (via words) within the scope of a specific language. It was originally conceived at the University of Trento, Italy and has since been continuously developed and enriched in terms of the number of natural languages aligned to the UKC and the varied services offered by the UKC in terms of language and knowledge technology research (see, for example, \cite{US1,US2,US3,ehr}). The UKC is composed of two characteristically distinct but functionally interrelated components, namely, the Concept Core and the Language Core. In the UKC, the meaning of words are represented not only with sets of synonymous words, i.e., synsets \cite{PWN,PWN2} (via the Language Core), but also using uniquely identifiable supralingual concepts (via the Concept Core) clustering together the synsets which, in different natural languages, codify the same meaning. We now concentrate in greater detail on the two aforementioned components of the Concept Core and the Language Core, respectively.

\textbf{Concept Core (CC)}: The Concept Core of the UKC is the language-independent ontological representation of the world as it explicitly represents the shared conceptualization as to what exists via the representation of concepts. The CC is structured as an evolving directed acyclic graph of supralingual concepts. Each such concept is identified by a unique identifier, the UKC Global IDentifier (GID), distinguishing it from any other concept in the hierarchy. The hierarchy of the CC is structured via the following backbone ontological relations which relate the meanings of concepts between distinct nodes in the CC hierarchy: 
\begin{itemize}
    \item \texttt{IS A}, representing the subsumption relation at the concept level,
    \item \texttt{PART OF}, representing meronymy at the concept level.
\end{itemize}
It is important to note that a strict constraint for the creation and existence of a concept in the CC is that there must be at least one natural language in the LC (described below) in which it is lexicalized.

\textbf{Language Core (LC)}: The Language Core of the UKC is the lexical representation of the world as it explicitly represents the shared conceptualization as to what we can name and define within a natural language via the representation of the set of words, senses, synsets, glossaries, and examples supported by the UKC, wherein:
\begin{itemize}
     \item a \textit{word} is a single term associated to a specific natural language
     \item a \textit{synset} is a set of synonymous words in a specific natural language. Every synset can have:
     \begin{itemize}
         \item a \textit{gloss} which describes the meaning of the synset in a particular natural language.
         \item \textit{examples} exemplifying the usage of the words of the synset. 
     \end{itemize}
     \item \textit{sense} which codifies the connection between words and synsets.
\end{itemize}
The LC, for a natural language, is structured as an evolving directed acyclic graph of synsets and, to that end, it is similar to the lexical knowledge organization of the Princeton WordNet (PWN) \cite{PWN,PWN2}. However, the first key difference of the UKC design from the PWN design is that each synset in the LC is identified and disambiguated via its linkage to a unique concept in the CC, thereby, distinguishing it from any other synset in that LC hierarchy. The second key difference is that the connection between different natural languages which constitute the overall LC are done only in the CC given the possibility that synonymous synsets across natural languages map to the same supralingual concept. It is also important to note three key observations. First, in an LC, similar to the design of the PWN, each synset is uniquely associated to a natural language and, within that language, with at least one word. Second, given the multilingual character of the LC, there is an inevitable one-to-many relationship between a concept and its linked synsets. Third, it may occur that a word in a particular (source) natural language can't be directly translated in another (target) natural language. This is represented in the UKC as a \textit{lexical gap} \cite{LG} and it is modelled by providing a description of the synset in the (target) natural language even if there are no examples.

\begin{figure}[htp]
\centering
\includegraphics[width=12cm,height=10cm]{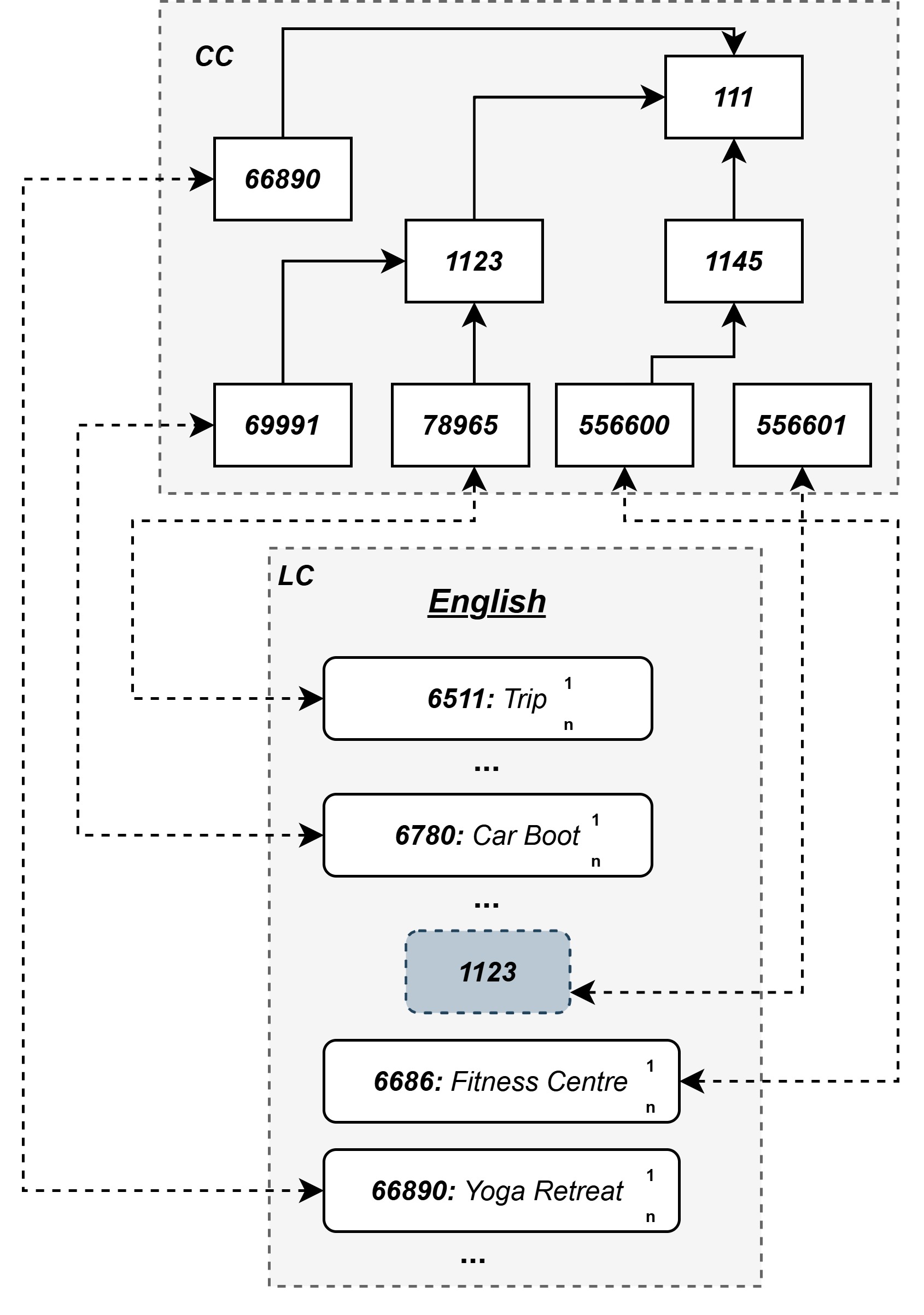}
\caption{An example illustration of a fragment of the UKC.}
\label{ukc-b}
\end{figure}

An illustrative example of the UKC architecture in terms of its component Concept Core and Language Core for the case of a fragment of the natural language English is presented in figure \ref{ukc-b}. In the figure, the Concept Core of the UKC (upper-half of the figure denoted by \texttt{CC}) is the language-independent ontological representation of the world as it explicitly represents the shared conceptualization as to what exists via the representation of supralingual concepts (illustrated as rectangles with a number) identified by their respective UKC GIDs, e.g., \texttt{111, 66890, 69991}. Note that the CC is structured as a directed acyclic graph with the directional connectors in the CC illustrating the \texttt{IS A} semantic relation which represent the subsumption relation at the concept level, e.g., between the concepts \texttt{111} and \texttt{66890}. 

The Language Core (lower-half of the figure denoted by \texttt{LC}) is the lexical representation of the world as it explicitly represents the shared conceptualization as to what we can name and define within the natural language English. It is modelled in terms of synsets illustrated as rounded rectangles with a number and a label, e.g., \texttt{6511: Trip} and \texttt{6686: Fitness Centre}, with the prefixed identifier denoting the natural language-specific identifier of the synset (e.g., \texttt{6511}, \texttt{6686}), the label denoting the preferred word employed to refer to the synset (e.g., \texttt{Trip}, \texttt{Fitness Centre}) and the index range \(1(1)n\) illustrating the fact that a synset is composed of \texttt{n} number of synonymous words. It is important to note two observations. First, notice that each synset (e.g., \texttt{Trip}) is unequivocally linked (illustrated in the figure with dashed bidirectional arrows) to a unique supralingual concept (e..g., \texttt{78965} for \texttt{Trip}) in the CC based on their mutually matching sense. This sense-based linkage is key to ground synsets in natural language to their language-independent ontological concept. Second, note that the coloured rounded rectangle identified by \texttt{1123} is an example illustration of a lexical gap with respect to the supralingual concept \texttt{556601} in the illustrated fragment of the English language UKC.

Finally, it is important to mention few important observations concerning the UKC which are relevant to the overall research context in which the thesis is positioned. Firstly, the foundational design principle underlying the construction of the UKC is the clear distinction between what is being described, i.e., the \textit{perceived world} and the \textit{language} used to describe the perceived world. For detailed considerations on this principle underlying the foundation of the UKC, please refer to the publications: \cite{SNCS-2021,2021-CAOS,2023-iconf,ECAI23,ISKO,IEEE}. Secondly, in the UKC, lexical gaps are still annotated with glosses. The rationale behind this principle is that such glosses can be seen as \textit{local }language-dependent descriptions of the missing synsets. This choice proved to be practically useful in, e.g., data integration scenarios, when one is interested in understanding the meaning of a lexical gap without knowing the language that produces the lexical gap \cite{UKC1,UKC2}. Thirdly, as a consequence of the language representation formalism of the UKC discussed above, there is a dedicated technology suite (processes, catalog, metadata, file formats) to generate, store and share concrete language resources conforming to the UKC language representation formalism. The LiveLanguage catalog\footnote{https://datascientiafoundation.github.io/LiveLanguage/} can be explored for more details on this aspect. Further, the fact that while, at present, the LC is primarily composed of lexical representations of different mainstream and under-resourced natural languages, there is a wealth of existing state-of-the-art informal language resources of different genres which can be potentially reused, reformatted and integrated within the UKC, thereby, extending the applicability of the UKC to new data science research scenarios, e.g., knowledge graph engineering \cite{itelos,itelos-1}. The last observation is a key research gap addressed in this chapter with the next section on informal language representation central to the solution proposed in this thesis to resolve the aforementioned research gap. It is also worthy to note that the UKC adheres to lexical-semantic quality requirements as prescribed in the guiding principles of \textit{Genus-Differentia} advanced by Aristotle \cite{GD} as well as the canons of classification proposed by Ranganathan \cite{srr}. Finally, notice that the hierarchy of the UKC CC and LC natively allows the representation and accommodation of conceptual unity/diversity and language unity/diversity, respectively. The above observation also holds valid for any UKC-conforming language representation, including those discussed hereafter.

\section{Topic}
In the context of this thesis, an informal language representation can be defined as a language representation which is not represented and integrated within the Universal Knowledge Core. Therefore, there are two characteristics which are key to the notion of an informal language representation. Firstly, it is a language representation which does not conform to the lexical-semantic representation formalism adopted by the Universal Knowledge Core. To that end, an informal language representation is not a representation composed of two components, namely, the fragment of the CC representing the language-independent ontological hierarchy of supralingual concepts and the fragment of the LC representing the language-dependent lexical hierarchy of language-specific synsets, wherein, each synset is linked to their supralingual concept in the CC given their mutually matching sense. Secondly, as a consequence of the first characteristic, an informal language resource is not generated following the dedicated processes and file formats which are native to a UKC language resource. A key highlight of the above characteristic is the fact that the synsets in an informal language resource are not annotated and disambiguated with a unique UKC GID. Informal language resources can exist with a focus on natural languages, e.g., informal lexical resources, or on specialized controlled vocabularies, e.g., namespaces, classifications and/or standards. The focus of this thesis is on the latter as the former is concerned with research in computational linguistics and, therefore, out of scope to the research presented here. 

Given the notion of an informal language representation and informal language resources, a \textbf{\textit{Topic}} is defined as an informal label under which informal controlled vocabularies belonging to a number of cognate universes of discourse may be (continuously) aggregated and collected. Therefore, a topic is composed of:
\begin{itemize}
    \item one or more W3C\footnote{W3C stands for the World Wide Web Consortium.} namespaces, and/or,
    \item one or more classification schemes, and/or,
    \item one or more (terminological) standards.
\end{itemize}
An illustrative example of a topic can be that of \texttt{Society and Territory} which can be composed of the following informal resources:
\begin{itemize}
    \item one or more W3C namespaces relevant to Society and Territory such as the Friend Of A Friend (FOAF) namespace\footnote{http://xmlns.com/foaf/spec/}, the W3C namespace of temporal concepts\footnote{https://www.w3.org/2006/time\#}, etc., and/or,
    \item one or more classification schemes relevant to Society and Territory such as the OpenStreetMap (OSM) classification of geosocial data\footnote{https://osmfoundation.org/wiki/Main\_Page}, etc., and/or,
    \item one or more standards relevant to Society and Territory such as General Transit Feed Specification (GTFS)\footnote{https://gtfs.org/}, Infrastructure for Spatial Information in Europe (INSPIRE)\footnote{https://knowledge-base.inspire.ec.europa.eu/index\_en}, etc.
\end{itemize}
Another illustrative example of a topic can be that of \texttt{Healthcare} which can be composed of the following informal resources:
\begin{itemize}
    \item one or more W3C namespaces relevant to Healthcare such as the healthcare extension of the schema.org namespace\footnote{https://schema.org/docs/meddocs.html}, etc., and/or,
    \item one or more classification schemes relevant to Healthcare such as International Statistical Classification of Diseases and Related Health Problems (ICD)\footnote{https://icd.who.int/en}, International Classification of Health Interventions (ICHI)\footnote{https://icd.who.int/dev11/l-ichi/en}, etc., and/or,
    \item one or more standards relevant to Healthcare such as Logical Observation Identifiers Names and Codes (LOINC)\footnote{https://loinc.org/}, Systematized Nomenclature of Medicine--Clinical Terms (SNOMED CT)\footnote{https://www.snomed.org/snomed-ct/what-is-snomed-ct}, etc.
\end{itemize}
It is important to note at this stage that a Topic, for a particular universe of discourse, is not necessarily composed of all the three types of informal language resources as exemplified above. To that end, a topic, at any given time, might be composed of only informal W3C namespaces or only classifications or only standards or a variable combination of any of the three types of resources above. Let us now examine and illustrate in some detail each of the three types of informal language resources listed above.  

\begin{figure}[htp]
\centering
\includegraphics[width=16cm,height=10cm]{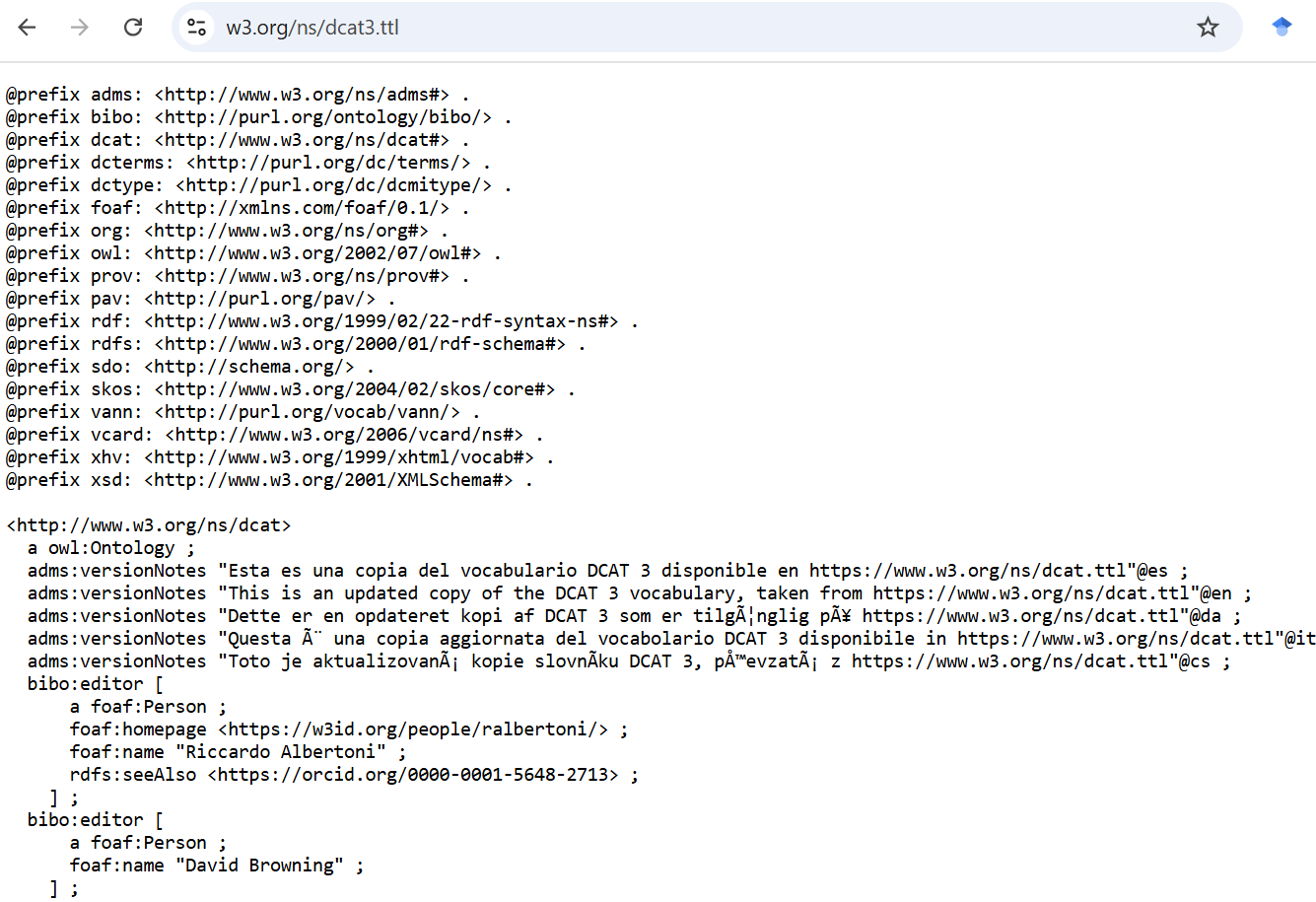}
\caption{A snapshot of the W3C DCAT3 namespace.}
\label{dcat3}
\end{figure}

\textbf{Namespace:} A (W3C) namespace\footnote{https://www.w3.org/2001/tag/doc/nsDocuments-2005-12-01/} can be defined as an informal language resource, identified by an internationalized resource identifier (IRI) reference, encoding any hierarchical collection of names which can be from amongst the options below:
\begin{itemize}
    \item a collection of element names, and/or,
    \item a collection of attribute names, and/or,
    \item a collection of properties (e.g., FOAF), and/or,
    \item a collection of concepts (e.g., as in a WordNet), and many other uses are likely to arise.
\end{itemize}
Notice that there is no requirement that the names in a namespace should necessarily be of a single type; elements and attributes can both come from the same namespace or from any other homogeneous or heterogeneous collection one can imagine. The names in a namespace can, in theory at least, be defined to identify any thing or any number of things. Further, note that namespaces can be encoded in a wide variety of file formats with variations of Extensible Markup Language (XML) encoding being a mainstream choice for sharing and resuing (W3C) namespaces. Finally, a criticial aspect of any namespace is the namespace \textit{prefix} shorthand which is prepended before any name from the namespace to denote the fact that the name and its meaning is unique to that namespace. The lookup service prefix.cc\footnote{https://prefix.cc/} provides the option to locate, look up and reuse mainstream namespace prefixes.

Let us briefly examine the Data Catalog Vocabulary (DCAT) - Version 3 namespace\footnote{https://www.w3.org/ns/dcat3.ttl}, a snapshot of which is visualized in Figure \ref{dcat3}. As we can observe from the figure that the namespace is encoded in the Turtle (\texttt{.ttl}) format and imports classes, properties and attributes from a host of other namespaces like \texttt{adms, bibo, foaf, prov}, etc., a list of whose prefixes and IRIs are provided at the beginning of the namespace declaration. An examination of the documentation of the DCAT3 namespace\footnote{https://www.w3.org/TR/vocab-dcat-3/} reveals that it is essentially a hierarchical collection of names defining classes (e.g., \texttt{dcat:Dataset, dcat:Distribution, dcat:DataService}), properties (e.g., \texttt{dcat:inSeries, dcat:servesDataset}) and attributes (e.g., \texttt{dcat:hasVersion, dcat:keyword, dcat:landingPage}) for representing various facets of data catalogs. 

\begin{figure}[htp]
\centering
\includegraphics[width=16cm,height=10cm]{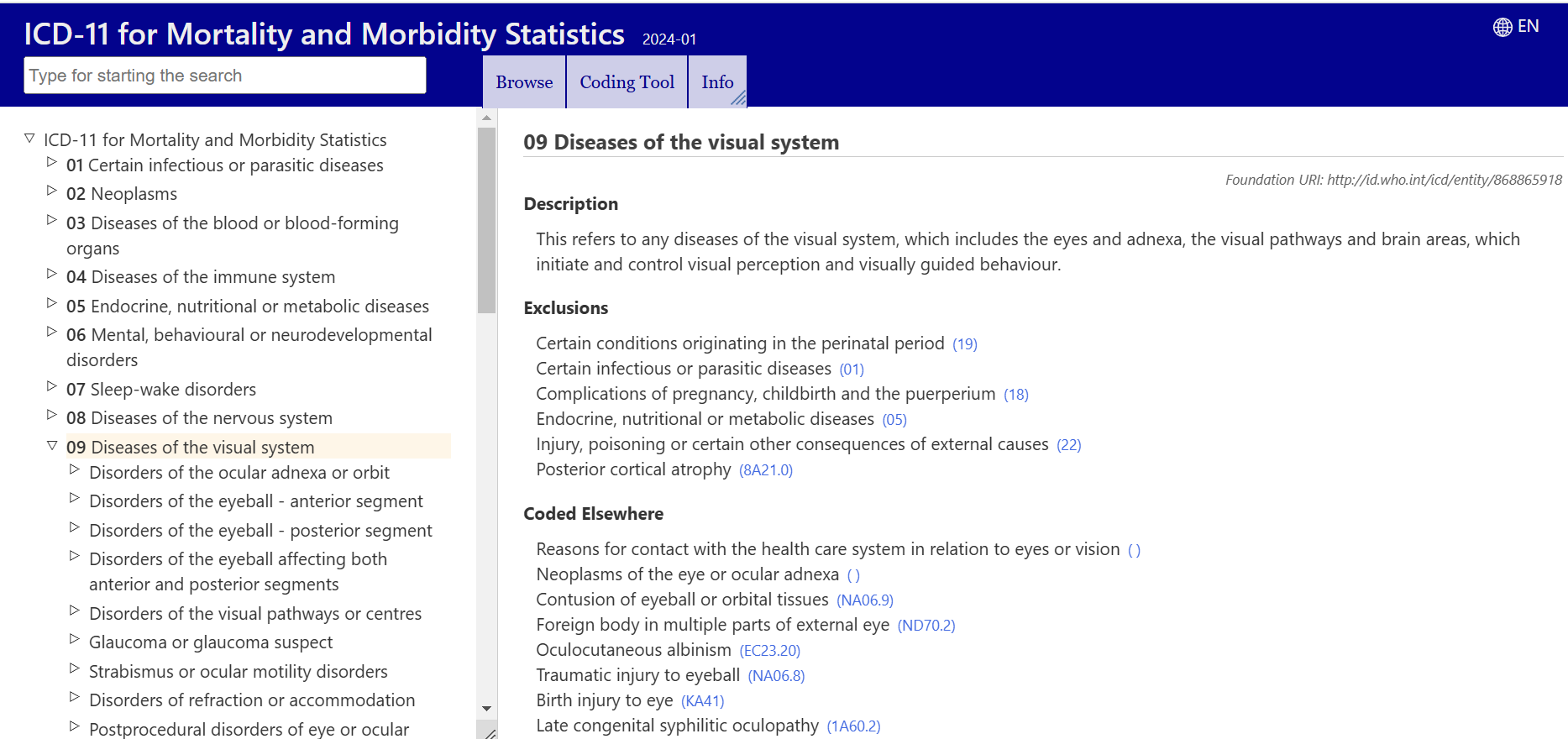}
\caption{A snapshot of the ICD-11 classification hierarchy.}
\label{icd}
\end{figure}

\textbf{Classification:} A classification scheme is an informal language resource encoding a hierarchy of terms representing concepts in the context of a specific subject or a specific universe of discourse or a specific specialized purpose. Such a scheme often includes synonyms, definitions, explanations and example of each term in its hierarchy so as to refer to the semantics of the concept with least possible ambiguity. Classification schemes are widely employed in library and information institutions (see, for example, \cite{joudrey2017organization}) to organize and catalogue information sources (e.g., books) in terms of, e.g., subjects (such as history, law, education, philosophy, etc.). Classification hierarchies are also employed in specialized domains, e.g., healthcare, to classify and organize terms and concepts relating to different purposes, e.g., diseases, health conditions and health interventions.

Let us briefly examine a fragment of the ICD-11 classification hierarchy for mortality and morbidity statistics\footnote{https://icd.who.int/en}, a snapshot of which is visualized in Figure \ref{icd}. The classification hierarchy is presented on the left-hand side of the snapshot wherein diseases and disorders are organized into their more general (e.g., \texttt{Diseases of the visual system}) and specific concepts (e.g., \texttt{Glaucoma or glaucoma suspect}). The right-half of the snapshot gives extra information on the chosen concept within the classification hierarchy like a gloss of the concept, exclusions considered and whether the concept is coded elsewhere. Note that the extent of such extra information varies across classification schemes.

\begin{figure}[htp]
\centering
\includegraphics[width=16cm,height=10cm]{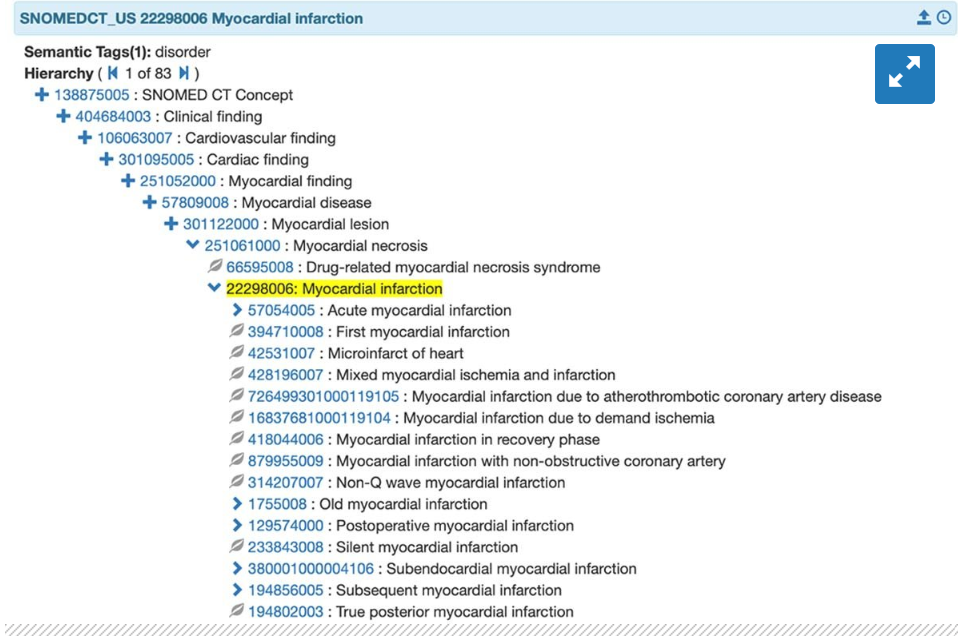}
\caption{A snapshot of the SNOMED-CT standard.}
\label{snomed}
\end{figure}

\textbf{Standard:} A terminology standard can be defined as an informal language resource encoding a hierarchy of terms which addresses a key requirement for unambiguous data and information interoperability, i.e., the ability to represent concepts in an unambiguous manner between a sender and receiver of information. A standard is always developed and continuously maintained by a leading organization. They are widely employed, e.g., in a domain, for data and information interoperability and they are often \textit{adapted} to suit local requirements, e.g., in terms of language translations, adding or deprecating terms. It is important to note that most communication between information systems relies on structured hierarchical vocabularies, terminologies and specialized code sets to represent concepts. To that end, terminology standards provide a foundation for interoperability by improving the effectiveness of data and information exchange as they both provide a means for organizing information and define the semantics of information using consistent and computable mechanisms. Some notable examples of terminology standards employed, for instance, in the healthcare sector include SNOMED-CT, LOINC, etc.

Let us briefly examine a fragment of the SNOMED-CT healthcare standard\footnote{https://www.snomed.org/snomed-ct/what-is-snomed-ct}, a snapshot of which is visualized in Figure \ref{snomed}. SNOMED CT is used for the electronic interoperability of clinical healthcare information and it is owned and maintained by SNOMED International\footnote{https://www.snomed.org/}, a not-for-profit association. Notice that the example focuses on the sub-hierarchy of \texttt{22298006: Myocardial infarction} with several specialized concepts modelled such as \texttt{42531007: Microinfarct of heart, 1755008: Old myocardial infarction}, etc. Further, note the clinical concept underlying each term is uniquely identified via the SNOMED-CT identifier.

There are two key observations for the above elucidation. First, notice that an informal language resource, whether a W3C namespace or a classification or a standard, can exist in a wide variety of formats ranging from mainstream computer-processable file formats (e.g., some flavour of XML) to even a detailed documentation describing the controlled vocabulary the resource embodies. This also implies that these resources, in their native formats, are not amenable to be formalized and integrated within the overall UKC language representation. Second, as a consequence of the first observation, there is a need for a dedicated process which transforms the informal language resources expressed in an informal language representation to a formal language resource in a formal language representation conforming to the UKC.

\section{UKC Annotation}
\begin{figure}[htp]
\centering
\includegraphics[width=16cm,height=6cm]{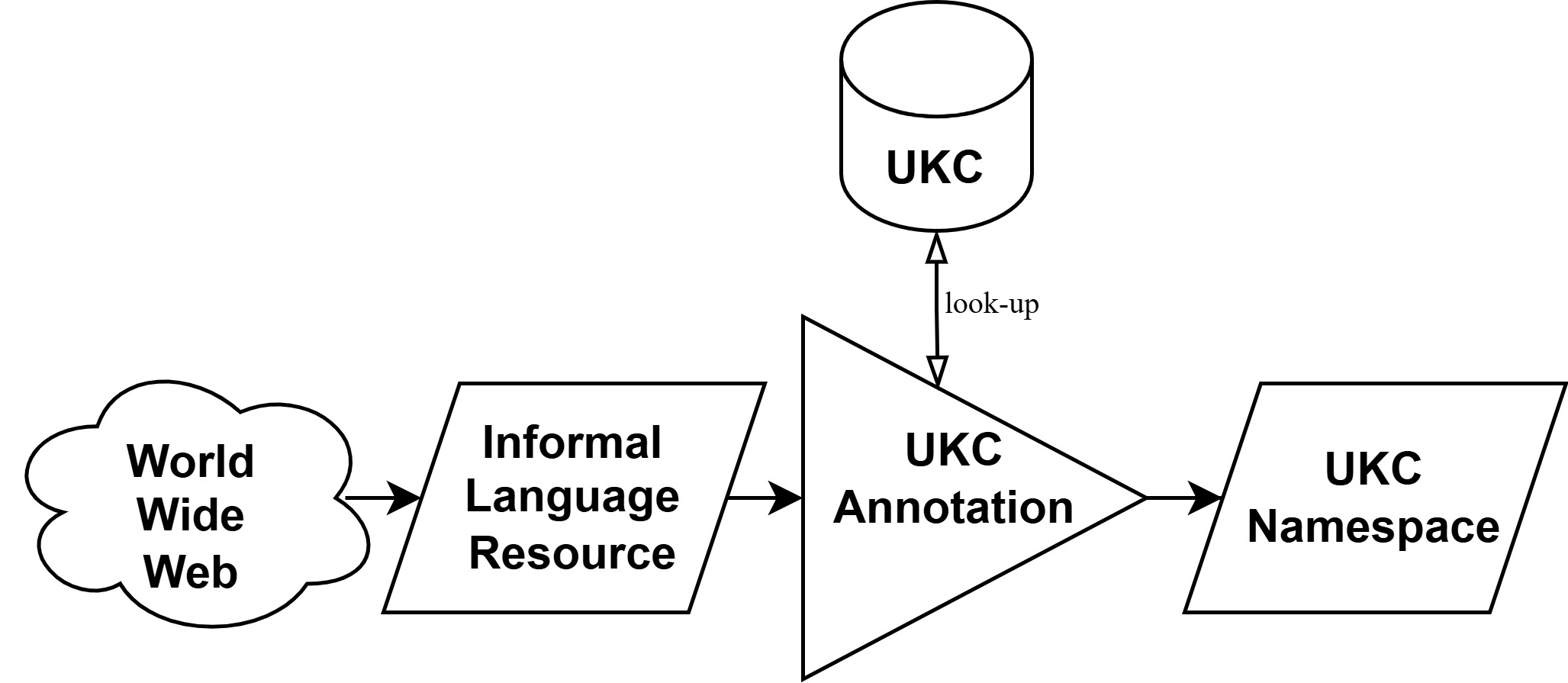}
\caption{A high level view of the UKC annotation process.}
\label{ukcan}
\end{figure}

Let us now focus on the dedicated process to facilitate the transformation and integration of an informal language resource to the formal language representation formalism adopted by the UKC. A high level view of the annotation pipeline is illustrated in Figure \ref{ukcan}. The pipeline starts with selection of a popular informal language resource within a topic from the world wide web. Then, the UKC annotation steps are executed over the (hierarchy of) terms of the resource in interaction with an instance of the UKC. Finally, the UKC namespace is generated as an output of the annotation steps. Let us now see the UKC annotation process in some greater detail.


The high-level pipeline of the UKC annotation process \cite{ERPHD} begins with selection of an informal language resource (represented as a parallelogram in Figure \ref{ukcan}) from a relevant information source in the world wide web (represented as a cloud in Figure \ref{ukcan}). There are two key observations. First, the selection criteria is based on the popularity of the informal language resource considered and it is calculated variously in various contexts, e.g., a criteria can be the popularity of reuse of terms of the language resource within datasets or as classes/properties within state-of-the-art applied ontologies considered within a specific topic. Further, the information sources for obtaining the resources can, again, be various in different contexts, e.g., from the documentation section of the official portals of standard organizations or from namespace look-up portals or from portals which share classifications in different file formats.

\begin{figure}[htp]
\centering
\includegraphics[width=16cm,height=6cm]{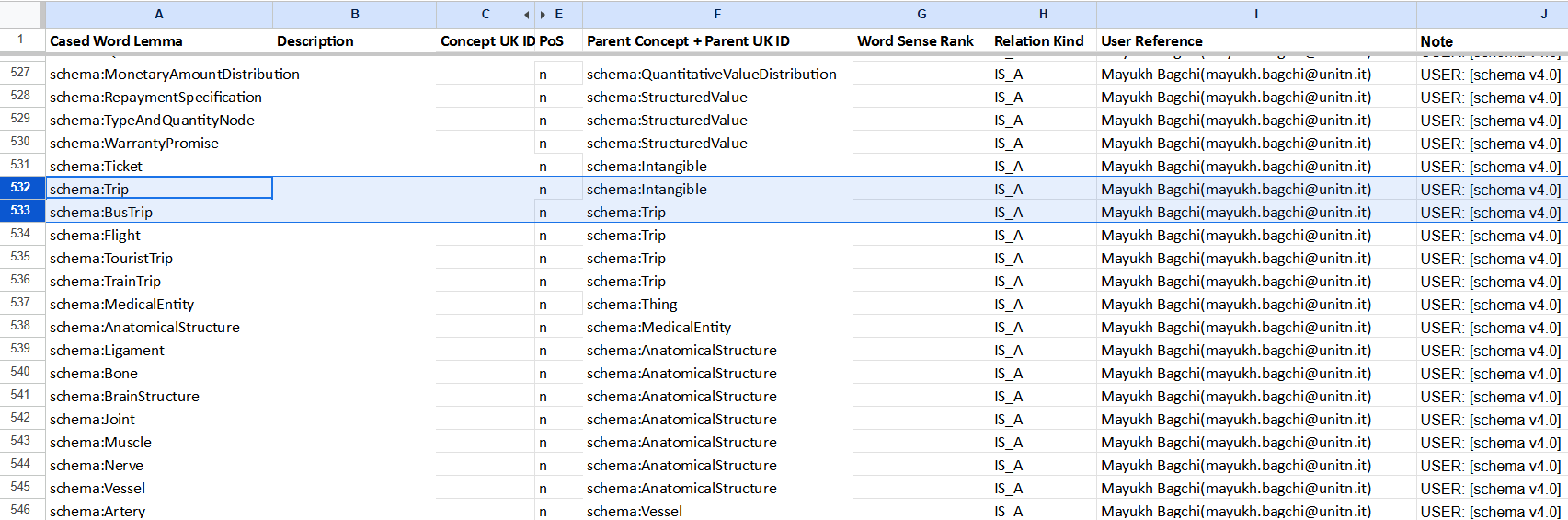}
\caption{A fragment of the intermediate spreadsheet format of schema.org}
\label{ilr}
\end{figure}

Second, as from the Figure \ref{ukcan}, the step from the world wide web to informal language resource involves a transformation of the raw format in which the resource is obtained to an intermediate spreadsheet format which is suitable for executing the next stage of the UKC annotation process. An illustrative example of an intermediate spreadsheet file of a fragment of schema.org\footnote{https://schema.org/} is provided in Figure \ref{ilr}. The first column \texttt{Cased Word Lemma} lists sequentially (the hierarchy of) concept label(s) of the selected informal language resource which are to undergo UKC annotation. The fourth and the fifth columns record the part of speech category and the parent concept of the selected concept, respectively. The seventh column records the ontological relation existing between the selected concept label and its parent. Finally, the eighth and the ninth columns provide the user reference information and the version associated to each selected concept label. Notice also the fact that the second, third and sixth columns are blank at this stage and would be filled up during the next step of UKC annotation. In the figure, an exemplification of the values for the aforementioned columns are highlighted for \texttt{schema:Trip} and \texttt{schema:BusTrip}.

Let us now elucidate the step-by-step UKC annotation process (illustrated as a tringle in Figure \ref{ukcan}) implemented on the input informal language resource encoded in the intermediate spreadsheet format. Notice that this stage involves interaction of the annotator with the UKC (illustrated as a cylinder in Figure \ref{ukcan}) via a look-up service (illustrated as a bidirectional labelled edge between the UKC annotation and the UKC in Figure \ref{ukcan}) which allows the annotator to navigate the lexical-semantic hierarchy of the UKC and search for the existence of synonymous terms. The steps are as follows:
\begin{enumerate}
    \item The annotation should ideally be performed by an annotator under the guidance and supervision of a \emph{domain expert} knowledgeable in the topic of the informal language resource. Each concept from the selected informal language resource is either annotated with its GID if it is already present in the CC, or a new concept with a new GID is created in the CC if the concept is new with respect to the CC. Concretely, the sub-steps are as follows - 
        \begin{enumerate}
        \item Each concept (term) from the informal language resource’s term hierarchy is considered (sequentially; one term at a time) in a top-down order, and its sense is understood from the gloss provided in the for the term and/or from its position relative to its immediate parent in the classification hierarchy.
        \item The concept is semantically searched in the UKC CC via the LC interface matching with the natural language in which the concept label is expressed. The search results in one of the following two scenarios:-
        \begin{enumerate}
            \item \emph{(S1): Synonymous Match} between the concept and an existing concept found in the UKC CC.
            \item \emph{(S2): No Synonymous Match} between the concept and any potential concept in the UKC CC.
        \end{enumerate}
        \item Concurrently with the above steps, a UKC-compatible partially annotated spreadsheet capturing requisite information is generated. The principle information to be recorded in the spreadsheet are as follows:-
        \begin{enumerate}
            \item \emph{(S1)}: In case of a \emph{Synonymous Match}, the GID and the Word Sense Rank of the concept have to be recorded in the spreadsheet, alongside its parent concept (and its GID).
            \item \emph{(S2)}: In case of \emph{No Synonymous Match}, a negative integer is recorded in place of the concept's GID. For all such concepts, the recorded negative integer should follow a decremental (negative integer) sequence starting from `-1'. In addition, the parent concept (and its GID) is also recorded. Further, for this scenario in specific, the gloss of the concept should also be recorded in the spreadsheet (respecting \emph{Genus-Differentia} paradigm \cite{GD}). Notice that the spreadsheet at this stage is termed as partially annotated due to the fact that it encodes a mix of terms uniquely identified with GIDs and terms with temporarily assigned negative integers.
        \end{enumerate}
    \end{enumerate}
    \item UKC enrichment: The partially annotated spreadsheet is imported into the UKC CC via the excel importer Application Programming Interface (API), wherein after successful import, two highlights are crucial:
    \begin{itemize}
        \item all the new concepts (annotated with negative integers) now have their own, unique GIDs, and
        \item  the entirety of the term hierarchy is formalized within the directed acyclic graph encoding the overall lexical-semantic hierarchy of the UKC. At this stage, the exporter service of the UKC can be utilized to generate a fully annotated spreadsheet which is termed as the UKC namespace. The notion of a UKC namespace and a domain language is discussed and illustrated in the next section of this chapter.
    \end{itemize}
\end{enumerate}

\begin{figure}[htp]
\centering
\includegraphics[width=16cm,height=6cm]{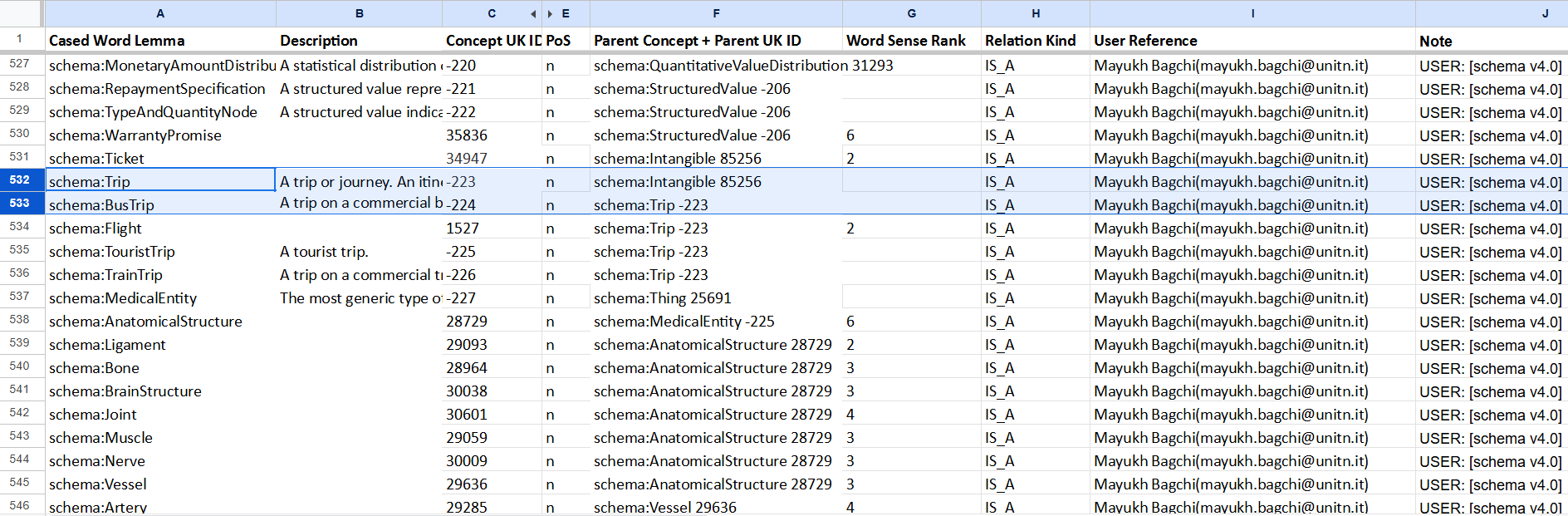}
\caption{A fragment of the partially annotated spreadsheet of schema.org}
\label{sch}
\end{figure}

An illustrative example of a partially annotated spreadsheet of a fragment of schema.org is provided in Figure \ref{sch}. The first column \texttt{Cased Word Lemma} lists sequentially (the hierarchy of) concept label(s) of the selected language resource which underwent (partial) UKC annotation. The second and the third columns which were previously kept blank in the intermediate spreadsheet is now filled during the UKC annotation with a gloss of the relative term and the concept GID (referred to as the \texttt{Concept UK ID}) following the step (1) of the UKC annotation process described above. The fourth and the fifth columns record the part of speech category and the parent concept as well as the parent concept GID of the selected concept/term, respectively. The sixth column records the word sense rank relative to the chosen concept. The seventh column records the ontological relation existing between the selected concept label and its parent. Finally, the eighth and the ninth columns provide the user reference information and the version associated to each selected concept label. In the figure, an exemplification of the values for the aforementioned columns are highlighted for \texttt{schema:Trip} and \texttt{schema:BusTrip}. For the specific case of these two terms, the concept IDs are temporarily specified as sequentially negative integers as no synonymous match could be found with respect to the UKC instance with which the UKC annotation was executed. To that end, as detailed in the above annotation process, the gloss of these two terms are also recorded in the spreadsheet and the word sense rank is blank for them as per UKC requirements.

\begin{figure}[htp]
\centering
\includegraphics[width=16cm,height=6cm]{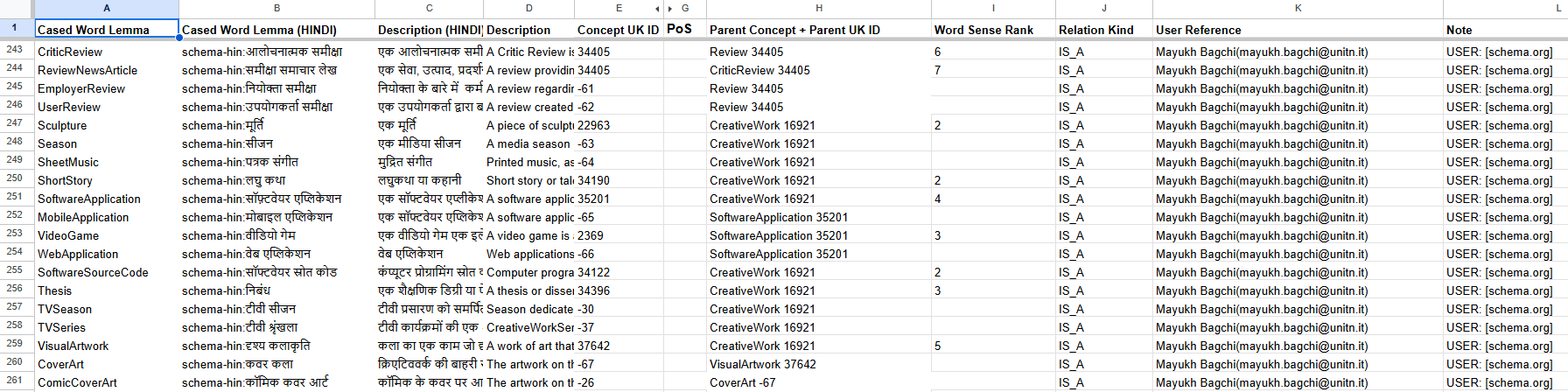}
\caption{A part of a preliminary translation spreadsheet of schema.org in Hindi.}
\label{hin}
\end{figure}

\noindent Let us briefly note a few observations which are important to properly appreciate the importance of the above pipeline. Firstly, via the UKC annotation, we concretely annotate an informal language resource with GIDs via the relevant LC interface and, thereby, accommodate both conceptual unity/diversity and language unity/diversity. Secondly, in the spreadsheet, we capture each term in the form \emph{`namespace prefix: term'} (e.g., \texttt{schema:Trip}) to identify the informal language resource from where the term originates, and, within a synset, to reflect the fact that domain terminology is grounded in natural language. We standardize the namespace prefix by referring to prefix.cc\footnote{prefix.cc} wherever available. Thirdly, due to the very organization of the UKC, we've multilinguality of terms for free and the potentiality to execute the UKC annotation process for informal language resources grounded in different natural languages. Please refer to Figure \ref{hin} for an illustration of a preliminary effort to translate schema.org namespace to the Hindi language for (partially) executing the UKC annotation. As of current date, this is an open issue and a motivation for future research on language representation as described in this chapter. Finally, note that there is a dedicated internal specification and mechanism to keep track of UKC provenance and metadata, e.g., metadata concerning the version and instance of the UKC employed for a specific UKC annotation exercise.

\section{UKC Namespaces and Domain Languages}
Let us now define the notion of a UKC namespace. A UKC namespace is a formal language representation, generated top-down, which encodes a hierarchically related set of uniquely identifiable concept names, via unique UKC GIDs, in which each concept name has a single unambiguous meaning and represents potential entity types and properties in one or multiple allied topics. A consequence of the above definition is the fact that a UKC namespace is fully aligned to the lexical-semantic hierarchy of the UKC and conforms to the language representation formalism of the UKC. The concepts composing a UKC namespace can have its origins either in a legacy (web) namespace (e.g., a W3C namespace), a standard, a classification scheme or an informal specification refactored into one of the former informal language representations. UKC Namespaces may be themselves, on requirement, be monotonically extended to extend the UKC in terms of new concepts and words occurring in a (new) universe of discourse and/or allied topics. Notice that a UKC namespace is an intermediate language representation employed to enrich and expand the UKC concepts and terminological base.

\begin{figure}[htp]
\centering
\includegraphics[width=16cm,height=10cm]{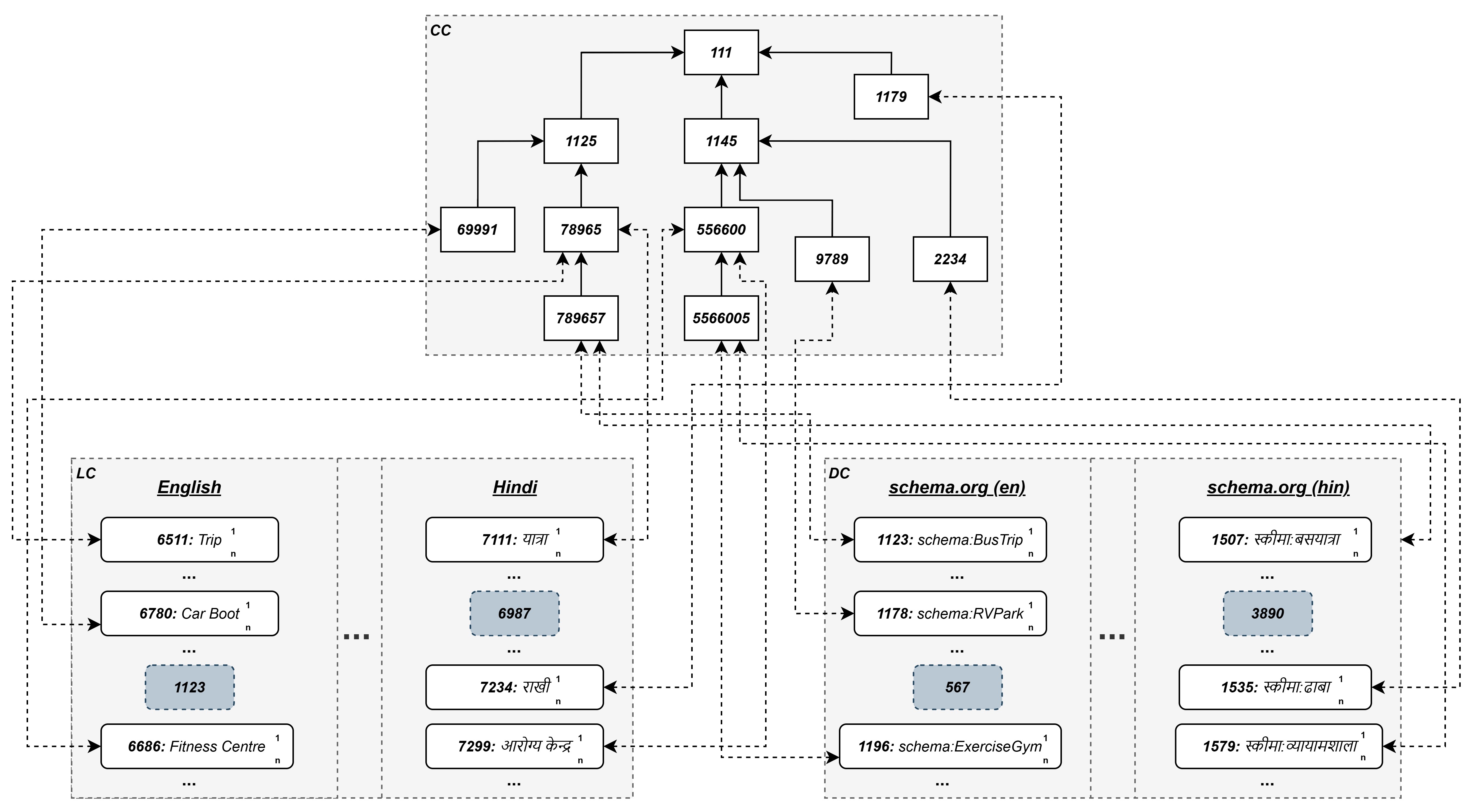}
\caption{An illustrative example of a fragment of a UKC Namespace.}
\label{ukcns}
\end{figure}

An informal diagrammatic example of a fragment of a UKC namespace is provided in Figure \ref{ukcns} which is generated after the UKC enrichment step of the UKC annotation process. In the figure, the Concept Core of the UKC (upper-half of the figure denoted by \texttt{CC}) is the language-independent ontological representation of the world as it explicitly represents the shared conceptualization as to what exists via the representation of supralingual concepts (illustrated as rectangles with a number) identified by their respective UKC GIDs, e.g., \texttt{1125, 69991}. Note that the the CC is structured as a directed acyclic graph with the directional connectors in the CC illustrating the \texttt{IS A} semantic relation which represent the subsumption relation at the concept level, e.g., between the concepts \texttt{1125} and \texttt{69991}. Further, the Language Core (lower-half of the figure denoted by \texttt{LC}) is the lexical representation of the world as it explicitly represents the shared conceptualization as to what we can name and define within the natural language (identified as English and Hindi within the \texttt{LC} component on the lower left-half of the figure). Finally, the Domain Core (DC) is the lexical representation of the world as it explicitly represents the shared conceptualization as to what we can name and define within specific/allied topics as exemplified by the UKC namespace schema.org (identified as English (\texttt{en}) and Hindi (\texttt{hin}) within the \texttt{DC} component on the lower right-half of the figure), both in the English language and the Hindi versions, respectively. It is modelled and visualized in terms of synsets illustrated as rounded rectangles with a number and a label, e.g., \texttt{6511: Trip} and \texttt{1123: schema:BusTrip}, with the prefixed identifier denoting the language-specific identifier of the synset (e.g., \texttt{6511}, \texttt{1123}), the label denoting the preferred word employed to refer to the synset (e.g., \texttt{Trip}, \texttt{schema:BusTrip}) and the index range \(1(1)n\) illustrating the fact that a synset can be composed of \texttt{n} number of synonymous words. It is important to note that the figure also illustrates how the UKC seamlessly integrates and interlinks both natural languages and UKC namespaces into a unified formal language representation.

\begin{figure}[htp]
\centering
\includegraphics[width=16cm,height=6cm]{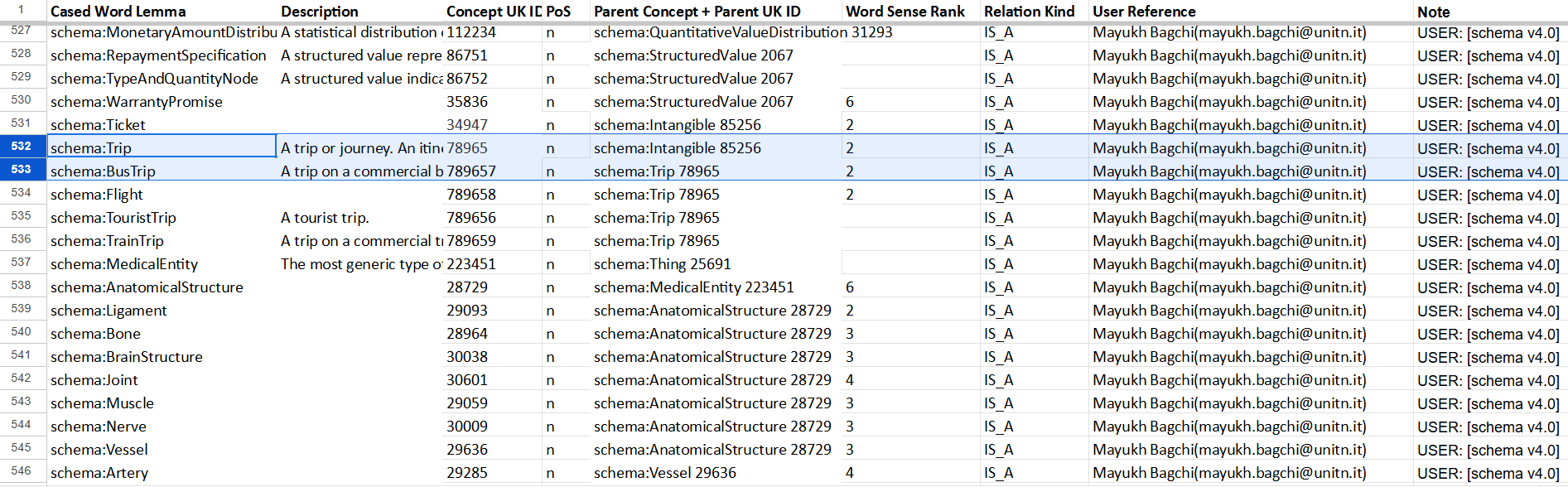}
\caption{A fragment of the fully annotated spreadsheet of schema.org.}
\label{flr}
\end{figure}

Further, an illustrative example of a fully annotated spreadsheet view of a fragment of schema.org UKC namespace is provided in Figure \ref{flr} which is generated after the UKC enrichment step of the UKC annotation process. The first column \texttt{Cased Word Lemma} lists sequentially (the hierarchy of) concept label(s) of the selected language resource which underwent UKC annotation. The second and the third columns which were previously kept blank in the intermediate spreadsheet is now filled during the UKC annotation with a gloss of the relative term and the concept GID following the UKC annotation process described before. The fourth and the fifth columns record the part of speech category and the parent concept as well as the parent concept GID of the selected concept/term, respectively. The sixth column records the word sense rank relative to the chosen concept. The seventh column records the ontological relation existing between the selected concept label and its parent. Finally, the eighth and the ninth columns provide the user reference information and the version associated to each selected concept label. In the figure, an exemplification of the values for the aforementioned columns are highlighted for \texttt{schema:Trip} and \texttt{schema:BusTrip}. Differently from what is illustrated in Figure \ref{sch}, the two terms \texttt{schema:Trip} and \texttt{schema:BusTrip} now have unique GIDs \texttt{78965} and \texttt{789657}, respectively. To that end, as detailed in the UKC annotation process, the gloss as well as the word sense rank of these two terms are now fully recorded in the spreadsheet as per UKC requirements.

\begin{figure}[htp]
\centering
\includegraphics[width=16cm,height=18cm]{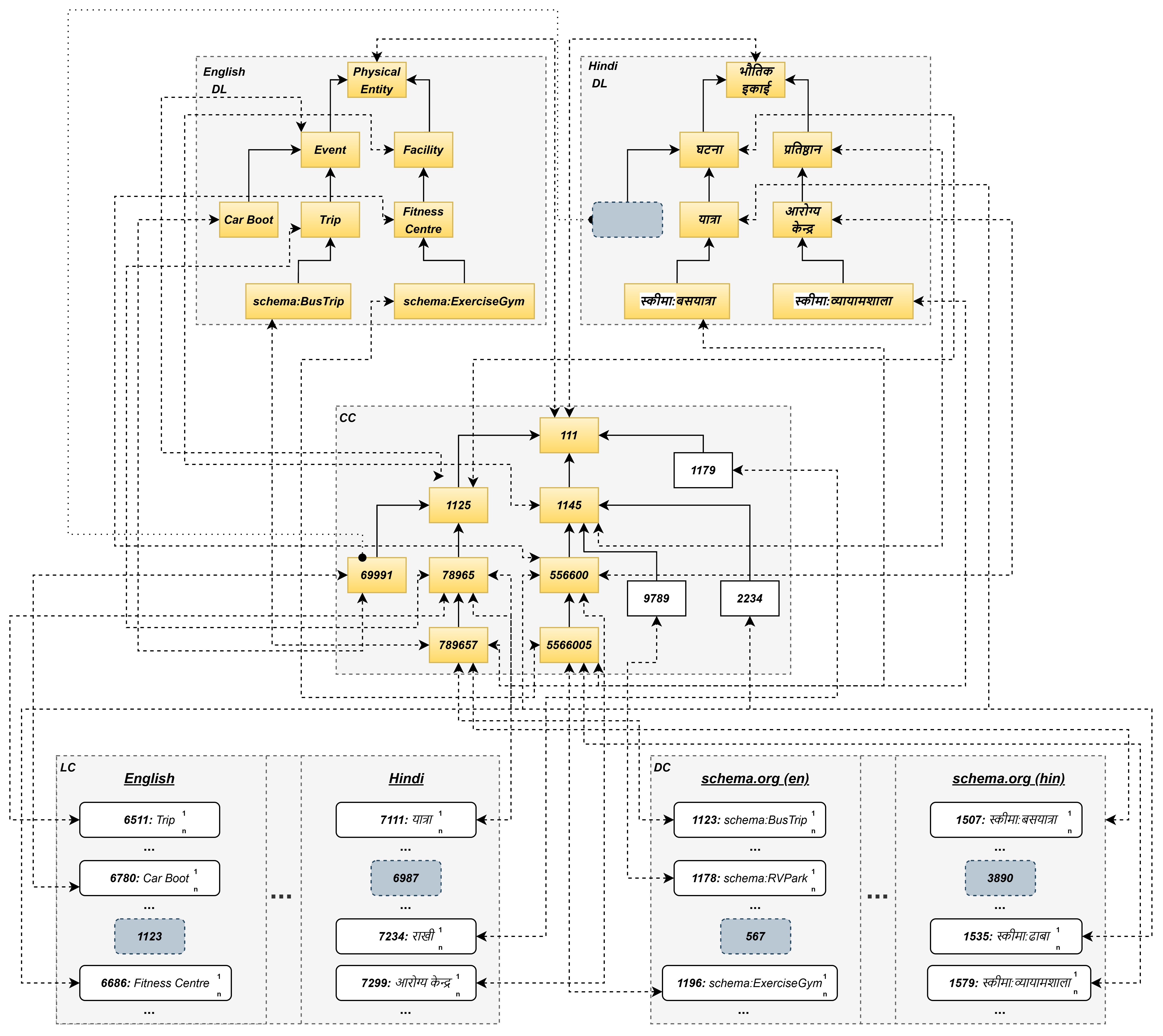}
\caption{An illustrative example of a fragment of a UKC Domain Language.}
\label{ukcdl}
\end{figure}

Given the design and integration of UKC namespaces within the UKC lexical-sematic hierarchy, a domain language is defined as a formal language representation encoding hierarchically related set of uniquely identifiable concept names, via unique UKC GIDs, in which each concept name has a single unambiguous meaning and represents potential entity types and properties across one or multiple natural languages and/or UKC namespaces relevant to a specific and/or allied topic(s). It is a sub-hierarchy of the UKC lexical-semantic hierarchy, i.e., it represents what exists and what can be named in one or more topic(s) out of the totality of the UKC (depending on the requirements of an application scenario). Domain Languages may be composed of (parts of) different topics and therefore include (parts of) diverse UKC namespaces and/or natural languages. It can also include external words which are native to an application scenario within a topic and not part of any existing UKC namespace. In essence, a domain language can be seen as a UKC-based formal topic with an objective to formalize informal terminology. 

An informal diagrammatic example of a fragment of a UKC domain language is provided in Figure \ref{ukcdl} which is same as the figure \ref{ukcns} extended with examples of domain languages on the top-half of the figure. Notice that in the figure, the Concept Core of the domain language is identified via the representation of a sub-hierarchy (illustrated as rectangles in yellow ochre with a number) of the overall hierarchy of supralingual concepts (illustrated as rectangles with a number) identified by their respective UKC GIDs. The two domain languages on the upper portion of the figure, namely, the English domain language (identified in the figure by \texttt{English DL}) and the Hindi domain language (identified in the figure by \texttt{Hindi DL}) is developed based on the aforementioned CC and involves a mix of terms from both natural languages and UKC namespaces. For example, the English domain langauge is modelled and visualized in terms of synsets illustrated as yellow-ochre coloured rectangles with a label, e.g., \texttt{Trip}, \texttt{schema:BusTrip}, with the unique identification of the terms recorded via a dedicated internal provenance recording mechanism.

\section{Related Work}
In terms of related work on language representations, let us first concentrate on state-of-the-art lexical resources. The Princeton WordNet (PWN) \cite{PWN,PWN2} is a lexical resource in the English language which models synsets across part of speech categories, e.g., nouns, which are, in turn, linked through semantic relations to determine word definitions. Based on the fundamental design of PWN, several multilingual lexical resources have been developed such as BabelNet \cite{BNET} as well as the EuroWordNet \cite{EWN} and the IndoWordNet \cite{IWN} for European and Indian languages, respectively. While the design of the lexical hierarchies in the UKC LC is similar to that of the aforementioned resources, the UKC CC provides a major advance over the aforementioned resources in facilitating a language-independent top-level ontological representation of concepts which can natively tackle conceptual and language heterogeneity in language representations. There is also considerable research work in information sciences on knowledge organization-based classifications \cite{CLS}, (domain) standards such as in healthcare \cite{STN}, controlled vocabularies \cite{CVC} and controlled natural languages \cite{CNL}, all of which are important informal language representations and, on a case-by-case basis, can be reused and transformed into UKC-based formal language representations like UKC namespaces and domain languages in order to specifically tackle conceptual and language heterogeneity.

Next, let us focus on the landscape of foundational ontologies (also known as top-level ontologies (TLOs)). The principles and distinctions inherent in foundational ontologies and their crucial role in rendering ontological commitments explicit and ontologically well-founded have been widely explored (see, for e.g., \cite{LO,2009-Guarino,OC,FO-Guizzardi}). The review in \cite{TLO-mascardi} analyze seven upper ontologies according to a select set of software engineering criteria, amongst which we brief below the most used ones. The TLO advanced in \cite{DOLCE} is a comprehensive upper \emph{``ontology of particulars"} with a clear cognitive bias grounded in the fundamental distinction between \emph{endurants} and \emph{perdurants}. The TLO proposed in \cite{BFO} is a bi-ontological theory with two components - a \emph{Snap} ontology of endurants and a \emph{Span} ontology of perdurants, and is majorly employed within the biomedical ontology community. Next, the work in \cite{SUMO} develops a TLO which acts a foundation for domain ontologies and contains about thousand terms grounded in the distinction between physical and abstract entities. The TLO presented in the work \cite{GFO} is geared towards biomedical applications and is organized into specialized modules, for instance, for formally representing biological functions. Amongst the more recent TLOs, the work in \cite{UFO} proposes a foundational ontology which is heavily grounded in philosophy, linguistics and cognitive psychology, and is composed of three distinct ontology fragments - endurants, perdurants and social entities. For the work advanced in this thesis, the top-level language-independent ontological representation of concepts in the UKC CC is relevant over the aforementioned TLOs as the UKC CC, by its very design, allows for accommodation of representation heterogeneity across concepts and multiple languages.

\section{Summary}

To summarize, given the ubiquitous problem of addressing and accommodating representation heterogeneity at the conceptual and the language level, this chapter proposed a language representation for modelling the meaning of words and how they relate to each other. To that end, the chapter discussed and exemplified the multilingual lexical-sematic resource named the Universal Knowledge Core (UKC), the notion of topic, the UKC annotation process, the notion of a UKC namespace and domain language. The next chapter will concentrate on describing and illustrating knowledge representation based on the UKC language representation as outlined in this chapter.

    \chapter{KNOWLEDGE REPRESENTATION}

\section{Introduction}
In the context of the thesis, knowledge representation can be understood as the formal representation and definition of entity types (classes of entities that share common properties), object properties (interrelating entity types) and data properties (describing entity types) using terms defined from UKC domain languages. As highlighted in chapter (2) on the formalism developed for the stratification of representation, knowledge representation is key to model the underlying knowledge unity/diversity (see figure \ref{fig:RDN3}) which, in turn, is core to address and accommodate the overarching issue of representation heterogeneity. To that end, this chapter elucidates and exemplifies the design of knowledge representations in terms of:
\begin{itemize}
    \item Language teleontology, modelling the uniquely identifiable concept names relevant to model entity types and properties, and,
    \item Knowledge teleontology, modelling the entity types and properties using the entity type names and property names as modelled in its reference language teleontology.
\end{itemize}
Note that throughout this chapter and in later illustrations, the figures visualizing fragments of language teleontologies and knowledge teleontologies are screenshots generated using relevant tabs in the Protégé\footnote{https://protege.stanford.edu} free and open-source ontology editor and framework. Let us now explain and exemplify each of the above aspects in greater detail.

\section{Language Teleontology}
A Language Teleontology is defined as a knowledge representation formally representing, reusing the semantically disambiguated words of domain language(s) modelled in the Language Representation phase, the uniquely identifiable concept names relevant to model entity types and properties relevant to the scope of a knowledge graph engineering project\footnote{For numerous examples of such projects implemented within an academic setting, please visit: http://knowdive.disi.unitn.it/teaching/kdi/}. A language teleontology uses three primary representational constructs: 
\begin{itemize}
    \item entity type names for modelling the concept names which will, at a later stage, be defined as entity types,
    \item object property names for modelling the concept names which will, at a later stage, be defined as entity types, and
    \item data property names for modelling the concept names which will, at a later stage, be defined as entity types.
\end{itemize}
It is important to note some initial observations about the notion of a language teleontology. First, a language teleontology should be aligned to the lexical-semantic hierarchy of the UKC and, as a consequence, a language teleontology is grounded in a natural language as well as commits to a language-independent ontology. Second, a language teleontology can be constructed as a combination of one or more domain languages already integrated into the UKC. It can also be the case that a domain language, in addition to domain language words, also encode natural language words and very specific contextual words which are not integrated into the UKC LC knowledge base. In the latter case, however, such context specific words should mandatorily undergo the UKC annotation process in order for them to be semantically disambiguated with a GID and, thereby, fully integrated into the UKC LC. Third, a language teleontology is essentially an intermediate artifact which acts as a bridge between language representation in the UKC and knowledge representation, namely, knowledge teleontologies (which will be described in the next section of this chapter). The file format we adopt for encoding a language teleontology in the context of this thesis is \texttt{OWL RDF/XML}\footnote{https://www.w3.org/TR/rdf10-xml/}. Finally, each element in a language teleontology has a KnowDive internationalized resource identifier with the generic expression \texttt{http://knowdive.disi.unitn.it/etype\#element\_name}. Let us now consider in greater detail each of the three representation constructs which mutually define a language teleontology.

\begin{figure}[htp]
\centering
\includegraphics[width=16cm,height=8cm]{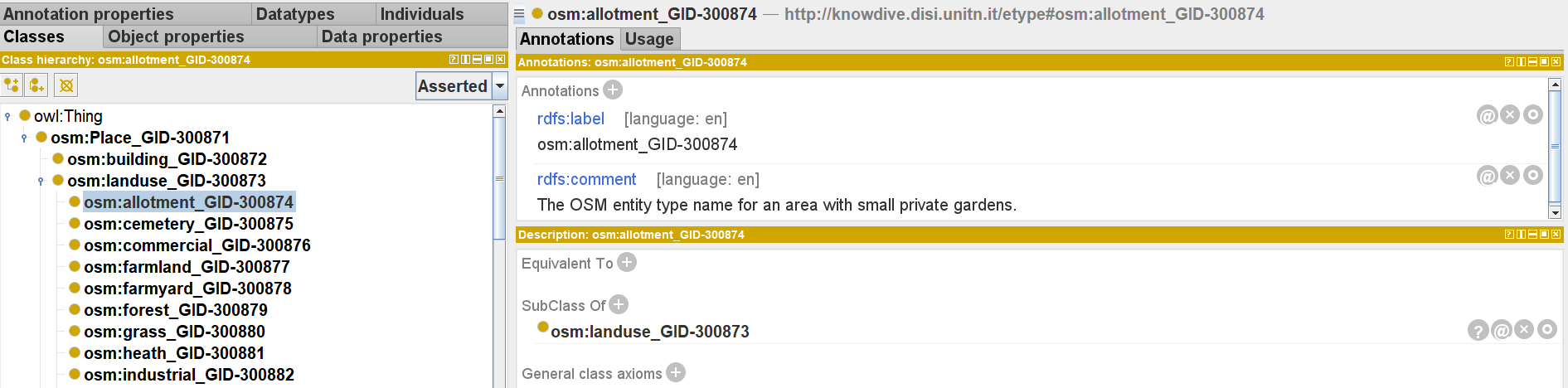}
\caption{A fragment of the entity type names of the OSM language teleontology.}
\label{lte}
\end{figure}

\textbf{Entity Type Name:} A Language Teleontology represents, using (partially) the words of one or more domain language(s) modelled in the Language Representation phase, the uniquely identifiable concept names relevant to model entity types relevant to represent the various classes of entities that share common attributes within and/or across the scope of a topic and/or allied topics. The entity type names are organized as a hierarchy represented in the syntax of \texttt{OWL RDF/XML}. It is important to notice some key characteristics of entity type names in a language teleontology. First, the meaning of each entity type name is uniquely identified and disambiguated by a UKC GID appended to the label of the name. In majority of the cases, the label of the entity type names also display the prefix of the UKC namespace (and, therefore, the domain language) from which it has been reused. Second, the entity type name should preferably be annotated with information such as comments (including textual definitions/descriptions), examples, external references, etc., which themselves can be modelled as annotation properties. Third, the entity type name should not be defined, i.e., there should be no associated axiomatic assertions and/or property constraint, e.g., disjointness assertions, class expressions via asserted property constraints, etc. The last characteristic is in line with the fact that a language teleontology enables unity out of diversity of topic-specific entity type names and does not focus on defining them with property constraints and axioms.

Let us illustrate the above elucidation of the entity type names in a language teleontology using the example of a fragment of the entity type names of the OSM language teleontology visualized in the Figure \ref{lte}. The entity type names are organized as a hierarchy as visualized from the left-half of the figure. Further, the meaning of each entity type name (e.g., \texttt{osm:allotment}) is uniquely identified and disambiguated by a UKC GID (e.g., \texttt{300874}) appended to the label of the name (e.g., \texttt{osm:allotment\_GID-300874}). The label of the entity type names also display the prefix (e.g., \texttt{osm}) of the UKC namespace from which it has been reused. The entity type name is annotated with information such as an \texttt{rdfs:comment} on the right-half of the figure modelled as an annotation property. Further, notice the entity type name is not defined, i.e., there are no associated axiomatic assertions and/or property constraint, e.g., disjointness assertions, class expressions via asserted property constraints, etc., associated to the selected entity type name \texttt{osm:allotment\_GID-300874}. 

\begin{figure}[htp]
\centering
\includegraphics[width=16cm,height=8cm]{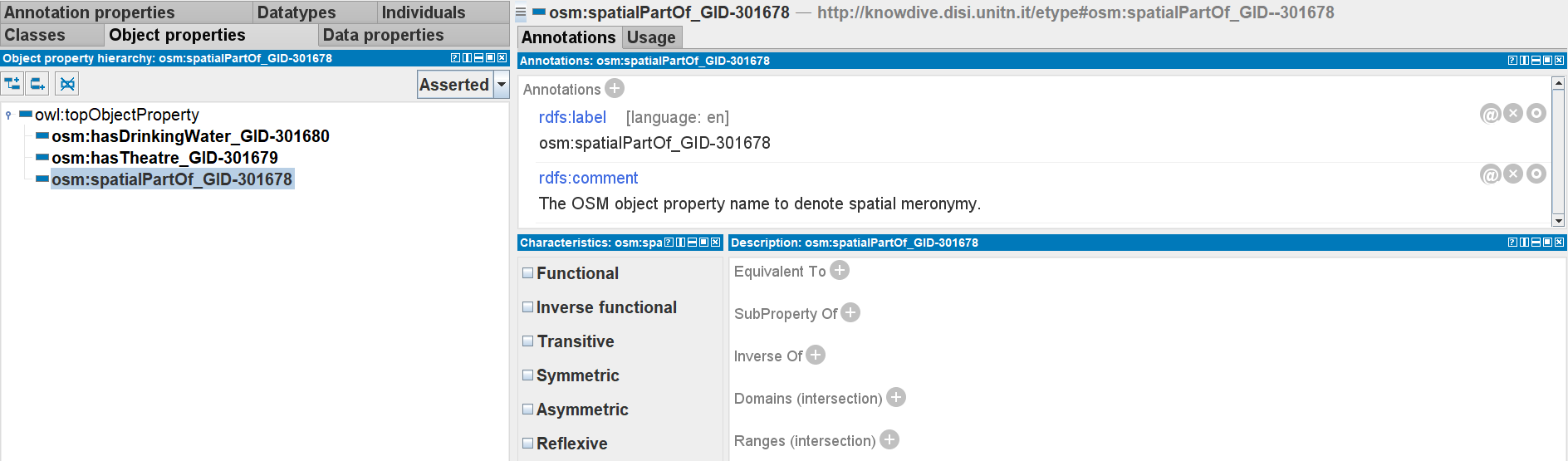}
\caption{A view of the object property names of the OSM language teleontology.}
\label{lto}
\end{figure}

\textbf{Object Property Name:} Similar to the above, a Language Teleontology represents, using (partially) the words of one or more domain language(s) modelled in the Language Representation phase, the uniquely identifiable concept names relevant to model object properties relevant to represent the interrelationships between entity types within and/or across the scope of a topic and/or allied topics. The object property names, similarly as before for the entity type names, are organized as a hierarchy represented in the syntax of \texttt{OWL RDF/XML}. Further, it is important to notice some essential features of object property names in a language teleontology. First, the meaning of each object property name, similar to the case of entity type names, is uniquely identified and disambiguated by a UKC GID appended to the label of the name. In majority of the cases, the label of the object property names also display the prefix of the UKC namespace (and, therefore, the domain language) from which it has been reused. Second, the object property names, similar to the case of entity type names, should preferably be annotated with relevant information such as comments (including textual definitions/descriptions), examples, external references, etc., which themselves can be modelled as annotation properties. Third, the object property names should not be associated to axiomatic assertions, e.g., characteristic assertions, domain and/or range assertions, etc. The last feature is in line with the fact that a language teleontology enables unity out of diversity of topic-specific object property names and does not focus on describing them with further axiomatic assertions.

Let us illustrate the above description of the object property names in a language teleontology using the example of a fragment of the object property names of the OSM language teleontology visualized in the Figure \ref{lto}. The object property names are organized as a hierarchy as visualized from the left-half of the figure. Further, the meaning of each object property name (e.g., \texttt{osm:spatialPartOf}) is uniquely identified and disambiguated by a UKC GID (e.g., \texttt{301678}) appended to the label of the name (e.g., \texttt{osm:spatialPartOf\_GID-301678}). The label of the object property names also display the prefix (e.g., \texttt{osm}) of the UKC namespace from which it has been reused. The object property name is also annotated with information such as an \texttt{rdfs:comment} on the right-half of the figure modelled as an annotation property. Finally, it is interesting to note that the object property name is not associated to axiomatic assertions, e.g., characteristic assertions, domain and/or range assertions, etc, associated to the selected entity type name \texttt{osm:spatialPartOf\_GID-301678}. 

\begin{figure}[htp]
\centering
\includegraphics[width=16cm,height=8cm]{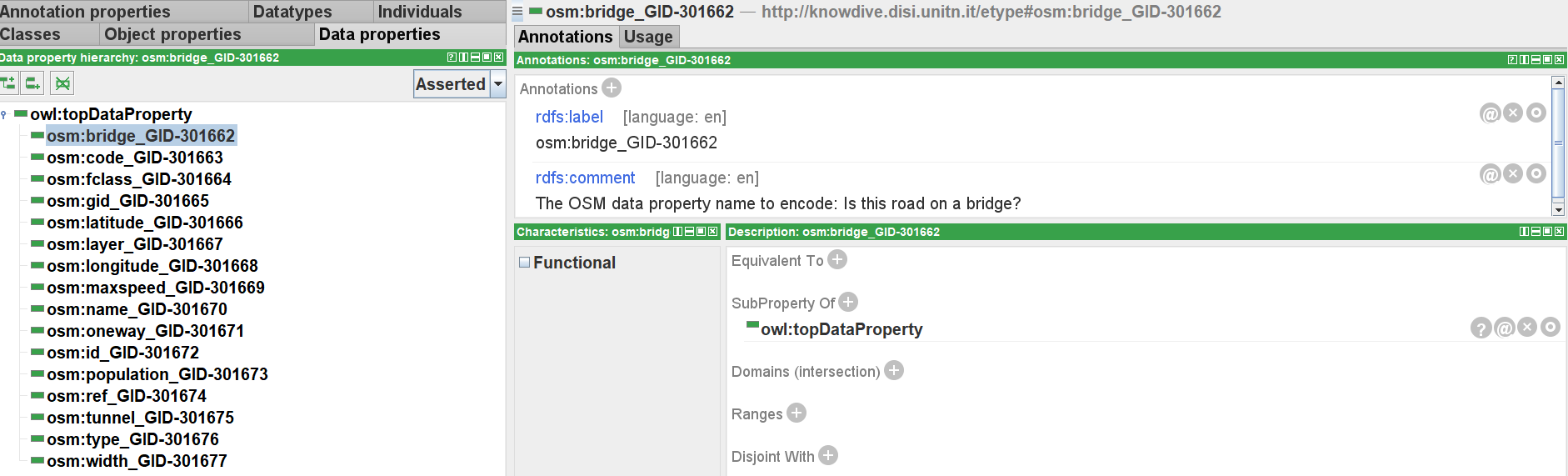}
\caption{A view of the data property names of the OSM language teleontology.}
\label{ltd}
\end{figure}

\textbf{Data Property Name:} Finally, a Language Teleontology represents, using (partially) the words of one or more domain language(s) modelled in the Language Representation phase, the uniquely identifiable concept names relevant to model data properties relevant to describe the attributes of entity types relevant within and/or across the scope of a topic and/or allied topics. The data property names, similarly as before for the entity type names and object property names, are organized as a hierarchy represented in the syntax of \texttt{OWL RDF/XML}. Further, it is important to note some highlights about data property names in a language teleontology. First, the meaning of each data property name, similar to the case of entity type names and object property names, is uniquely identified and disambiguated by a UKC GID appended to the label of the name. In majority of the cases, the label of the data property names also display the prefix of the UKC namespace (and, therefore, the domain language) from which it has been reused. Second, the data property name, should preferably be annotated with relevant information such as comments (including textual definitions/descriptions), examples, external references, etc., which themselves can be modelled as annotation properties. Third, the data property names, similar to the cases described before, should also not be associated to axiomatic assertions, e.g., characteristic assertions, disjointness assertions, domain and/or range assertions, data type assertions, etc. The last feature is in line with the fact that a language teleontology enables unity out of diversity of topic-specific data property names and does not focus on describing them with further axiomatic assertions.

Let us illustrate the above description of the data property names in a language teleontology using the example of a fragment of the data property names of the OSM language teleontology visualized in the Figure \ref{ltd}. The data property names are organized as a hierarchy as visualized from the left-half of the figure. Further, the meaning of each data property name (e.g., \texttt{osm:bridge}) is uniquely identified and disambiguated by a UKC GID (e.g., \texttt{301662}) appended to the label of the name (e.g., \texttt{osm:bridge\_GID-301662}). The label of the data property names also display the prefix (e.g., \texttt{osm}) of the UKC namespace from which it has been reused. The data property name is also annotated with information such as an \texttt{rdfs:comment} on the right-half of the figure modelled as an annotation property. Lastly, it is interesting to note that the data property name is not associated to axiomatic assertions, e.g., characteristic assertions, disjointness assertions, domain and/or range assertions, data type assertions, etc, associated to the selected entity type name \texttt{osm:bridge\_GID-301662}. 

\section{Knowledge Teleontology}
A Knowledge Teleontology is defined as a knowledge representation which, reusing the concept names formalized in a language teleontology, defines entity types, object properties modelling the interrelationships between entity types and and data properties encoding descriptions of an entity type, within the scope of a topic which can be exploited in potential knowledge graph engineering projects relevant to that topic or allied topics. To that end, a knowledge teleontology is defined using four primary representational constructs: 
\begin{itemize}
    \item entity types which define the various classes of entities that share common attributes,
    \item object properties which define the interrelationships amongst different entity types,
    \item data properties which describe the attributes of different entity types, and,
    \item data types which asserts the acceptable range of data properties within a knowledge teleontology.
\end{itemize}
It is important to note some initial observations about the notion of a knowledge teleontology. First, a knowledge teleontology should be compliant to the lexical-semantic hierarchy of the UKC by construction via language teleontology, and, as a consequence, a knowledge teleontology is grounded in a natural language as well as commits to a language-independent ontology. Second, a knowledge teleontology can be modelled using a top-down approach by (partially) reusing domain knowledge of domain experts, consulting (entity) dataset schemas within a topic, reusing state-of-the-art applied ontologies and/or reusing one or more relevant language teleontologies. Third, a knowledge teleontology defines the entity types, object properties and data properties relevant to a specific purpose by encoding definitional assertions, e.g., it asserts the domain and range for a particular object property or it asserts the domain and data type of a particular data property. The file format we adopt for encoding a knowledge teleontology in the context of this thesis is also \texttt{OWL RDF/XML} (as a future work, the encoding will be in a customized language \cite{azp}). Similar to that in the case of a language teleontology, each element in a knowledgee teleontology has a KnowDive internationalized resource identifier with the generic expression \texttt{http://knowdive.disi.unitn.it/etype\#element\_name}. Let us now consider in greater detail each of the representation constructs which mutually define a knowledge teleontology.

\begin{figure}[htp]
\centering
\includegraphics[width=16cm,height=8cm]{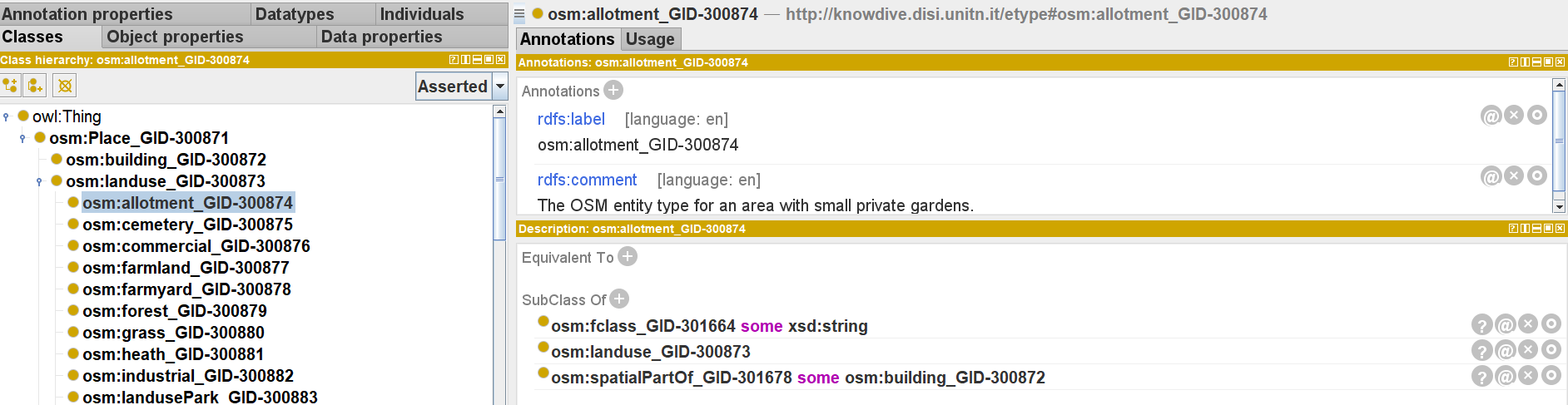}
\caption{A fragment of the entity type hierarchy of the OSM knowledge teleontology.}
\label{kte}
\end{figure}

\textbf{Entity Type:} A knowledge teleontology represents, using (partially) the entity type names of one or more language teleontology, the uniquely identifiable and dereferenceable entity types relevant to represent the various classes of entities that share common attributes within and/or across the scope of a topic and/or allied topics. The entity types are organized as a hierarchy represented in the syntax of \texttt{OWL RDF/XML}. It is important to notice some key characteristics of entity types in a knowledge teleontology. First, the meaning of the concept behind each entity type is uniquely identified and disambiguated by a UKC GID appended to the label of the entity type. In majority of the cases, the label of the entity types also display the prefix of the UKC namespace from which it has been reused. The above set of information are reused from the language teleontology from which the entity type names relative to the entity type originate. Second, the entity type should preferably be annotated with added textual annotations such as comments (including textual definitions/descriptions), external references, etc., which themselves can be modelled as annotation properties. Third, the entity types should be well-defined defined, i.e., there should be relevant axiomatic assertions and/or property constraints associated to the entity type, e.g., disjointness assertions, class expressions via asserted property constraints, etc. The last characteristic is in line with the fact that a knowledge teleontology enables unity out of diversity of topic-specific entity types and thereby focus on defining them with property constraints and axioms.

Let us illustrate the above description of the entity types in a knowledge teleontology using the example of a fragment of the OSM knowledge teleontology visualized in the Figure \ref{kte}. The entity types are organized as a hierarchy as visualized from the left-half of the figure. Further, the meaning of the name of each entity type (e.g., \texttt{osm:allotment}) is uniquely identified and disambiguated by a UKC GID (e.g., \texttt{300874}) appended to the label of the name (e.g., \texttt{osm:allotment\_GID-300874}). The label of the entity types also display the prefix (e.g., \texttt{osm}) of the UKC namespace from which it has been reused. The entity types are annotated with information such as an \texttt{rdfs:comment} on the right-half of the figure modelled as an annotation property. Further, notice the entity types are defined, i.e., there are axiomatic assertions and/or property constraint associated to the entity type using the object and data properties defined. For example, for the entity type \texttt{osm:allotment\_GID-300874}, the axiomatic descriptions include, e.g., data restriction constraint (e.g., \texttt{osm:fclass\_GID-301664 \textbf{some} xsd:string}), object restriction constraint (e.g., \texttt{osm:spatialPartOf\_GID-301678 \textbf{some} \\ osm:building\_GID-300872}), etc.

\begin{figure}[htp]
\centering
\includegraphics[width=16cm,height=8cm]{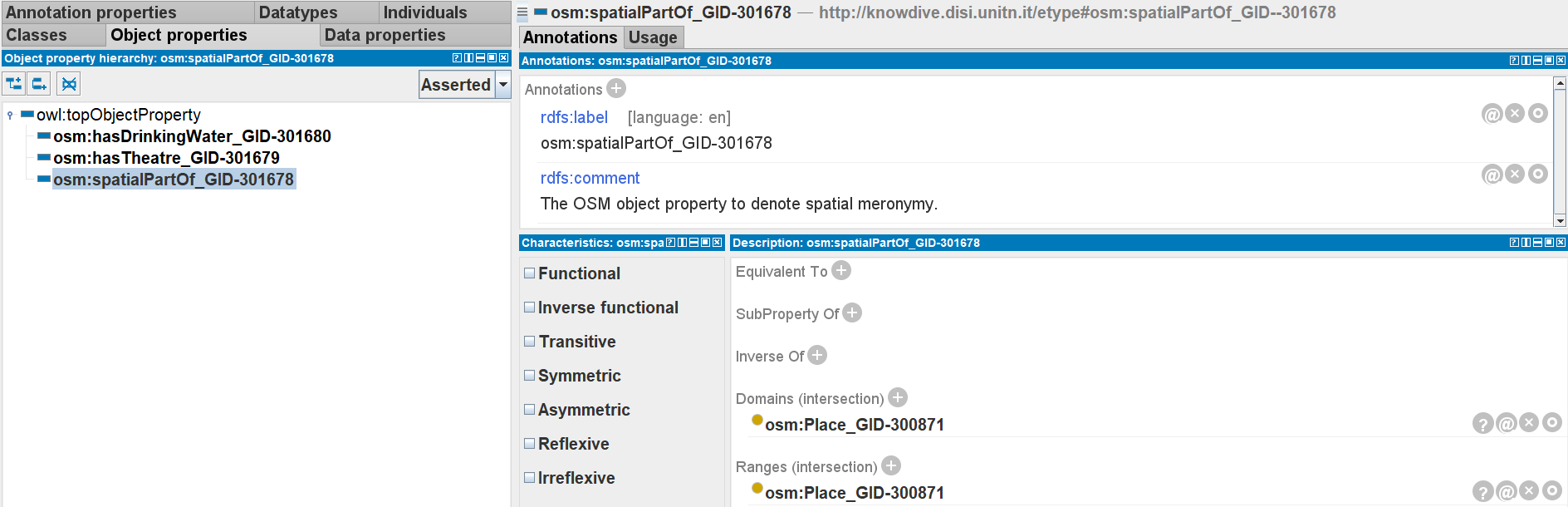}
\caption{A view of the object property hierarchy of the OSM knowledge teleontology.}
\label{kto}
\end{figure}

\textbf{Object Property:} Similar to the above, a knowledge teleontology represents, using (partially) the object property names from one or more language teleontology, the uniquely identifiable and dereferenceable object properties relevant to represent the interrelationships between entity types within and/or across the scope of a topic and/or allied topics. The object properties, similarly as before for the entity types, are organized as a hierarchy represented in the syntax of \texttt{OWL RDF/XML}. Further, the meaning of the name of each object property, similar to the case of entity types, is uniquely identified and disambiguated by a UKC GID appended to the label of the name. The label of the object property also display the prefix of the UKC namespace from which it has been reused. The above set of information are reused from the knowledge teleontology from which the object property names relative to the object property originate. The object property themselves, like the entity types, should preferably be annotated with relevant information such as comments (including textual definitions/descriptions), etc., which themselves can be modelled as annotation properties. Finally, the object properties should be associated to axiomatic assertions, e.g., characteristic assertions, disjointness assertions, domain and/or range assertions, etc. The last characteristic is in line with the fact that a knowledge teleontology enables unity out of diversity of topic-specific object properties and thereby focus on defining them with constraints and axioms.

Let us illustrate the above description of the object properties in a knowledge teleontology using the example of a fragment of the OSM knowledge teleontology visualized in the Figure \ref{kto}. The object properties are organized as a hierarchy as visualized from the left-half of the figure. Further, the meaning of each object property (e.g., \texttt{osm:spatialPartOf}) is uniquely identified and disambiguated by a UKC GID (e.g., \texttt{301678}) appended to the label of the name of the object property (e.g., \texttt{osm:spatialPartOf\_GID-301678}). The label of the object properties also display the prefix (e.g., \texttt{osm}) of the UKC namespace from which it has been reused. The object properties are also annotated with information such as an \texttt{rdfs:comment} on the right-half of the figure modelled as an annotation property. Last but not the least, it is interesting to note that the object property is well-defined using axiomatic assertions, e.g., domain assertion (\texttt{osm:place\_GID-300871}) and range assertion (\texttt{osm:place\_GID-300871}), etc, associated to the selected object property \texttt{osm:spatialPartOf\_GID-301678}. 

\begin{figure}[htp]
\centering
\includegraphics[width=16cm,height=8cm]{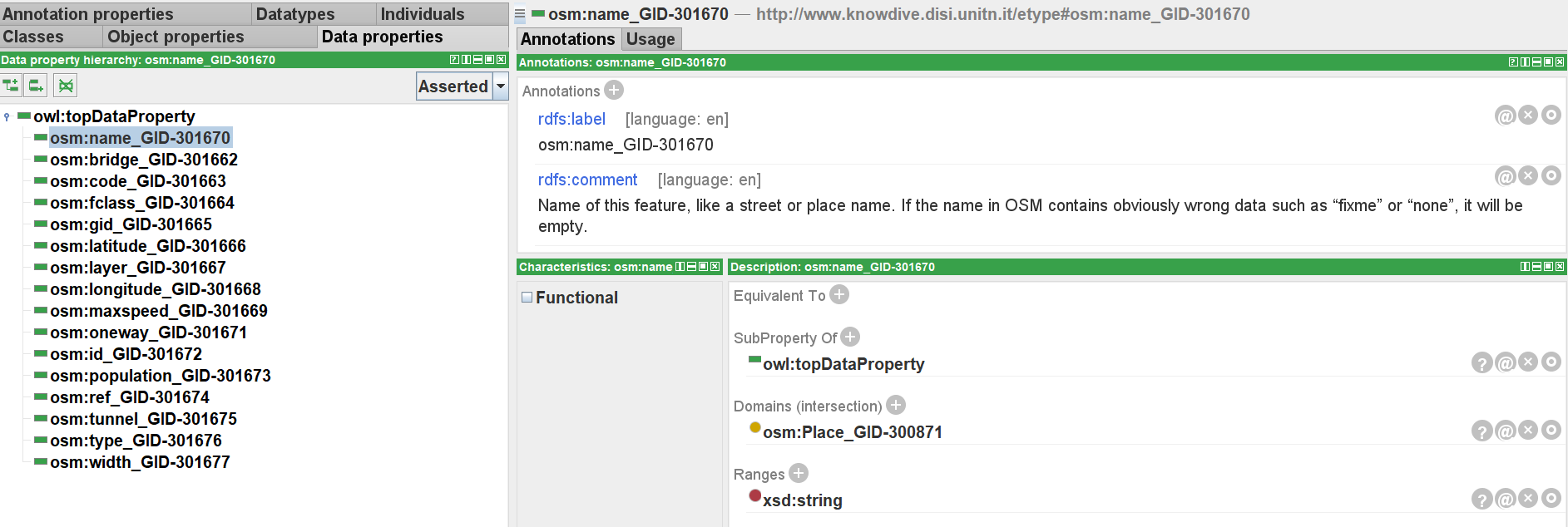}
\caption{A view of the data property hierarchy of the OSM knowledge teleontology.}
\label{ktd}
\end{figure}

\textbf{Data Property:} Finally, a knowledge teleontology represents, using (partially) the data property names from one or more language teleontology, the uniquely identifiable and dereferenceable data properties relevant to describe entity types within and/or across the scope of a topic and/or allied topics via relevant attributes. The data properties are organized as a hierarchy represented in the syntax of \texttt{OWL RDF/XML} and the meaning of the name of each data property is uniquely identified and disambiguated by a UKC GID appended to the label of the name. The label of the data property also display the prefix of the UKC namespace from which it has been reused. The above set of information are reused from the knowledge teleontology from which the data property names relative to the data property originate. Further, the data property themselves should preferably be annotated with relevant information such as comments (including textual definitions/descriptions), etc., which themselves can be modelled as annotation properties. Finally, the data properties should be associated to axiomatic assertions, e.g., characteristic assertions, disjointness assertions, domain and/or range assertions (the latter explicitly using the OWL-mandated data types or topic-specific data range expressions), etc. The last characteristic is in line with the fact that a knowledge teleontology enables unity out of diversity of topic-specific data properties and thereby focus on defining them with constraints and axioms.

Let us illustrate the above description of the data properties in a knowledge teleontology using the example of a fragment of the OSM knowledge teleontology visualized in the Figure \ref{ktd}. The data properties are organized as a hierarchy as visualized from the left-half of the figure. Further, the meaning of each data property (e.g., \texttt{osm:name}) is uniquely identified and disambiguated by a UKC GID (e.g., \texttt{301670}) appended to the label of the name of the data property (e.g., \texttt{osm:name\_GID-301670}). The label of the data properties also display the prefix (e.g., \texttt{osm}) of the UKC namespace from which it has been reused. The data properties are also annotated with information such as an \texttt{rdfs:comment} on the right-half of the figure encoded as an annotation property. Lastly, notice that the data property is well-defined using axiomatic assertions, e.g., domain assertion \texttt{osm:place\_GID-300871} and range assertion \texttt{xsd:string} with string as the chosen data type, etc, associated to the selected data property \texttt{osm:name\_GID-301670}. 

It is fascinating to note two key observations concerning knowledge teleontologies. First, a knowledge teleontology should be compliant to the hierarchy of the UKC by construction, i.e., via its alignment to one or more language teleontologies and, thereby, to the lexical-semantic hierarchy of the UKC. Second, in the overall scenario of knowledge graph engineering projects \cite{KGSWC}, a knowledge teleontology is crucial for entity type alignment \cite{ETR,etr2} necessary to generate semantically interoperable teleologies \cite{T1,T2} and the final purpose-driven schema of the knowledge graph. 

\section{Related Work}
In terms of related work on knowledge representations, let us first concentrate on mainstream research in the closely related research communities of conceptual modelling and applied ontology, respectively. Within the conceptual modelling research community, the cumulative research in works like \cite{ER,EER,UML} concentrated on conceptual modelling and conceptual modelling formalisms. The approach of UKC-based knowledge representation via knowledge teleontologies proposed in this chapter, in sync with recent proposals in \cite{GGM} and \cite{TCM}, provides an explicit and transparent explanation of how  top-down conceptual models consolidate by linking, quoting \cite{TCM}, \emph{``a set of concepts that are presented by means of terms in a predefined linguistic format"}. On the other hand, the proposed work is orthogonal to the mainstream work done in the foundations of ontology-driven conceptual modelling (see, for instance, \cite{SCM} and its derivative works).

Further, within the applied ontology research community, a huge amount of work has been published in the general area of ontologies and ontology development. Here, it is worth citing the research in \cite{1998-FOIS}, which introduces the role of formal ontologies in information systems and the proposal in \cite{o}, which illustrates, via examples, the definition and characteristics of a computational ontology. The work in \cite{LO,OC} puts forth a general methodology for ontological analysis of the modelling decisions behind engineering an ontology, via, a suite of proposed metaproperties. It is especially focused on the \emph{polysemy} implicit in the concepts modelled in an ontology, and, thereby, the overall confusion inherent in \texttt{IS-A} hierarchies. The approach advanced in this chapter, while being orthogonal to that of the above approach, tackles both the problems. It tackles polysemy problem by unambiguously annotating each concept in a knowledge teleontology with a unique language-independent UKC ID. The confusion in hierarchies is tackled by modelling knowledge teleontologies in classificatory conformance with the lexical-semantic hierarchy modelling in the UKC-based language representations. Finally, the research proposed in the preceding chapter combined with this chapter also contributes to the recent research interest on terminology and language-driven conceptual models and knowledge representations (see, for example, \cite{ao12,ao13}). Notice that the research proposed in this chapter is uniquely positioned to completely tackle heterogeneity at all three levels, \textit{viz.}, concepts, language and knowledge, something which, by design and purpose, is partial in all of the aforementioned related works.

\section{Summary}
To summarize, given the ubiquitous problem of addressing and accommodating representation heterogeneity at the knowledge level, this chapter proposed a UKC-based knowledge representation for modelling entity types, object properties and data properties by reusing the UKC-based language representation as outlined in chapter (3). To that end, the chapter discussed and exemplified the notion of a language teleontology and a knowledge teleontology, respectively. The next chapter will concentrate on the LiveKnowledge catalog which has been adopted to share and facilitate iterative reuse of UKC-based language and knowledge representations as outlined in this and the preceding chapter.

    \chapter{LIVEKNOWLEDGE CATALOG}

\section{Introduction}
DataScientia\footnote{https://datascientia.disi.unitn.it/} is a community-driven open data initiative led by the KnowDive research group at the University of Trento, Italy. To reach its goal of facilitating digital-infrastructure based community-driven sharing and reuse of data, DataScientia continuously develops, maintains and expands on five core catalogs created within the initiative, namely: LiveMedia (LM) \footnote{The development of LiveMedia catalog is currently ongoing.}, LiveLanguage (LL)\footnote{https://datascientiafoundation.github.io/LiveLanguage/}, LiveKnowledge (LK)\footnote{https://datascientiafoundation.github.io/LiveKnowledge/}, LiveData (LD) \footnote{https://datascientiafoundation.github.io/LiveData/} and LivePeople (LP) \footnote{https://datascientiafoundation.github.io/LivePeople/}.

\begin{figure}[htp]
\centering
\includegraphics[width=16cm,height=8cm]{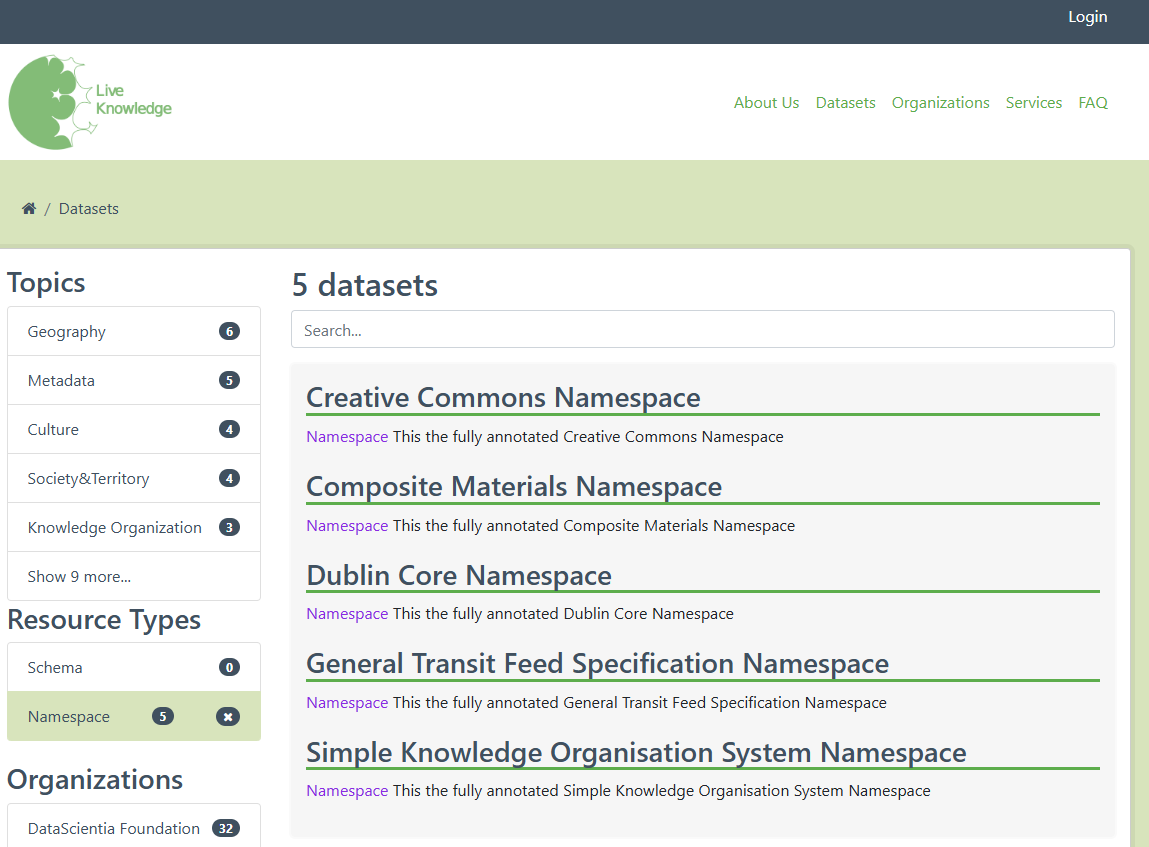}
\caption{A snapshot of the LiveKnowledge website.}
\label{lksite}
\end{figure}

The LiveKnowledge catalog (see Figure \ref{lksite}), as a central focus of the DataScientia intiative, is the core data catalog that supports both the shareability and iterative reuse of the resources generated during language representation (i.e., the UKC namespaces) and knowledge representation (i.e., language teleontology, knowledge teleontology), respectively. To that end, it presents users of knowledge graph engineering projects, whether a producer or an intermediary or a consumer, a unique gateway to find, reuse, analyse, share and distribute high quality formal language and knowledge resources conforming to the UKC-driven language and knowledge representation ecosystem. To achieve the sharing and reuse of the resources indexed by LiveKnowledge as per community-accepted data managament principles, the Live Knowledge catalog is created following the FAIR principles \cite{FAIR}, i.e., its resources are guaranteed to be \textit{Findable}, \textit{Accessible}, \textit{Interoperable} and \textit{Reusable}. More details on this and constituent technicalities on LiveKnowledge can be found in the dedicated thesis \cite{barb} co-supervised by the author of the current thesis. Further, the Live Knowledge catalog will inherit the more advanced services for data analysis and preparation that were offered by its predecessor - \textit{LiveSchema} \cite{2023-LS}, a catalog created with the purpose of being a portal to facilitate knowledge engineers in finding, analysing and preparing state-of-the-art applied ontologies they needed. This chapter presents an overview of LiveKnowledge in terms of its logical architecture, types of resources and some aspects of LK metadata.

\section{Logical Architecture}
\begin{figure}[htp]
\centering
\includegraphics[width=16cm,height=10cm]{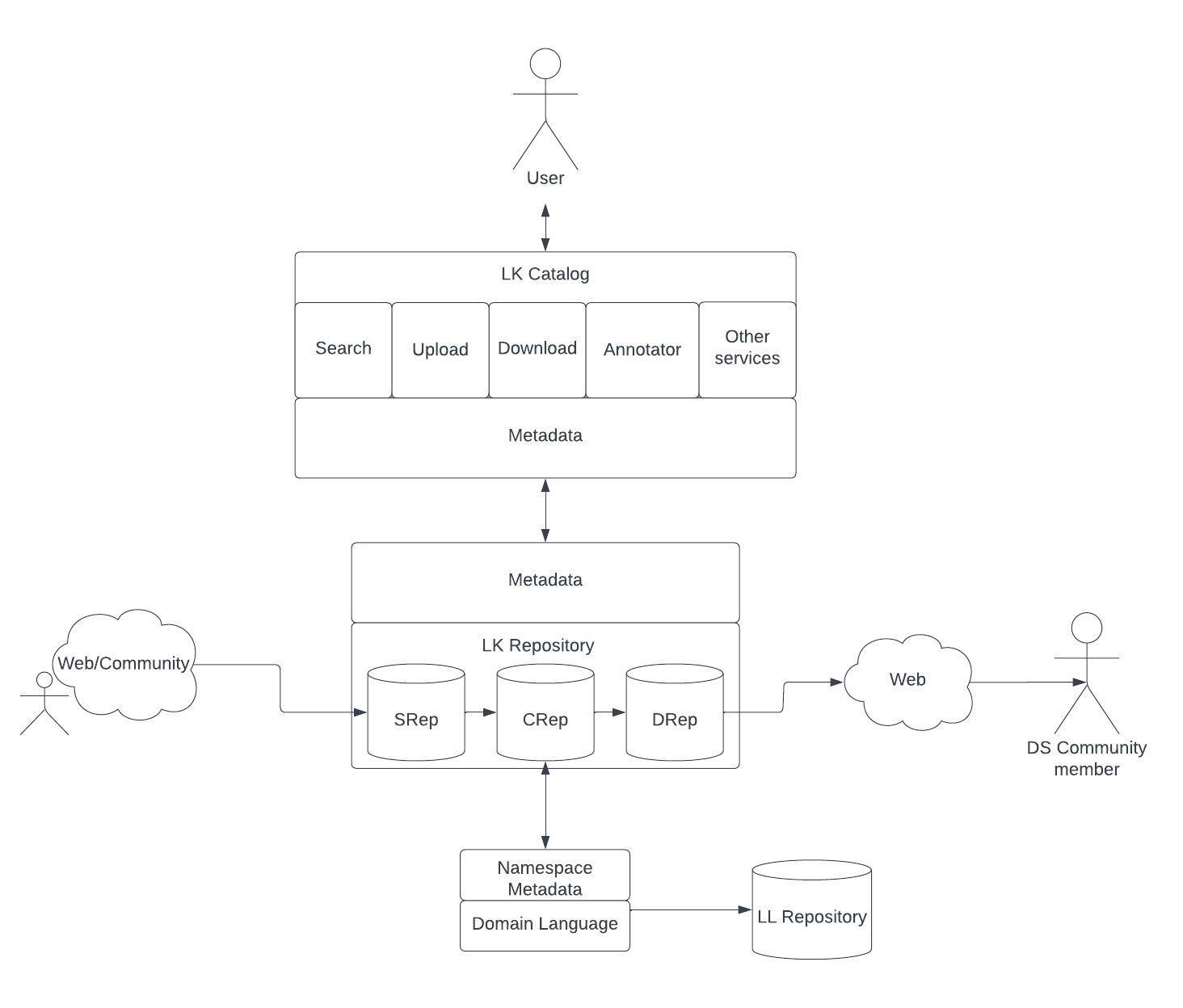}
\caption{LK Logical Architecture.}
\label{lka}
\end{figure}

The LiveKnowledge Catalog is responsible for exposing detailed metadata regarding different genres of language and knowledge resources such as UKC namespaces, language teleontologies and knowledge teleontologies. In addition to the above resources, the catalog also hosts legacy resources such as lightweight classification ontologies and informal schemas (external to the UKC ecosystem) as deemed relevant for various purposes within the broader DataScientia ecosystem. The catalog only publishes the metadata and not the actual resources which are stored in a separate internal LiveKnowledge repository (see Figure \ref{lka}).

The actual resources held in the LiveKnowledge repository (LK Repository) is organized by classifying the resources in three different sub-repositories (see Figure \ref{lka}): 
\begin{itemize}
    \item \textbf{Source Repository (SRep):} Whenever an informal language or knowledge resource is downloaded from the World Wide Web or when informal resources are received from the DataScientia community, they are stored in SRep.
    \item \textbf{Content Repository (CRep):} This is the core repository containing all the formal language and knowledge resource files validated with respect to quality requirements and conforming to the UKC-driven language and knowledge representation formalism.
    \item \textbf{Distribution Repository(DRep):} This is a partition of CRep that can be provided to the users registered in the DataScientia community as per requirements. 
\end{itemize} 

\noindent The purpose of the LK catalog is not just limited to exposing metadata but also to offer a set of services to reuse, analyze and share language and knowledge resources. The core services offered by LK are the search (for searching language and knowledge resources based on metadata-filters), download (for downloading language and knowledge resources following proper approval mechanism) and upload (for uploading language and knowledge resources following proper quality evaluation) of resources. The service to (semi)automate the UKC annotation process as described in the chapter on language representation is currently under development. The above services will cumulatively allow users to browse the resources published in the catalog by filtering them or through query-based search, to download the actual file described by the metadata upon due authorization, and, to upload a new file to be published and to produce annotated language and knowledge resources. 

\section{Type of Resources}
The type of resources that the LiveKnowledge catalog indexes can be classified as follows: 
\begin{itemize}
\item UKC namespaces,
\item Language Teleontologies,
\item Knowledge Teleontologies.
\end{itemize}
In addition to the above UKC-conforming language and knowledge resources, the LK catalog also manages several resources which are external to the direct scope of this thesis. This may include relevant teleologies which are based on the UKC-driven language and knowledge representation proposed in the thesis as well as several legacy resources which are external to the top-down UKC-driven language and knowledge representation described in this thesis. Some of them are as follows:
\begin{itemize}
\item lightweight classification ontologies, with a tree structure where each node is associated with a natural language label \cite{2008-LWO},
\item teleologies \cite{T1,T2} which are constructed bottom-up and are always grounded to a knowledge teleontology,
\item schemas, including external resources like state-of-the-art applied ontologies, e.g., from Linked Open Vocabulary \cite{LOV},
\item external namespaces such as W3C namespaces.
\end{itemize}

\section{Resource Quality Checklist}
Once a potential LK resource file is uploaded by the user, it is stored temporarily in a designated internal storage where it is maintained until a mandatory quality check by the LK catalog administrator (admin, in short). The admin will receive a notification via email for every new resource added to the aforementioned storage and has the possibility to consult the list of all the resources waiting for approval through the catalog itself in a protected area of the upload service, accessible only through authentication. From this list, the admin can download the actual resources and the associated resource metadata spreadsheets in order to perform a quality control check and validate or reject them accordingly. The quality criteria for some selected resource types that the admin should check during the quality control of the resources uploaded are listed below.

\subsubsection{UKC Namespaces}
\begin{itemize}
    \item The file should be in the spreadsheet format.
    \item For an individual word, there should be a Cased Word Lemma.
    \item For an individual word, there should be a natural language definition.
    \item For an individual word, there should be an assignment of a UKC GID.
    \item For an individual word, there should be an assigned Part of Speech (PoS).
    \item For an individual word, there should be an assignment of Parent (concept + UKC GID).
    \item For an individual word, there should be an assigned word sense rank.
    \item For an individual word, there should be an assignment of the kind of relation between it and its parents.
    \item For an individual word, there should be a reference to the user creating it.
    \item For an individual word, there can be a reference to the validator who validated it.
    \item For an individual word, there can be an assignment of timestamp.
    \item For an individual word, there can be a general notes description.
    \item The namespace should be associated with proper catalog metadata as per specification.
\end{itemize}

\subsubsection{Language Teleontologies}
\begin{itemize}
    \item There should be entity type name hierarchy/object property name hierarchy/data property name hierarchy in the form of a concept tree.
    \item Each entry in the entity type name hierarchy/object property name hierarchy/data property name hierarchy should have a meaningful natural language definition.
    \item For each entry in entity type name hierarchy/object property name hierarchy/data property name hierarchy, there should not be any additional axiom or constraint assertion (e.g., direct assertion, via property constraint).
    \item There should not be cycles in the entity type name hierarchy/object property name hierarchy/data property name hierarchy.
    \item The entries in the entity type name hierarchy/object property name hierarchy/data property name hierarchy should not be defined using polysemous terminology.
    \item There should not be multiple entity type name hierarchy/object property name hierarchy/data property name hierarchy with the same semantics.
    \item There should not, preferably, be an entry in the entity type name hierarchy/object property name hierarchy/data property name hierarchy having only a single child.
    \item There should not be any isolated entry in the entity type name hierarchy/object property name hierarchy/data property name hierarchy.
    \item Each entry in the entity type name hierarchy/object property name hierarchy/data property name hierarchy should be annotated with a unique UKC GID.
    \item The language teleontology should be associated with proper catalog metadata as per specification.
\end{itemize}

\subsubsection{Knowledge Teleontologies}
\begin{itemize}
    \item There should be entity type hierarchy/object property hierarchy/data property hierarchy in the form of a concept tree.
    \item Each entry in the entity type/object property/data property hierarchy should have a meaningful natural language definition.
    \item The domain and range of each object property/data property should be specified in one of the ways allowed within the OWL specification (e.g., direct assertion, via property constraint).
    \item There should not be cycles in the entity type hierarchy/object property hierarchy/data property hierarchy.
    \item The entries in the entity type hierarchy/object property hierarchy/data property hierarchy should not be defined using polysemous terminology.
    \item There should not be multiple entity types/object properties/data properties with the same semantics.
    \item There should not, preferably, be an entry in the entity type hierarchy/object property hierarchy/data property hierarchy having only a single child.
    \item There should not be any isolated entry in the entity type hierarchy/object property hierarchy/data property hierarchy.
    \item Each entry in the entity type hierarchy/object property hierarchy/data property hierarchy should be annotated with a unique UKC GID.
    \item The knowledge teleontology should be associated with proper catalog metadata as per specification.
\end{itemize}

\subsubsection{Schemas (e.g., external applied ontologies)}
\begin{itemize}
    \item Each entry amongst classes, object properties, and data properties should have a meaningful natural language definition.
    \item The domain and range of each object property/data property should be specified in one of the ways allowed within the OWL specification (e.g., direct assertion, via property constraint).
    \item The entries in classes, object properties, and data properties should not be defined using polysemous terminology.
    \item There should not be multiple classes/object properties/data properties with the same semantics.
    \item There should not be any isolated entry amongst classes, object properties, and data properties .
    \item The schema should be associated with proper catalog metadata as per specification.
\end{itemize}
After the completion of the quality control process following the criteria as described above, the user who has requested the upload of such file is notified (by email) with its result. If negative, the motivation of the rejection will be explained in detail in the message. Otherwise, the resources will be uploaded in the appropriate catalog repository and published on the appropriate catalog with proper metadata. This message will be written by the administrator who performed the quality control for the resource.

\section{Resource Metadata}
The public metadata attributes describing a LiveKnowledge dataset are listed as follows (\texttt{ds} being the DataScientia namespace prefix).

\begin{figure}[htp]
\centering
\includegraphics[width=16cm,height=8cm]{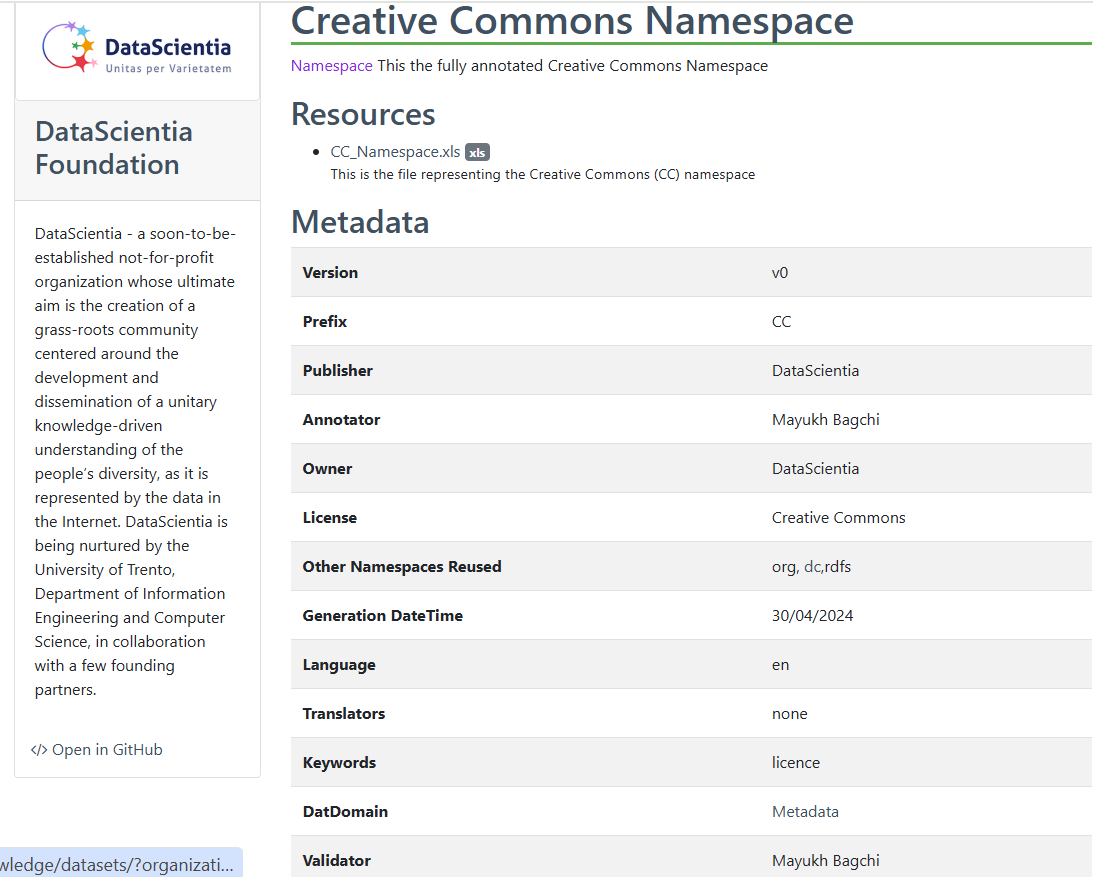}
\caption{A snapshot of the LiveKnowledge website.}
\label{lkns}
\end{figure}

\begin{enumerate}
    \item ds:DatLicense: this attribute encodes the license of the dataset, e.g., CC-BY-SA-4.0.
    \item ds:DatURL: this attribute encodes the dereferenceable URL of the dataset.
    \item ds:DatKeyword: this attribute encodes the keywords which can quickly convey the topic of the dataset.
    \item ds:DatPublisher: this attribute encodes the publisher of the dataset.
    \item ds:DatCreator: this attribute encodes the creator of the dataset.
    \item ds:DatOwner: this attribute encodes the owner of the dataset.
    \item ds:DatLanguage: this attribute encodes the natural language(s) in which the dataset information is represented.
    \item ds:DatLevel: this attribute encodes the representation level of the dataset, e.g., language, knowledge.
    \item ds:DatSize: this attribute encodes the byte size of the dataset.
    \item ds:DatName: this attribute encodes the name of the dataset in a natural language.
    \item ds:DatPublicationTimestamp: this attribute encodes the timestamp of the publication of the dataset in the respective catalog.
    \item ds:DatDescription: this attribute encodes the description about the dataset in a natural language.
    \item ds:DatVersion: this attribute encodes the version of the dataset.
    \item ds:DatFileFormat: this attribute encodes the file format of the dataset, e.g., OWL RDF/XML, Excel.
\end{enumerate}
An illustration of the metadata properties instantiated for the Creative Commons namespace (including those from the above list) is visualized in the figure \ref{lkns}.

\begin{figure}[htp]
    \centering
    \includegraphics[width=15cm, height=6cm]{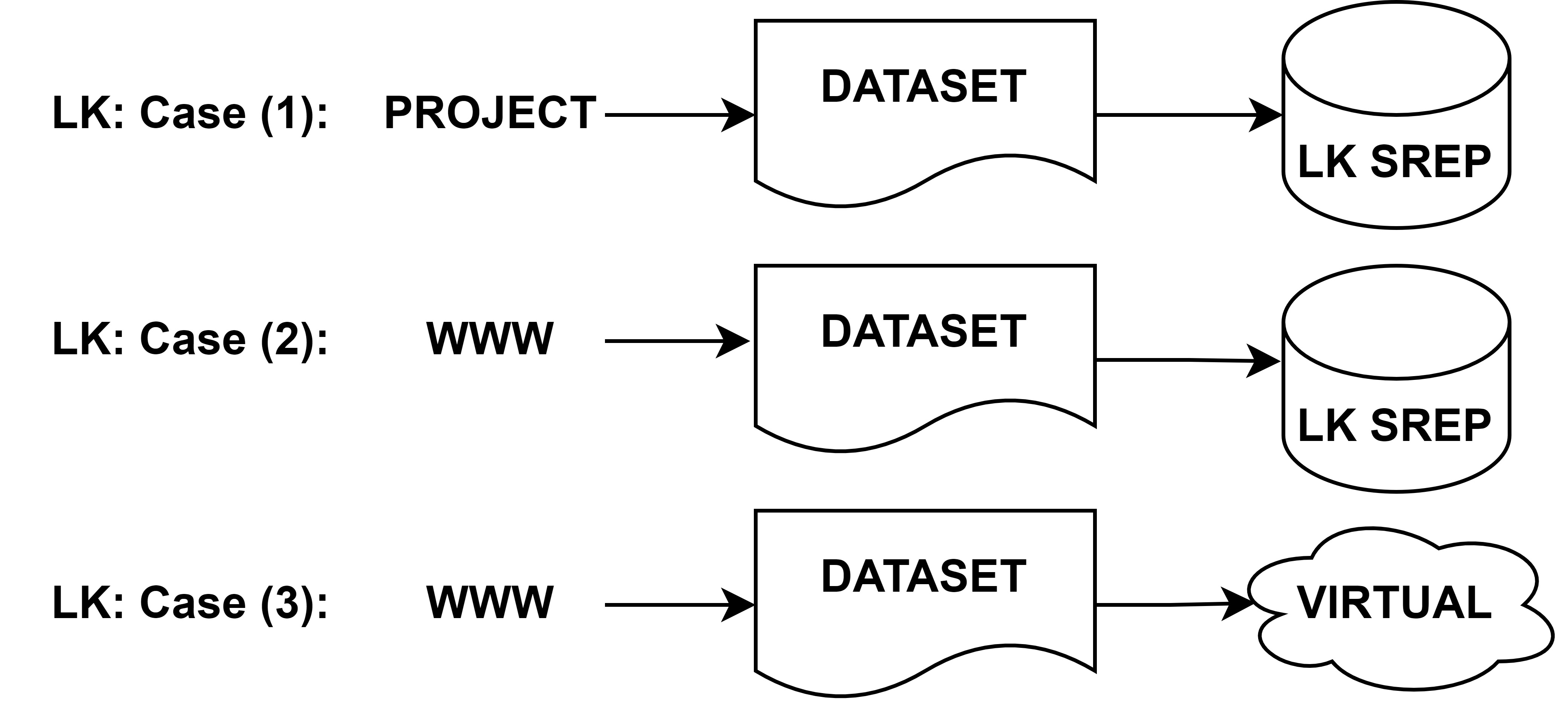}
    \caption{LK Resource Possible Inputs.}
    \label{ELP}
\end{figure}

\noindent Further, the general schema for the possible options of dataset input for LiveKnowledge is captured in the figure \ref{ELP}. They are briefed as follows:
\begin{enumerate}
    \item A Project generates a dataset which is stored in LK SREP repository and indexed by LK SREP catalog (Case 1).
    \item A copy of an external dataset from the World Wide Web (WWW) is stored in LK SREP repository and indexed by LK SREP catalog. This is essential to keep a base reference version of the dataset (Case 2).
    \item A \emph{virtual copy} (i.e., URL) of an external WWW dataset is indexed by LK SREP catalog. This is essential for referring to the latest version or application profile of the dataset if required for a specific purpose (Case 3).
\end{enumerate}

\noindent In connection to the three aforeementioned cases, the properties to be captured for the propagation of provenance for LiveKnowledge datasets across the three different sub-repositories are described as follows: 
\begin{enumerate}
    \item LK SRep
    \begin{enumerate}   
         \item CASE 1: Dataset: Project URL, Owner, Timestamp, License.
         \item CASE 2: Dataset: URL, Owner, Timestamp, License.
         \item CASE 3: Dataset: URL, Owner, Timestamp, License.    
    \end{enumerate}
    \item LK CRep
    \begin{enumerate}
        \item Dataset: Project URL, Owner, Timestamp, License. 
    \end{enumerate}
    \item LK DRep
    \begin{enumerate}
        \item Dataset: Project URL, Owner, Timestamp, License.
    \end{enumerate}
\end{enumerate}

\noindent Notice that LiveKnowledge only allows monotonic evolution of knowledge and there is no need to record the lineage of different versions of a LK resource and, thereby, there is no need of item-level provenance for LK resources.

Finally, the (propagation of) license proposed for LiveKnowledge datasets are listed as follows for each license type (assuming LK resources in input are licensed following Creative Commons (CC)\footnote{https://creativecommons.org/}): 
\begin{enumerate}
    \item WORLD
    \begin{enumerate}
        \item WORLD WIDE WEB (WWW)
        \begin{enumerate}
            \item Type (A): WWW Resource has CC0 license
            \item Type (B): WWW Resource has CC-BY license
            \item Type (C): WWW Resource has CC-BY-NC license
            \item Type (D): WWW Resource has CC-BY-SA license
            \item Type (E): WWW Resource has CC-BY-NC-SA license
        \end{enumerate}
    \end{enumerate}
    \item LK SREP
    \begin{enumerate}
        \item LK SREP Resource: Input from the WWW
        \begin{enumerate}
            \item Type (A): LK SREP Resource has CC0 license
            \item Type (B): LK SREP Resource has CC-BY license
            \item Type (C): LK SREP Resource has CC-BY-NC license
            \item Type (D): LK SREP Resource has CC-BY-SA license
            \item Type (E): LK SREP Resource has CC-BY-NC-SA license
            \item \textbf{NOTE (1):} ND licenses (CC-BY-ND and CC-BY-NC-ND) are unlikely to be applicable for LK SREP Resources
            \item \textbf{NOTE (2):} In CC, \emph{owner, creator, licenser, rights holder} are used interchangeably.
        \end{enumerate}
    \end{enumerate}
    \item LK CREP
    \begin{enumerate}
        \item LK CREP Resource:
        \begin{enumerate}
            \item Type (A): LK CREP Resource can take any appropriate license (no restrictions)
            \item Type (B): LK CREP Resource can take any appropriate license + attribution to LK SREP Resource creator
            \item Type (C): LK CREP Resource can take any appropriate license + attribution to LK SREP Resource creator + only for non-commercial use 
            \item Type (D): LK CREP Resource must take LK SREP Resource license + attribution to LK SREP Resource creator
            \item Type (E): LK CREP Resource must take LK SREP Resource license + attribution to LK SREP Resource creator + only for non-commercial use
        \end{enumerate}
    \end{enumerate}
    \item LK DREP
    \begin{enumerate}
        \item LK DREP Resource:
        \begin{enumerate}
            \item Type (A) [Single or Composed\footnote{Here, composed refer to a LK resource created by composing one or more LK resources.}]: LK DREP Resource can be distributed as allowed by LK CREP Resource license. 
            \item Type (B) [Single or Composed]: LK DREP Resource can be distributed as allowed by LK CREP Resource license.
            \item Type (C) [Single or Composed]: LK DREP Resource can be distributed as allowed by LK CREP Resource license. 
            \item Type (D) [Single or Composed]: LK DREP Resource can be distributed as allowed by LK CREP Resource license.
            \item Type (E) [Single or Composed]: LK DREP Resource can be distributed as allowed by LK CREP Resource license.
            \item Type (D) cannot be composed with Type (E) and vice-versa
            \item Composition of LK DREP resources (amongst Types (A,B,C,D,E)) should follow CC License Compatibility License Chart\footnote{https://wiki.creativecommons.org/wiki/Wiki/cc\_license\_compatibility}
        \end{enumerate}
    \end{enumerate}
\end{enumerate}

\noindent Notice that in case of composition of compatibly licensed LK resources, the composed resource should be distributed via the more expressive license. For example, if a dataset has license CC-BY and another dataset has CC-BY-SA, the resultant license of the composed datasets of these two datasets should have CC-BY-SA licensing. 

Finally, note that while the main intent behind developing the LiveKnowledge catalog is to facilitate iterative sharing and reuse of UKC-based language and knowledge representations, the actual decision to share such resources are made on a case-by-case basis due to different techno-legal considerations such as intellectual property rights, continuous update and refinement of versions of resources, and so on. Further, it is also important to highlight that the design of the LiveKnowledge ecosystem (inclusive of its constituent resources) adhere to the general guiding principles of lexical-semantics, classification and cataloguing as advanced by Aristotle and Ranganathan. This specific aspect is also an ongoing work in terms of developing newer guiding principles on aspects not covered by the principles mentioned above. Last but not the least, while related work with respect to LiveKnowledge catalog design, resources and metadata are out of scope for this thesis and can be found in the dedicated thesis \cite{barb}, it is worthy to mention that mainstream data catalogs and repositories sharing different types of language and knowledge representations, e.g., \cite{LXB,LOV,BPL}, are external to both the scope and focus of UKC-based language and knowledge representation and, therefore, are not relevant for facilitating shareability and iterative reuse of UKC namespaces, language teleontologies and knowledge teleontologies.

\section{Summary}
To summarize, this chapter proposed the adoption and use of the LiveKnowledge catalog to facilitate shareability and iterative reuse of the resources generated during language representation and knowledge representation, respectively as outlined in chapter (3) and chapter (4). To that end, the chapter discussed and exemplified the organization of the catalog and the type of resources it indexes by providing additional details regarding the quality control checklist and metadata dimensions considered for LK resources. The next chapter will concentrate on the top-down \textit{kTelos} methodology which methodologically integrates the solution components proposed in the current as well preceding chapters (chapters (2), (3) and (4)) to iteratively generate the language and knowledge representations absolving representation heterogeneity.

    \chapter{THE KTELOS METHODOLOGY}

\section{Introduction}
This chapter elaborates the top-down \textit{kTelos} methodology which methodologically integrates the solution components proposed in the preceding chapters, namely, the UKC (inclusive of the UKC namespaces and domain languages integrated within the UKC hierarchy), language teleontology and knowledge teleontology, to iteratively generate the language and knowledge representations absolving representation heterogeneity ready to be exploited in the iTelos KGC process (as from the motivation briefed in section (2.1) of chapter (2)). To that end, the chapter first elucidates the research background of how the \textit{kTelos} methodology evolved to its present form from a detailed study and delineation of state-of-the-art data-driven knowledge modelling methodologies in the two distinct research communities of Knowledge Representation (KR)\footnote{Notice that by ``\textit{Knowledge Representation (KR)}", we refer to state-of-the-art work extrinsic to the UKC-based knowledge representation proposed in this thesis. The written context would make it clear as to which knowledge representation paradigm is being referred to.} and Knowledge Organization (KO) \cite{KROB,KORB}. Then, the chapter elaborates the the top-down \textit{kTelos} methodology advanced in this thesis. 

\section{Background}
Knowledge Representation (KR) is the arena of Artificial Intelligence (AI) dealing with \emph{``how knowledge can be represented symbolically"} \cite{2004-KR} within intelligent systems. To that end, KR encompasses a wide spectrum of advanced \emph{technologies} (e.g., the Semantic Web technology stack \cite{2009-SWTS}) and \emph{methodologies} (e.g., \cite{METHONTOLOGY,DERA}) to facilitate generation of KR artifacts (e.g., ontologies \cite{o}, conceptual models \cite{2014-OAP}, Knowledge Graphs (KGs) \cite{2021-KG}). Such methodologies have been widely adopted or adapted for application scenarios such as, e.g., data integration \cite{2021-KGCW}. However, while KR methodologies have remained highly mature in terms of supporting technology and scalable in terms of technology-enabled services, a \emph{key} criticism has been that they have traditionally underemphasized \emph{modelling quality}, e.g., of ontologies or conceptual models (see, \cite{2005-FLAIRS,2013-QQUARE,2015-OAP}, for a few prominent studies), resulting in, often, representationally flawed and biased datasets designed according to such models.

Knowledge Organization (KO), on the other hand, is the arena of Information Science dealing with the cumulative set of activities concerning, quoting \cite{2008-KO}, the \emph{``description, indexing and classification"}  of information resources (e.g., books) in different kinds of \textit{`memory institutions'} (e.g., libraries). To that end, KO encompasses a wide spectrum of modelling systems \cite{2008-KOS}, e.g., classification schemes, taxonomies, catalogs, etc., and, different approaches \cite{2008-KO}, e.g., enumerative approach, facet-analytical approach, etc., which integrate different KO systems. In this paper, we concentrate exclusively on the facet-analytical KO approach originally proposed by Ranganathan \cite{SRR-64,srr}. For example, to develop the classification hierarchy for a subject, facet-analysis prescribes \textit{analysis}, i.e., analyzing and breaking down the different dimensions of the subject into hierarchical models of primitive atomic concepts (termed \textit{facets}), followed by \textit{synthesis}, wherein, different facets can be combined in a specific order to obtain, e.g., the final unique classification number of a book of that subject. Further, the classification number is reverse-engineered to generate the subject headings which, with other bibliographic attributes, form the description of the book within the catalogue record which is used by a library user. Noticeably, while the activities in (faceted) KO are predominantly intellectual in nature and less technology-driven, they have, in order to support the aforementioned activities, developed a huge number of \emph{guiding principles} for (conceptual) modelling, termed \emph{canons}, following which \emph{high-quality models}, .e.g., conceptual hierarchies, conceptual models and model-based \emph{datasets}, can be generated. 

Notice two dimensions which are apparent from the brief overviews on KR and KO given above. First, there is a similarity between KR and KO in terms of their high-level functionality, i.e., roles, artefacts, and activities, to methodologically design and develop models which can organise and represent knowledge for a domain. Second, KR and KO have complementary strengths and weaknesses, with KO being strong on guiding principles and weak in terms of technology-driven activities and services and vice versa for KR.

\subsubsection{KR Methodology}
\begin{figure}[htp]
\includegraphics[width=16cm,height=8cm]{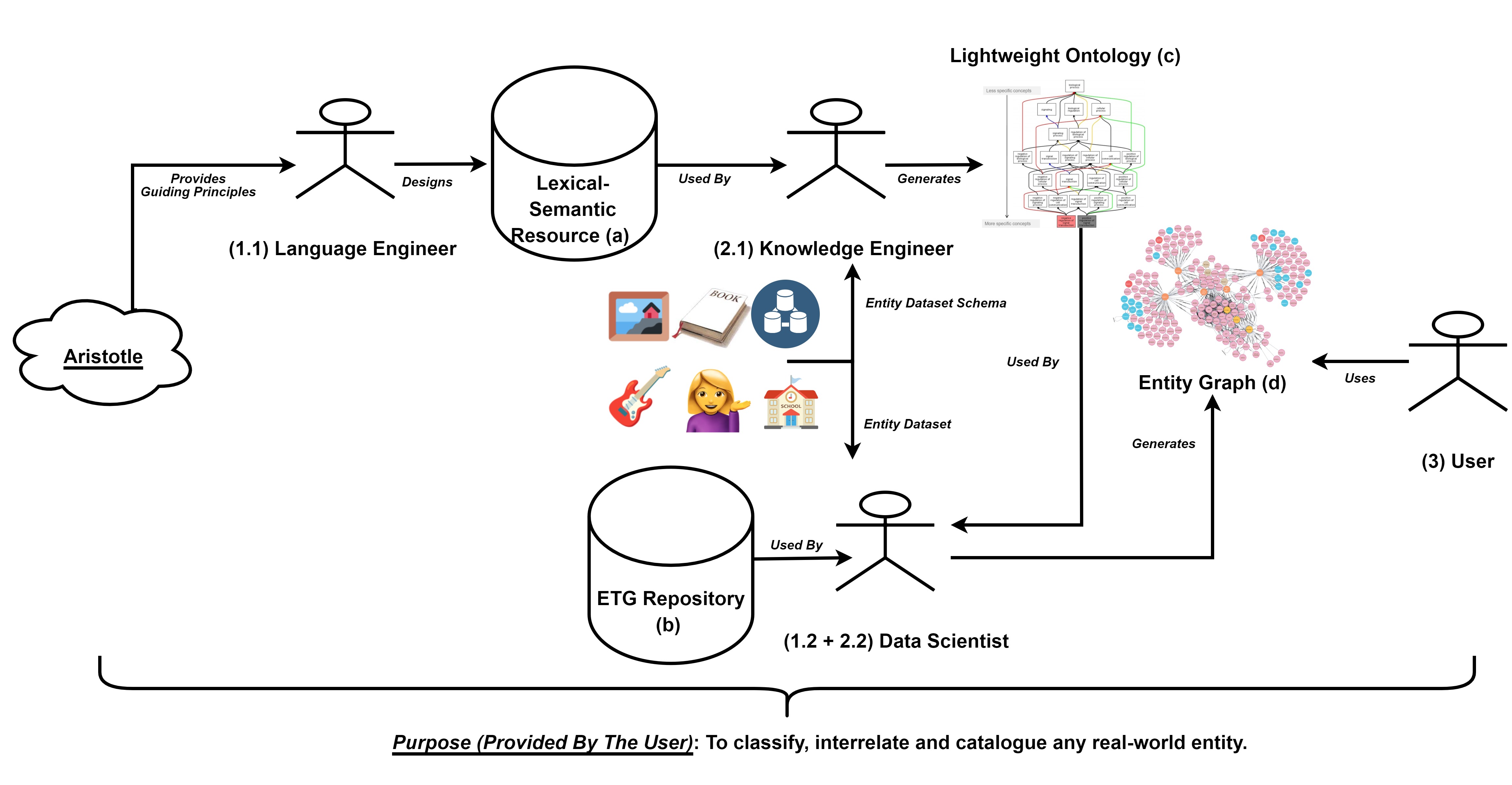}
\caption{A high-level view of the KR Methodology.}
\centering
\label{I1}
\end{figure}

\noindent Let us first concentrate on a detailed exposition of the various phases, and within each phase, the various roles, activities and artifacts, which together compose to form a standard KR methodology. See Figure \ref{I1} for a high-level view of the KR methodology. The methodology can be seen as being constituted of the following four distinct phases (the terminologies being detailed later):
\begin{enumerate}
    \item The first phase initiating with guiding principles and concluding with the generation of the Lightweight Ontology\footnote{Notice that the lightweight ontology is a conceptual predecessor of language teleontology (which, by the very scope of this section, is an evolution over lightweight ontology).}.
    \item The second phase concerning the development of the Entity Type Graph (ETG) repository \footnote{Notice that the ETG repository is a conceptual predecessor of LiveKnowledge (which, by the very scope of this section, is an evolution over ETG repository).}.
    \item The third phase of the methodology concentrating on how the data scientist takes in three different inputs, namely, the Lightweight Ontology, Entity Dataset(s) and ETGs from the ETG repository, and, suitably integrates them to generate the Entity Graph (EG).
    \item Finally, the fourth and the final phase concentrating on the different ways in which a user can use and exploit the EG. 
\end{enumerate}
The diagrammatic symbols of Figure \ref{I1} include, amongst others, the various roles (visualized via \textit{actor} icon), activities (visualized as edge \textit{labels}) and artifacts (visualized variously via other (coloured) icons). Notice also that the numbers and lowercase alphabets which identify roles and artifacts, respectively, in Figure \ref{I1}, are employed later for functional mapping purposes. Let us now consider each phase of the methodology individually.

The first phase, as already noted, commences with guiding principles articulated as the intensional definition-building paradigm of \textit{Genus-Differentia} \cite{GD} proposed by Aristotle (visualized via a cloud) over two millennia ago. According to the paradigm, the linguistic definition of any real-world object \cite{T2} expressed, e.g., as a noun, is formulated in terms of two constituent definitions: \textit{Genus} and \textit{Differentia}. While \textit{Genus} defines an \emph{a priori} set of properties shared across distinct objects, e.g., the property of being a stringed musical instrument, \textit{Differentia} defines a novel set of properties used to differentiate objects having the same \textit{Genus}, e.g., the properties of musical instruments having six strings or thirteen strings. Therefore, as illustrated in Figure \ref{I1}, the \textit{Genus-Differentia} guidelines are taken in input and adhered to by the \textit{Language Engineer} to create machine-processable language data, e.g., WordNet-like lexical-semantic hierarchies of synsets \cite{PWN} codifying word meanings in different natural languages and in different domains \cite{2004-WDH} represented in a machine processable format, e.g., Lexical Markup Framework \cite{2014-LMF}. 

Such machine-processable language data are stored and managed as part of a \textit{Lexical-Semantic Resource} (see Figure \ref{I1}) which is usually designed as a collection of WordNet-like machine-processable lexical hierarchies and, in some cases \cite{UKC1,UKC2}, with an additional language-independent semantic layer unifying different language-specific lexical hierarchies. Given the design of the \textit{Lexical-Semantic Resource}, the final activity of this phase shifts to the \textit{Knowledge Engineer} (see Figure \ref{I1}) who has to now generate the \textit{Lightweight Ontology} (Figure \ref{I1}) which is an intermediate machine-processable formal hierarchy \emph{``consisting of backbone taxonomies"} \cite{2008-LWO} that are being considered for representing knowledge in the context of a specific purpose provided by the user. To that end, the \textit{Knowledge Engineer} has to take in two important inputs. Firstly, (s)he has to take in input the appropriate lexical-semantic hierarchy of words in a specific language which will inform the syntax and modelling of the taxonomical hierarchy of the \textit{Lightweight Ontology} she will generate. In addition, (s)he also takes in input the (dataset) \textit{schema} of the entities which (s)he wants to model in the \textit{Lightweight Ontology}, this, providing her with the \textit{exact} way in which the subsumption hierarchy of parent and child concepts, pertaining to the entities (and their datasets), should be organized\footnote{thereby, significantly speeding up the process of merging the datasets with the KR model at a later stage of the methodology.}.

The second phase, as illustrated in Figure \ref{I1}, concerns the design and development of a repository (exposed via a catalog) of reusable Entity Type Graphs (ETGs) \cite{2023-LS}, wherein, ETGs are defined as machine-processable \textit{ontological} representations formalizing entity types \cite{ETR} which capture the semantics inherent in (dataset) entities. Notice the fact that, ETGs, being \textit{ontological} representations, also encode \textit{object properties} (modelling how an entity is related to other entities) and \textit{data properties} (modelling the attributes which describe an entity) as is standard to any KR model. This repository is crucial to the methodology in all the four dimensions - \textit{F}indability, \textit{A}ccessibility, \textit{I}nteroperability, and \textit{R}eusability - advanced by the \textit{FAIR} paradigm \cite{FAIR} of scientific data management. Especially, the repository would facilitate not only interoperability amongst ETGs (as they are modelled following the same technological standards) but also enable ETG-based interoperability of data (when ETGs from the repository are reused, in conjunction with lightweight ontology, to produce the final EG). It would also promote the circular reuse of ETGs, for instance, the reuse of the same ETG but for different use case scenarios of KR. A crucial observation. Notice that while there are several technical advantages of designing an ETG repository as elucidated above, the methodology \textit{does not} specify any activity path or roles for enforcing guiding principles which ensure the high-quality of the ETG models constituting the repository (noticeable by the lack of any activity preceding the ETG repository in Figure \ref{I1}).

The third phase concentrates on how the \textit{Data Scientist} (see Figure \ref{I1}) exploits the outputs of the previous two phases and the concrete (entity) datasets of the use case at hand to generate the final Entity Graph (EG). An EG is a type of Knowledge Graph (KG) which is: (i) taxonomically structured via a lightweight ontology, and, (ii) interrelated and described with object properties and data properties from the relevant ETG. To that end, the \textit{Data Scientist} receives in input the lightweight ontology (output of the first phase), wherein, the use-case specific concepts are hierarchically modelled following the appropriate lexical-semantic hierarchy. The lightweight ontology is then grounded in the relevant classificatorily-compliant ETG from the ETG repository (output of the second phase). This activity has three key advantages. Firstly, the ETG endows the lightweight ontology-ETG combined KR artifact with crucial object properties (which interrelate, at a conceptual level, the entities it encode) and data properties (which encode the descriptive attributes of entities). Secondly, the grounding also ensures the completeness of the KR model in terms of grounding use-case specific concepts (modelled bottom-up) into general (domain) concepts (modelled top-down). Thirdly, the above grounding also ensures the fact that, later, the data encoded via the specific lightweight ontology becomes interoperable with data encoded via other lightweight ontologies which commit to the same ETG. Given the lightweight ontology-ETG combined KR artifact, the \textit{Data Scientist} takes in input the entity datasets and semi-automatically maps them to the combined artifact (see, \cite{KGSWC}, for the detailed process), one dataset at a time, to generate the final Entity Graph (EG) which is a KG modelled by populating the nodes of the combined KR artifact with data.

Finally, the fourth phase of the methodology concentrates on the different ways in which different \textit{users} can exploit the Entity Graph (EG). This, in turn, informs as well as depends on the different type of basic and specialized \textit{services} which can be developed to explore the EG. A set of basic services can include the option to download a (part of a) KG in different formats, the option to query a KG (see, \cite{2023-LS}, for an enumeration of potential services).


\subsubsection{KO Methdology}
\begin{figure}[htp]
\includegraphics[width=16cm,height=8cm]{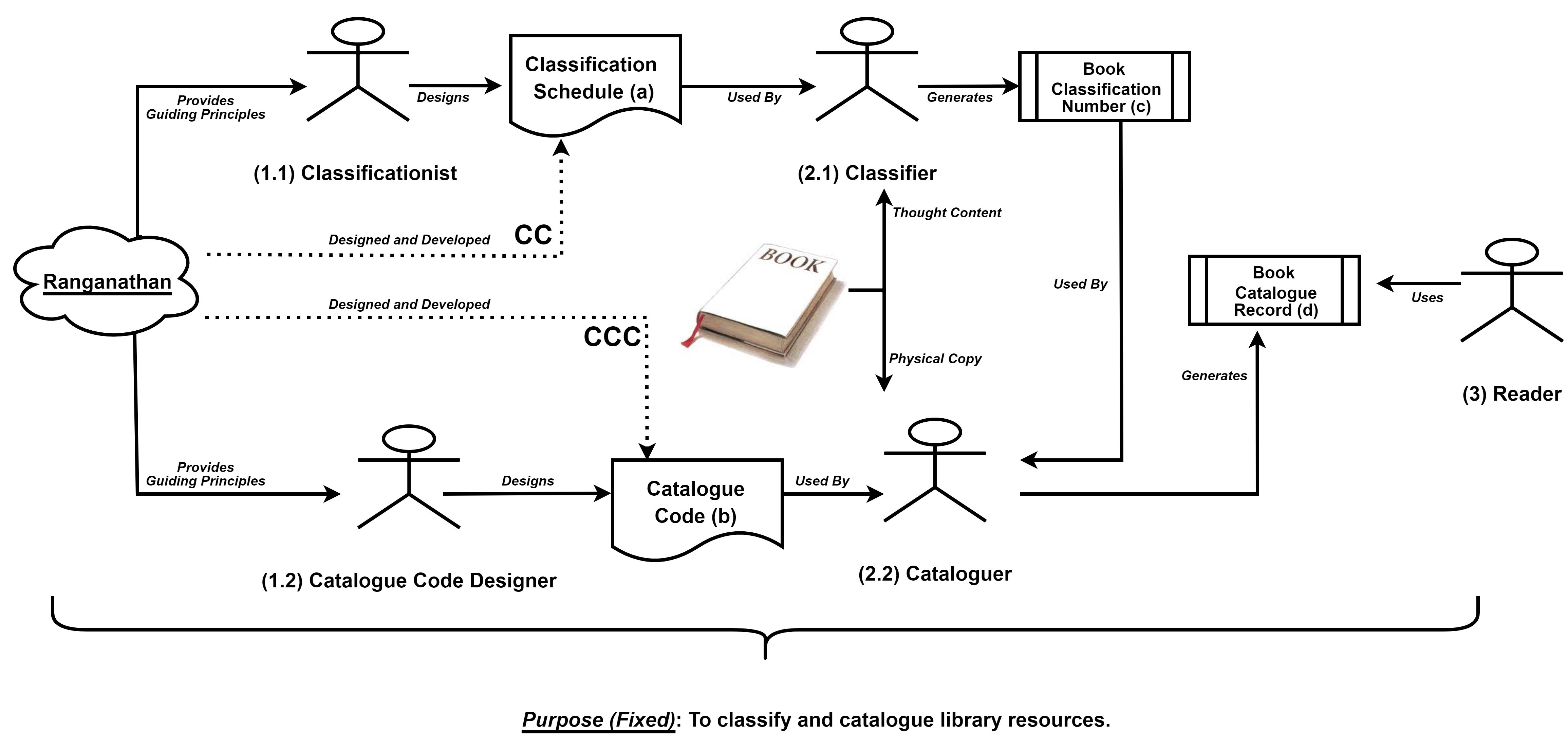}
\caption{A high-level view of the facet-analytical KO Methodology.}
\centering
\label{I2}
\end{figure}

\noindent In sync with the previous section on the KR methodology, let us now concentrate on a detailed exposition of the various phases which together compose to form a standard KO methodology within the facet-analytic tradition of KO. See Figure \ref{I2} for a high-level view of the KO methodology. The methodology can be seen as being constituted of the following four distinct phases (the terminologies being detailed later):
\begin{enumerate}
    \item The first phase initiating with the guiding principles for \emph{Classification} \cite{srr} and concluding with the generation of the book classification number.
    \item The second phase concerned with the guiding principles for \emph{Cataloguing} \cite{SRR-64} and concluding with the design of the catalogue code.
    \item The third phase of the methodology concentrating on how the cataloguer takes in three different inputs, namely, the book classification number, the physical copy of the book and the catalogue code, and, suitably integrates them to generate the book catalogue record. 
    \item Finally, the fourth and the final phase concentrating on the different ways in which a reader visiting a library can use and exploit the book catalogue record.
\end{enumerate} 
The diagrammatic symbols of Figure \ref{I2} are similar as before. We now consider each phase of the methodology individually.

The first phase, as already noted, commences with guiding principles, termed \textit{canons of classification}, postulated by Ranaganathan \cite{srr} to oversee the generation of \textit{high-quality} classification hierarchies for any subject. Notice that, by \textit{subject}, Ranganathan meant not only the universe of \emph{macro thought} (e.g., disciplines such as Mathematics, Chemistry) but also, crucially, the universe of \textit{micro thought} \cite{SRR1954} (e.g., depth classification of differential equations). We now enumerate below an overview of the three groups of \textit{canons of classification} (detailed in \cite{srr}) which are core to the methodology illustrated in Figure \ref{I2}:
\begin{enumerate}
    \item \textit{Canons of Idea Plane}, which focus on the modelling of concepts in a classification hierarchy based on their perceivable properties. They include:
    \begin{enumerate}
        \item canons about \textit{characteristics} based on which concepts should be differentiated with respect to a single level in the classification hierarchy, e.g., canon of \textit{relevance}, canon of \textit{ascertainability}.
        \item canons about \textit{succession of characteristics}, i.e., how differentiating characteristics should succeed one after the other with respect to a classification hierarchy, e.g., canon of \textit{relevant succession}.
        \item canons about \textit{arrays} based on which concepts, within the same horizontal level in the classification hierarchy, should be modelled, e.g., canon of \textit{exhaustiveness}.
        \item canons about \textit{chains} based on which concepts, within a single hierarchical path in the classification hierarchy, should be modelled, e.g., canon of \textit{modulation}.
    \end{enumerate}
   \item \textit{Canons of Verbal Plane}, which focus on the proper linguistic rendering of the concepts modelled following the canons of the Idea Plane, e.g., canon of \textit{reticence}.
   \item \textit{Canons of Notational Plane}, which focus on assigning a unique numerical identifier for each linguistically labelled concept in the classification hierarchy, e.g., canon of \textit{synonym} and canon of \textit{homonym}. 
\end{enumerate}

\noindent Thereafter, as illustrated in Figure \ref{I2}, the \textit{canons of classification} as briefed above are taken in input by the \textit{Classificationist} to design (a set of) faceted classification schedules for either general knowledge organization usage, e.g., the Colon Classification (CC) \cite{CC} designed and developed by Ranganathan himself (indicated with dashed lines), or, for specialized knowledge organization usage, e.g., Uniclass\footnote{https://www.thenbs.com/our-tools/uniclass}, specialized classification for the construction sector. Given the design of the classification schedule(s), the final activity of this phase shifts to the \textit{Classifier} (see Figure \ref{I2}) who has to now generate the unique \textit{Book Classification Number} (Figure \ref{I2}), e.g., Colon Number, for the subject matter of a book. To that end, the \textit{Classifier} has to take in two important inputs. Firstly, (s)he has to take the \emph{a priori} designed classification schedules which provide him/her with the concept hierarchy (with each concept uniquely identified via an identifier) as well as the formula (i.e., the \textit{facet formula}) in which relevant concepts should be combined to generate the book classification number. Secondly, (s)he has to take in input the \textit{thought content} of the book to be classified. Finally, the \textit{Classifier} follows Ranganathan's analytico-synthetic classification number generation procedure \cite{CC} to generate the \textit{Book Classification Number} which uniquely identifies its subject matter. Notice that the above process is valid not only for books but also for any library resource.

The second phase, as illustrated in Figure \ref{I2}, concerns the design and development of a catalogue code. It commences with the \textit{canons of cataloguing} as guiding principles postulated by Ranaganathan \cite{SRR-64} to oversee the generation of \textit{high-quality} description of any book (or, any library resource). Some of the canons are notable to be briefed. For example, the canon of \textit{sought heading} mandates that the metadata attributes which should be captured about a book in a catalogue should be strictly based on the \textit{likelihood} of how a user might approach the catalogue. To that end, all unnecessary metadata should be excluded from the catalogue record. Further, the canon of \textit{consistence} prescribes that, for a specific type of (library) resource, the set of metadata which constitute its catalogue record should be consistent, unless otherwise prescribed by the canon of \textit{context}. Another crucial principle is that of \textit{local variation} which allows flexibility in the description of a catalogue record if need arises due to a very typical resource specific to a context. These canons of cataloguing, amongst many others detailed in \cite{SRR-64}, are taken as input guidelines by the \textit{Catalogue Code Designer} to finally design a \textit{Catalogue Code} - a body of specifications on how and what metadata should be encoded in a catalogue record for a specific bibliographic resource type. In fact, Ranganathan himself designed and developed one such catalogue code termed the \textit{Classified Catalogue Code} (or, CCC; shown in Figure \ref{I2} via dashed lines).

The third phase concentrates on how the \textit{Cataloguer} (see Figure \ref{I2}) exploits the outputs of the previous two phases and the concrete physical copy of the resource (e.g., book) at hand to generate the final \textit{Book Catalogue Record}. A bibliographic catalogue record encodes metadata which are termed as \textit{access points}, e.g., title, author, year of publication, subject headings, etc., which might be help the reader in her quest to search and identify the actual copy of the book (or, resource). To that end, the \textit{Cataloguer} receives in input the book classification number (output of the first phase), wherein, the subject matter of the book is modelled following the appropriate classification schedule (and facet formula). Given the book classification number, the procedure of \textit{Chain Indexing} is performed, whereby, the classification number is reverse-engineered through a series of specified steps (see \cite{SRR-64} for details) to generate subject headings (\textit{tags} in modern parlance) which can serve as access points in the catalogue record. Further, the \textit{Cataloguer} also receives two other inputs: the concrete copy of the book which contains all its imprint details, and, the catalogue code (e.g., CCC) which strictly specifies which and how such imprint information should be modelled in the catalogue record. Thereafter, the \textit{Cataloguer} integrates the subject headings together with the requisite imprint attributes and the classification number (together with the \textit{call number} to identify the book's exact place in the shelves) to generate the final \textit{Book Catalogue Record}.

Finally, the fourth phase concentrates on how a library \textit{reader} can exploit the book catalogue record. In addition to the usual ways of using the catalogue (see, e.g., \cite{SRR-64}), Ranganathan's \texttt{APUPA} principle \cite{DACC} facilitates a reader in finding very related and somewhat related books/resources on either side of the particular resource one is searching for). Further, a reader can use the catalogue record of an Online Public Access Catalog (OPAC) which is enhanced with \textit{library discovery services} \cite{LRD}, thereby, going beyond traditional means of library search to include web-scale \textit{exploratory search} and recommendations.


\subsubsection{From KR to KO and Back}
\begin{table}[]
\resizebox{\textwidth}{!}{%
\begin{tabular}{|l|l|l|l|}
\hline
\textbf{Phase} &
  \textbf{KR} &
  \textbf{KO} &
  \textbf{KO-Enriched KR} \\ \hline
1 &
  \begin{tabular}[c]{@{}l@{}}Language Engineer\\ (1.1)\end{tabular} &
  \begin{tabular}[c]{@{}l@{}}Classificationist\\ (1.1)\end{tabular} &
  \begin{tabular}[c]{@{}l@{}}Language Engineer\\ (1.1)\end{tabular} \\ \hline
2 &
  \begin{tabular}[c]{@{}l@{}}Knowledge Engineer\\ (2.1)\end{tabular} &
  \begin{tabular}[c]{@{}l@{}}Classifier\\ (2.1)\end{tabular} &
  \begin{tabular}[c]{@{}l@{}}Knowledge Engineer\\ (2.1)\end{tabular} \\ \hline
3 &
  \begin{tabular}[c]{@{}l@{}}Data Scientist\\ (1.2 + 2.2)\end{tabular} &
  \begin{tabular}[c]{@{}l@{}}Catalogue Code Designer\\ (1.2)\\ \\ Cataloguer\\ (2.2)\end{tabular} &
  \begin{tabular}[c]{@{}l@{}}Ontology Engineer\\ (1.2 + 2.3)\\ \\ Data Scientist\\ (2.2)\end{tabular} \\ \hline
4 &
  \begin{tabular}[c]{@{}l@{}}User\\ (3)\end{tabular} &
  \begin{tabular}[c]{@{}l@{}}Reader\\ (3)\end{tabular} &
  \begin{tabular}[c]{@{}l@{}}User\\ (3)\end{tabular} \\ \hline
\end{tabular}%
}
\\
\caption{Functional mapping between roles of KR, KO and KO-Enriched KR.}
\label{T1}
\end{table}

\noindent In the previous two sections, we've examined a detailed elucidation of the principal roles, activities and artifacts of both the Knowledge Representation and the facet-analytical Knowledge Organization methodology, respectively. In this section, before showing how to complete the loop \textit{from KR to KO and back}, let us first concentrate on a \textit{functional mapping} of, chiefly, the roles and the artifacts, of the two aforementioned methodologies (see the first three columns of Table \ref{T1}). Notice two things. Firstly, via functional mapping, we can ascertain whether or not two roles or artifacts perform the \textit{same or synonymous} broad function, irrespective of differences in their syntax, semantics or form. Secondly, the Table \ref{T1} illustrates only the roles as they embody the major functional differences between the two methodologies and not the artifacts (duely elucidated in the following description) which are functionally synonymous. The description constantly refer to Table \ref{T1}, and, to Figure \ref{I1} and Figure \ref{I2} as required, in the following discussion.

At the very outset, let us proceed the functional mapping on the basis of the four informal phases via which we detailed each individual methodology. In the first phase, there is a mapping between the roles of \textit{Language Engineer (1.1)} and that of \textit{Classificationist (1.1)}. This is clearly due to the fact that both these roles, based on input guiding principles, generate lexical-semantic classification hierarchies. Next, we also note that the artifacts that the above two roles produce, namely, the \textit{Lexical-Semantic Resource (a)} (see Figure \ref{I1}) and the \textit{Classification Schedule (a)} (see Figure \ref{I2}) are also in functional mapping to each other, given that both are essentially constituted of (a set of) lexical-semantic classification hierarchies focused on different subjects, domains, etc. Further, the roles \textit{Knowledge Engineer (2.1)} and \textit{Classifier (2.1)} are also mapped to each other because their central function is to use prescribed lexical-semantic hierarchies and classify the \textit{subject} of resources according to their relevant concept within the hierarchy. To that end, the output artifacts they produce, i.e., \textit{Lightweight Ontology (c)} (see Figure \ref{I1}) and \textit{Book Classification Number (c)} (see Figure \ref{I2}) also serve the synonymous function of encoding the categorization of either the entity dataset schema or the book subject matter.  

In the second phase, the \textit{ETG Repository (b)} (see Figure \ref{I1}) and \textit{Catalogue Code (b)} (see Figure \ref{I2}) are also functionally synonymous in the sense that both of these artifacts endow their input artifacts with reusable specifications of how to describe concepts via attributes within a classification hierarchy. The first major break in functional mapping occurs with respect to the role of \textit{Data Scientist (1.2 + 2.2)}. The Data Scientist role subsumes two roles, namely, that of the \textit{Catalogue Code Designer (1.2)} which exclusively deals with the specification of the descriptive (data) attribute schema in addition to that of the \textit{Cataloguer (2.2)} whose function is to generate the book catalogue record. Further, notice the functional synonymity between the two artifacts: \textit{Entity Graph (d)} (see Figure \ref{I1}) and \textit{Book Catalog Record (d)} (see Figure \ref{I2}), both of whose function is to provide the end user with an integrated view of top-down and bottom-up knowledge. Finally, the roles of a \textit{User (3)} and a \textit{Reader (3)} are also mapped as both function as end-users using the respective output artifacts of their methodologies in different ways. Notice also the fact that the activities across both the methodologies are also functionally mapped (see the labelled edges on Figure \ref{I1} and Figure \ref{I2}, respectively) with the only exception of the gap prior to the ETG repository in Figure \ref{I1}.

\begin{figure}[htp]
\includegraphics[width=16cm,height=8cm]{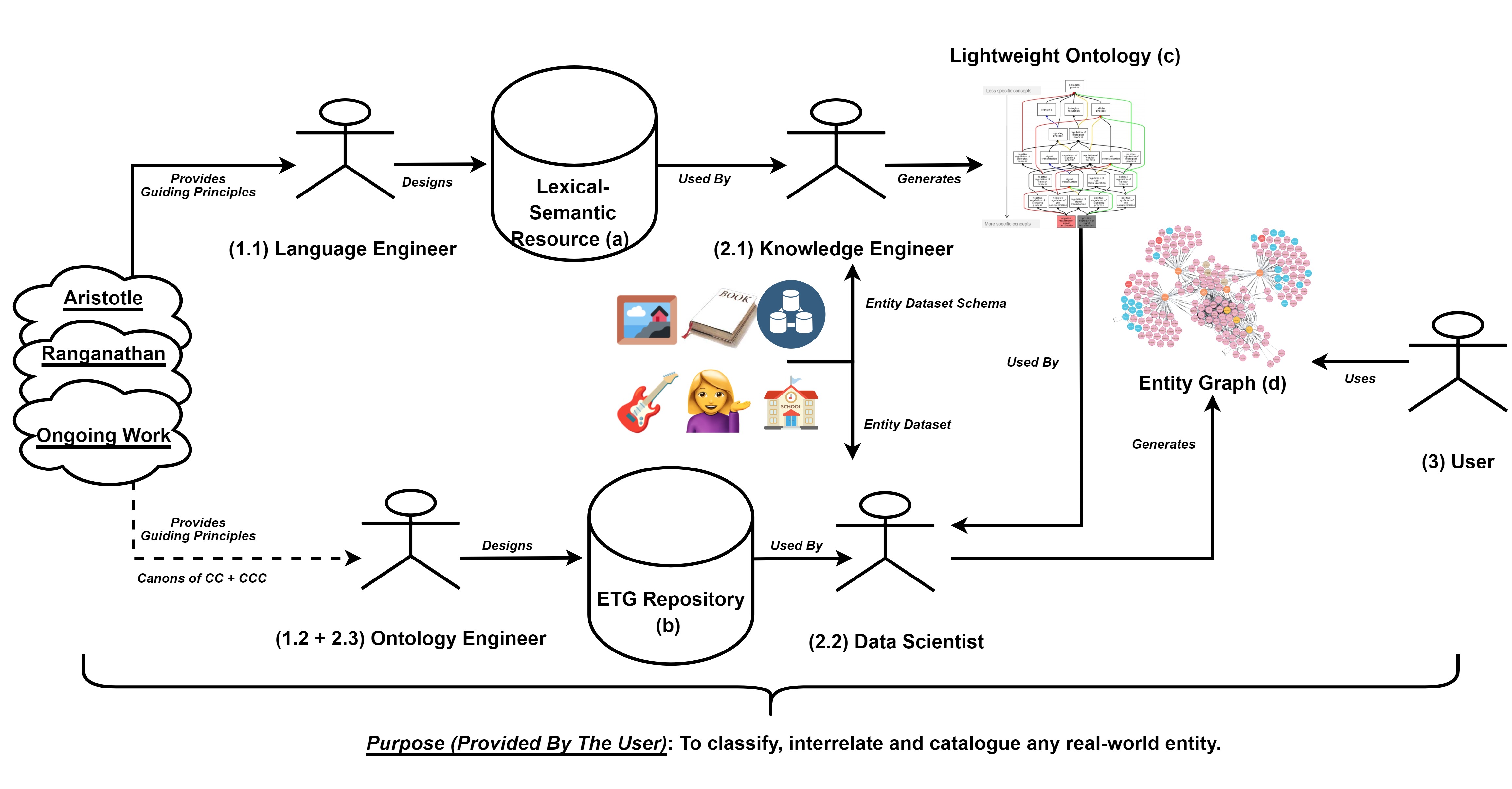}
\caption{A high-level view of the KO-Enriched KR Methodology.}
\centering
\label{I3}
\end{figure}

\noindent Given the functional mapping, let us now concentrate on how the guiding principles, i.e., \textit{canons}, advanced by the facet-analytic KO approach can be incorporated within the KR methodology to ultimately result in a KO-enriched KR methodology. To that end, let us concentrate on the various phases (harmonized from both the KR and KO methodologies via the functional mapping) which together compose to form the KO-enriched KR methodology (see Figure \ref{I3}). The methodology can be seen as being constituted of the following four distinct phases:
\begin{enumerate}
    \item The first phase initiating with guiding principles and concluding with the generation of the Lightweight Ontology.
    \item The second phase initiating with guiding principles (\textit{canons}) guiding the development of \textit{high-quality} Entity Type Graphs (ETGs) within the ETG repository.
    \item The third phase of the methodology concentrating on how the data scientist generates the Entity Graph (EG), and,
    \item The final phase concentrating on the different ways in which a user can use and exploit the EG.
\end{enumerate}
The diagrammatic symbols of Figure \ref{I3} are similar as before. We now consider each phase of the methodology briefly.

The first phase (see Figure \ref{I3}) is exactly the same as that of the original KR methodology (see Figure \ref{I1}). Briefly, the \textit{Language Engineer} designs a \textit{Lexical-Semantic Resource} composed of machine-processable language data, i.e., lexical-semantic hierarchies, designed by following the \textit{Genus-Differentia} guiding principle of Aristotle. These lexical-semantic hierarchies are then used by the \textit{Knowledge Engineer} in conjunction with the entity dataset schemas to generate the \textit{Lightweight Ontology}. The second phase of the methodology, i.e., the development of the ETG repository, is markedly different from that in Figure \ref{I1} and constitutes the core of the \textit{back} loop in \textit{from KR to KO and back}. It initiates with the still ongoing work (represented via dashed line in Figure \ref{I3}) on the adaptation of Ranganathan's guiding canons of both classification and cataloguing for the development of high-quality ETGs. Notice that the current effort (and the final goal) is to adapt all the relevant canons for classification (i.e., the different canons from the Idea Plane, Verbal Plane and Notational Plane) to generate taxonomically well-founded ETG hierarchies and the canons for cataloguing to mandate guidelines as to how (data) properties should be modelled to describe (conceptual) entities in such hierarchies. To that end, the methodology introduces a new role, that of an \textit{Ontology Engineer}, who will \textit{ensure} the development of the ETGs, and thereby, the ETG repository, in full conformance with the quality guidelines prescribed by the canons of Ranganathan suitably adapted. Notice that, with the support of Ranaganathan's canons of classification, the \textit{Ontology Engineer} would be able to generate ETGs not only restricted to macro-domains but also focused on \textit{fine-grained} micro domain of interests.

The third and the fourth phase of the methodology (see Figure \ref{I3}) is also the same as that of the original KR methodology (see Figure \ref{I1}). Briefly, the Data Scientist takes in input the lightweight ontology, the entity datasets and the relevant \textit{quality-enriched ETG} and suitably integrates them, in an iterative fashion, to ultimately generate an EG. Finally, the fourth phase concentrates on the ways in which the user can exploit the output EG. 

Notice also, from Table \ref{T1}, how the roles within the integrated KO-enriched KR methodology functionally compares with that of the original KR and facet-analytical KO methodology. The role of the \textit{Language Engineer (1.1)}, \textit{Knowledge Engineer (2.1)} and \textit{User (3)} in the integrated methodology is functionally similar to that of their counterparts in the KR and KO methodology (which have been functionally mapped before). The key difference, however, is reflected in the role of \textit{Ontology Engineer (1.2 + 2.3)}, which, builds the ETG hierarchy adhering to the \textit{canons} (a completely new role absent in the other methodologies) and additionally specifies the data attributes within the ETG (part of the role of \textit{Data Scientist} in the KR and the role of \textit{Catalog Code Designer} in the KO methodology). The \textit{Data Scientist (2.2)} in the KO-enriched KR methodology is functionally mapped to the role of \textit{Cataloguer} in the KO and to a part of the role of \textit{Data Scientist} in the KR methodology. This difference is because the former \textit{Data Sceintist (2.2)} role only integrates the previous three outputs to generate the EG, whereas, the latter role of \textit{Data Sceintist (1.2 + 2.2)}, in addition to the above function, also specifies data attributes (performed by the new role of \textit{Ontology Engineer} in the KO-enriched KR methodology). In terms of artifacts, the KO-enriched KR methodology is, functionally, the same as that of the KR methodology, and, thereof, to the KO methodology. Finally, in terms of activities, the KO-enriched KR methodology adds two extra activty flows over and above the KR methodology, namely: (i) the first activity flow from \textit{Ranganathan} and \textit{Ongoing Work} to the \textit{Ontology Engineer} in terms of adapting guiding principles of CC and CCC, and (ii) the activity flow from the \textit{Ontology Engineer} to the \textit{ETG Repository} in terms of designing the repository.

\section{The \textit{kTelos} Methodology}
\begin{figure}[htp]
\includegraphics[width=16cm,height=8cm]{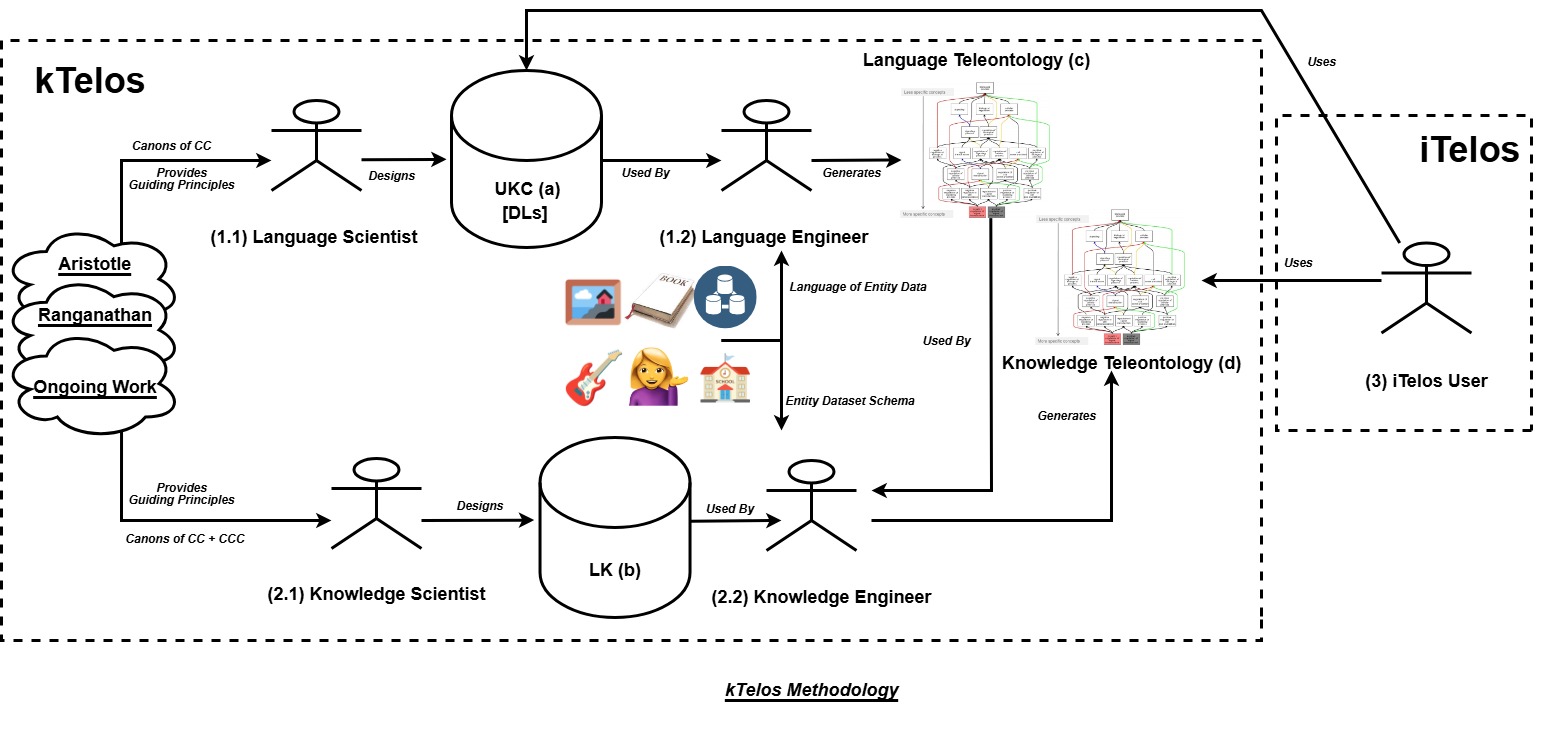}
\caption{A high-level view of the \textit{kTelos} methodology.}
\centering
\label{ktelos}
\end{figure}

\noindent Given a detailed delineation of the standard KR, KO and KO-enriched KR knowledge modelling methodologies, let us now concentrate on the top-down \textit{kTelos} methodology which methodologically integrates the solution components proposed in the thesis, i.e., the UKC (inclusive of the UKC namespaces and domain languages integrated within the UKC hierarchy), language teleontology and knowledge teleontology, to iteratively generate the language and knowledge representations absolving representation heterogeneity. The purpose of the methodology is to create, following a top-down approach, reusable language and knowledge representations, albeit informed by the representation heterogeneity underlying entities and real-world datasets in the context of a specific topic and/or similar topics. Notice that while the proposed \textit{kTelos} methodology is characteristically independent in terms of its input and output language and knowledge representation, the overall phase-wise process of the methodology has a functional symmetry and similarity with the KO-enriched KR methodology from which it is inspired and has evolved. In sync with its methodological predecessors as described before, the \textit{kTelos} methodology can also be seen as being constituted of the following four distinct phases:
\begin{enumerate}
    \item The first phase initiating with guiding principles to design the UKC (inclusive of the UKC namespaces and domain languages integrated within the UKC hierarchy) and concluding with the generation of the language teleontology.
    \item The second phase initiating with guiding principles informing the development of the LiveKnowledge catalog containing high-quality reusable language and knowledge resources.
    \item The third phase of the methodology concentrating on how the knowledge engineer generates the knowledge teleontology, and,
    \item The final phase concentrating on the different ways in which an iTelos user can use and exploit the domain language-enriched UKC and the knowledge teleontology.
\end{enumerate}
The diagrammatic symbols of the figure \ref{ktelos} are similar to the preceding methodologies from which it has been inspired. We now consider each phase of the \textit{kTelos} methodology briefly.

In the first phase (see Figure \ref{ktelos}), the \textit{Language Scientist} designs the \textit{UKC} composed of the lexical-semantic hierarchies of the UKC namespaces integrated as domain languages (DLs) within the overall UKC lexical-semantic hierarchy. The process underlying the design of the domain language-enriched UKC is that of the UKC annotation as described in section (3.4) of chapter (3). To that end, the guiding principles of \textit{Genus-Differentia} advanced by Aristotle as well as the canons of classification proposed by Ranganathan are natively satisfied in the very design of the domain languages and, therefore, don't need explicit discussion. The UKC namespaces as well as their final integration within the the UKC as domain languages are both machine-processable language data, i.e., a collection of lexical-semantic hierarchies of synsets \cite{PWN} codifying word meanings in different natural languages and/or domain languages represented in a machine processable format, with an additional language-independent conceptual layer unifying different language-specific lexical hierarchies.

Given the design of the \textit{UKC}, the final activity of this phase shifts to the \textit{Language Engineer} (see Figure \ref{ktelos}) who now generates the \textit{Language Teleontology} (Figure \ref{ktelos}) which is an intermediate machine-processable formal hierarchy encoding concept names for entity types and properties that are being considered for representing knowledge in the context of a specific topic. To that end, the \textit{Language Engineer} has to take in two important inputs. Firstly, (s)he has to take in input the appropriate domain language which will inform the terminology of the \textit{Language Teleontology} (s)he will generate. In addition, (s)he also consults the language of data \cite{bella2020exploring} encoded in representationally heterogeneous datasets of the relevant topic (referred to as ``language of entity data" in the figure) to generate the \textit{language teleontology}.

The second phase, as illustrated in Figure \ref{ktelos}, concerns the design and development of the LiveKnowledge (LK) repository (exposed via the LiveKnowledge catalog) by the \textit{Knowledge Scientist}, wherein, knowledge resources in terms of \textit{exisitng} knowledge teleontologies and/or \textit{exisitng} legacy schemas representing entity types \cite{ETR} and properties are available for reuse. Notice the fact that the above knowledge representations specialize the notion of properties into \textit{object properties} (modelling how an entity type is interrelated to other entity types) and \textit{data properties} (modelling the attributes which describe an entity type). Especially, the repository would facilitate not only interoperability amongst knowledge teleontologies but also promote the circular sharing and reuse of knowledge teleontologies (partially shown in the figure due to constraint of space) such as, for instance, the reuse of the same knowledge teleontology but for different use case scenarios. Note also the fact the design of the LiveKnowledge ecosystem (inclusive of its constituent resources) adhere to the general guiding principles of Aristotle and Ranganathan as depicted in the figure. This specific aspect is also an ongoing work in terms of developing newer guiding principles on aspects not covered by the principles mentioned above.

The third phase concentrates on how the \textit{Knowledge Engineer} (see Figure \ref{ktelos}) exploits the outputs of the previous two phases and \textit{consults} the representationally heterogeneous entity dataset schemas to generate the knowledge teleontology. To that end, the \textit{Knowledge Engineer} receives in input the language teleontology (output of the first phase) which is then grounded in the relevant lexical-semantically compliant reusable knowledge representation from the LiveKnowedge repository (output of the second phase). Given the language teleontology-LK resource combined knowledge representation, the \textit{Knowledge Engineer} consults the entity dataset schemas to ensure relevant common and core entity types, object properties and data properties (see \cite{KGSWC} are represented in the combined artifact and, thereby, generates the knowledge teleontology.

Finally, the fourth phase concentrates on the different ways in which different \textit{iTelos Users} can exploit language teleontologies and/or the knowledge teleontologies generated by the aforementioned process for the language definition and knowledge definition phases of the iTelos KG construction. Notice that the first three phases are core to the \textit{kTelos} methodology and the fourth phase, while being a part of the methodology, can only be delineated and expanded in detail as part of the iTelos process (which is outside of the main scope of this thesis).

There are two key observations relevant to the methodology as described above and the language and knowledge representations which are developed as its outputs. First, note that while language representations (i.e., UKC namespaces, domain languages) and knowledge representations (i.e., language teleontologies, knowledge teleontologies) can vary within and across topics/domains, what remains \textit{invariant} is the process underlying the \textit{kTelos} methodology which facilitates the design and development of such representations. Second, the language and knowledge representations developed following the methodology can be qualitatively evaluated in terms of whether the representations accommodate different levels of representation heterogeneity in terms of conceptual unity/diversity, language unity/diversity and knowledge unity/diversity.

\section{Related Work}
Let us now concentrate on relating the proposed methodology to state-of-the-art knowledge modelling methodologies, e.g., in ontology development. The survey in \cite{OEM-IJCAI} provides an analysis of several early generation methodologies (e.g., \cite{TOVE,OEM-Uschold,OEM-UGruninger,SENSUS}) with respect to parameters adapted from software life cycle process evaluation. METHONTOLOGY \cite{METHONTOLOGY} proposed a \emph{``life cycle to build ontologies based in evolving prototypes"} and employs it to develop an ontology in the domain of chemicals. Ontology Development 101 \cite{OD101}, instead, offered the flexibility of choosing top-down, bottom-up or middle-out approaches in engineering ontologies. KACTUS \cite{KACTUS} was one of the early methodologies grounded in reuse of knowledge via their notion of \emph{``levelling of ontologies"}. More recently, the NeOn methodology \cite{NeOn} offers a set of \emph{very generic} scenarios for reuse, re-engineering and merging of ontological resources. The work in \cite{DERA} proposed a faceted KR for the development of ontologies able to describe and reason about relevant entities of a domain. In a similar vein, the methodology proposed in \cite{yamo} advances a step-by-step process of building a formally defined large-scale faceted ontology.The eXtreme Design (XD) methodology \cite{XD}, on the other hand, is very specific, in the sense that it is grounded on reusage of content ontology design patterns for modelling new ontologies. MOMo \cite{MOMO}, the most recent methodology, builds on \cite{XD} by adding the conceptual and tooling support for \emph{``graphical schema diagrams"} for knowledge elicitation from experts. Differently from the above methodologies, but very closely related to them, is the \emph{Telos} conceptual modelling \emph{language} \cite{telos1,telos2}, which, being originally envisioned to support \emph{``the development of information systems throughout the software lifecycle"} \cite{telos1}, has had implementations in varied arenas such as data integration, cultural informatics and ontology engineering (detailed in \cite{telos3}). The methodology proposed in this chapter, while partially sharing aspects of the above methodologies, is unique in at least three dimensions. First, the \textit{kTelos} methodology is designed on the basis of an explicit formalism for stratification of representation into concepts, language and knowledge representations. Second, as a consequence of the first dimension, the methodology is designed to accommodate and address representation heterogeneity in terms of conceptual unity/diversity, language unity/diversity and knowledge unity/diversity towards resolving data unity/diversity in KGs. Third, the methodology is also designed to emphasize an ecosystem of iterative reuse of language and knowledge representations with an overarching objective of resolving existent heterogeneity with each iteration.

\section{Summary}
To summarize, this chapter discussed the top-down \textit{kTelos} methodology to iteratively generate the UKC-based language and knowledge representations iteratively addressing representation heterogeneity. To that end, it discussed the evolution of the \textit{kTelos} methodology starting from an initial delineation of state-of-the-art KO and KR methodologies. The next chapter will discuss and exemplify a proof-of-concept for the UKC-based language and knowledge representations proposed in this thesis in the context of the DataScientia (data catalogs) initiative.

    \chapter{CASE STUDY ON DATA CATALOGS}

\section{Introduction}
As the first proof-of-concept, the thesis briefly illustrates a first version of language and knowledge representation produced in the DataScientia\footnote{https://datascientia.disi.unitn.it/} initiative following the UKC-based language and knowledge representation formalism and process advanced in the thesis. The DataScientia initiative is a virtual community-driven data science, research and innovation platform and ecosystem composed of several different components, including, international projects, members, participants and educational courses which generate datasets which are then populated and exposed via metadata-enriched data catalogs. An important effort within the DataScientia initiative is to produce an Entity Graph to model and integrate representationally heterogeneous data about different aspects of DataScientia, including about the aspects mentioned above. Considering this objective, the language and knowledge resources to ultimately enable the development and publication of DataScientia Entity Graphs are being created and updated by the University of Trento in collaboration with different research partners working within the DataScientia initiative. To that end, first, the section provides a brief informal background on some core concepts relevant to be modelled for the Entity Graphs within the DataScientia initiative. Second, the section illustrates and elucidates snapshots of a first version of the UKC-based language and knowledge representation developed as a part of the DataScientia initiative following the representation formalism and process proposed in the thesis. It also includes a brief discussion on how the language and knowledge representations produced within the initiative can be qualitatively evaluated in the context of the overarching problem of representation heterogeneity as proposed and considered in this thesis.

\section{Background}
This section concentrates on the informal background essential for representing some selected core concepts of the DataScientia initiative and describing them via appropriate (data) properties. Notice that the prefix ``\texttt{ds}" prepended in front of each concept or property is to indicate the fact that each entry within this informal body of concepts and associated properties are developed with the objective of (later) developing a dedicated DataScientia UKC namespace of which a first version is briefly described and illustrated in this chapter. First, some of the properties describing different roles within the DataScientia initiative are as follows.

\begin{enumerate}
    \item ds:member: The properties assigned to describe the concept of a \emph{Member} are generic in nature and can be used for any of the specific roles which are part of the DataScientia community ecosystem. The (public) properties for members are as follows:
    \begin{enumerate}
        \item ds:comIdentifier: this property encodes the DataScientia community identifier uniquely identifying a person within the DataScientia ecosystem.
        \item ds:firstName: this property encodes the first name of the person in a natural language.
        \item ds:lastName: this property encodes the last name of the person in a natural language.
        \item ds:email: this property encodes the email address of the person.
        \item ds:nationality: this property encodes the nationality of the person.
        \item ds:gender: this property encodes the gender of the person.
        \item ds:affiliation: this property encodes the organization to which the person is affiliated in a natural language.
        \item ds:personalWebpage: this property encodes the Uniform Resource Locator (URL) of the personal webpage of the person.
    \end{enumerate}
    \item ds:prjResearcher: includes all members of the DataScientia community ecosystem who are additionally a researcher within a particular DataScientia project. This role inherits all the properties for \emph{ds:member}, plus, the following properties:
    \begin{enumerate}
        \item ds:researcherProfile: this property encodes the URL of a research profile of the researcher (e.g., Google Scholar).
        \item ds:areasOfInterest: this property lists the areas of research interest of the researcher.
        \item ds:nameOfPrjsCoordinating: this property encodes the DataScientia projects which are being coordinated by the researcher.
    \end{enumerate}
    \item ds:prjParticipant: includes all members of the DataScientia community ecosystem who are additionally a participant within a particular DataSceintia project. This role inherits all the properties for \emph{ds:member}, plus, the following property:
    \begin{enumerate}
        \item ds:nameOfPrjsPaticipatingIn: this property encodes the DataScientia projects in which the participant is contributing to.
    \end{enumerate}
\end{enumerate}

\noindent Let us now concentrate on describing DataScientia projects. The different (public) properties describing a project are:

\begin{enumerate}
    \item ds:prjTitle: this property encodes the name of the DataScientia project in a natural language as a string.
    \item ds:prjURL: this attribute encodes the dereferenceable URL of the DataScientia project.
    \item ds:prjKeywords: this property encodes the various keywords in a natural language that can be utilized to quickly understand the theme of the project.
    \item ds:prjType: this property encodes the type of the DataScientia project. e.g., Knowledge Resource Generation, Knowledge Resource Annotation, etc.
    \item ds:prjDescription: this property can be used to provide a description of the DataScientia project in a natural language.
    \item ds:prjStartDate: this property encodes the date of the commencement of a DataScientia project. 
    \item ds:prjEndDate: this property encodes the date of conclusion of a DataScientia project.
    \item ds:prjFundingAgency: this property encodes the name of the agency or institution funding a DataScientia project.
    \item ds:prjInput: this property encodes the various inputs (e.g., datasets, specifications, etc.) with respect to a DataScientia project.
    \item ds:prjOutput: this property encodes the various outputs (e.g., datasets, domain languages, etc.) with respect to a DataScientia project.
    \item ds:prjCoordinator: this property encodes the name of the research coordinator in charge of a DataScientia project.
    \item ds:prjObservations: this property can be used to record any observations about a DataScientia project in a natural language.
\end{enumerate}

\noindent The (public) properties for representing a DataScientia dataset are listed as follows. Notice that these properties apply to all types of datasets (e.g., LiveLanguage Datasets, LiveKnowledge Datasets, etc.)

\begin{enumerate}
    \item ds:DatLicense: this property encodes the license of the dataset, e.g., CC-BY-SA-4.0 
    \item ds:DatURL: this property encodes the dereferenceable URL of the dataset. 
    \item ds:DatKeyword: this property encodes the keywords which can quickly convey the topic of the dataset.
    \item ds:DatPublisher: this property encodes the publisher of the dataset.
    \item ds:DatCreator: this property encodes the creator of the dataset.
    \item ds:DatOwner: this property encodes the owner of the dataset.
    \item ds:DatLanguage: this property encodes the natural language(s) in which the dataset information is represented.
    \item ds:DatLevel: this property encodes the representation level of the dataset.
    \item ds:DatSize: this property encodes the byte size of the dataset. 
    \item ds:DatName: this property encodes the name of the dataset in a natural language.
    \item ds:DatPublicationTimestamp: this property encodes the timestamp of the publication of the dataset in the respective catalog.
    \item ds:DatDescription: this property encodes the description about the dataset in a natural language.
    \item ds:DatVersion: this property encodes the version of the dataset. 
    \item ds:DatFileFormat: this property encodes the file format of the dataset.
\end{enumerate}

\noindent The (public) properties for a LiveLanguage dataset are listed as follows. Notice that all properties for a DataScientia dataset in general also applies to a LiveLanguage dataset.

\begin{enumerate}
    \item ds:LLDatCategory: refers to the category the dataset belongs to from the list as follows: UKC Lexicons, Raw Monolingual datasets, Cross-lingual datasets. 
    \item ds:LLDatMaintainer: maintainer is the personnel responsible for the modifying dataset published on  catalog. 
    \item ds:LLDatMaintainerEmail: email address of the maintainer.
    \item dds:LLDatTags: tags associated with the datasets. Example: Dataset Mongolian WordNet, Tags: WordNet, Mongolian. 
    \item ds:LLDatDownloadAccessLevel: access Level for the end-user to download the dataset. Example: Open Access.
    \item ds:LLDatLandingPage: refers to a web page that provides access to the Dataset, its Distributions and/or additional information.
    \item ds:LLDatDateOfCollection: the date signifying when the data was collected.
    \item ds:LLDatWords: statistical metadata with the number of words present in the lexicon.
    \item ds:LLDatSynsets: statistical metadata with the number of synsets present in the lexicon.
    \item ds:LLDatSenses: statistical metadata with the number of senses present in the lexicon.
    \item ds:LLDatSynsetRelations: statistical metadata with the number of synset relations present in the lexicon.
    \item ds:LLDatSensesRelations: statistical metadata with the number of sense relations present in the lexicon.
    \item ds:LLDatMoreInformation: link to the webpage which provides more information about the dataset.
    \item ds:LLDatLastUpdated: timestamp giving information about the date and time the dataset was last updated.
    \item ds:LLDatReleaseDate: timestamp giving information about the date and time the dataset was released.
    \item ds:LLDatISO3LanguageCode: unique language code representing the language of the lexicon in the dataset. Example: Ari UKC lexicon code: aac.
    \item ds:LLDatLanguageType: type of the languages represented in the dataset. Ex: Multi-lingual / Mono-lingual / Cross-lingual.
\end{enumerate}

\noindent The one specialized (public) property for a LiveKnowledge dataset is listed below. Notice that all metadata properties for a DataScientia dataset in general also applies to a LiveKnowledge dataset.

\begin{enumerate}
    \item ds:LKDatType: type of the LK dataset, e.g., UKC namespaces, language teleontology, knowledge teleontology.
\end{enumerate}

\noindent There are two observations to be noted for the aforementioned elucidation of the informal background of the proof-of-concept for language and knowledge representation for the DataScientia initiative. First, notice that the informal background is being continuously developed and updated as newer constituent concepts and newer properties of existing constituent concepts within the DataScientia initiative are being deliberated upon and finalized in collaboration with members from different international research partners. To that end, the above description focuses on concepts and properties related to members, projects and datasets and skips detailing concepts/properties related to courses, interest groups, member profiles, etc., which are being developed, refined and updated constantly. Second, note that the above listing is restricted only to properties which can be exposed publicly via the DataScientia community platform and excludes private properties which are essential only for back-end administration and database management purposes.

\section{Language Representation}
\begin{figure}[htp]
\includegraphics[width=16cm,height=6cm]{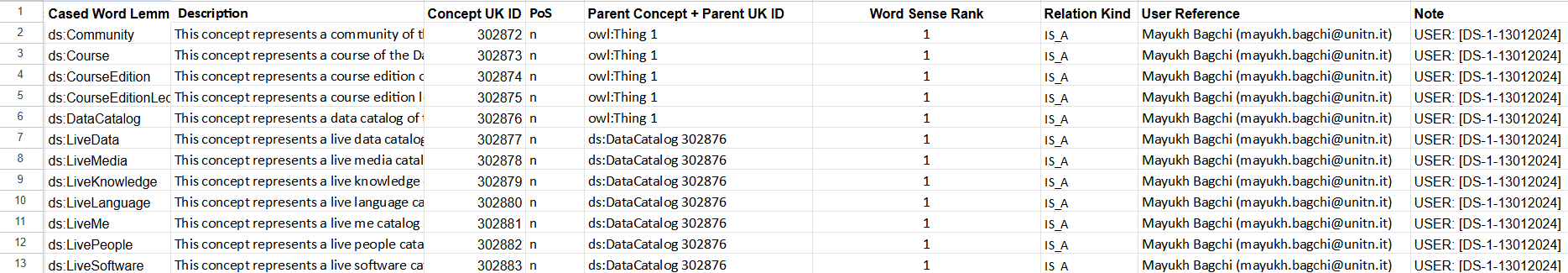}
\caption{A fragment of the DataScientia UKC namespace.}
\centering
\label{dsns}
\end{figure}

Let us now concentrate on the first version of the UKC-based language representation developed for the DataScientia initiative following the formalism and process proposed in the thesis. An illustrative example of a portion of the fully annotated spreadsheet of the DataScientia UKC namespace is provided in Figure \ref{dsns} which has been generated following an end-to-end execution of the UKC annotation process described in section (3.4) of chapter (3). Note that the namespace prefix \texttt{ds} for the DataScientia UKC namespace has been adopted from the informal background as specified above. Further, notice also the fact that since the UKC annotation process has been executed, the terms in this version of the DataScientia UKC namespace are integrated into the lexical-semantic hierarchy of the UKC as a domain language.

\noindent The first column \texttt{Cased Word Lemma} lists sequentially (the hierarchy of) concept label(s) which underwent UKC annotation (e.g., \texttt{ds:Course}). The second and the third columns encodes a gloss of the relative term (e.g., \texttt{\textit{This concept represents a course of the DataScientia initiative}}) and the concept GID (e.g., \texttt{302873}) following the UKC annotation process. The fourth and the fifth columns record the part of speech category (e.g., \texttt{n (noun)}) and the parent concept as well as the parent concept GID (e.g., \texttt{owl:Thing 1}) of the selected concept/term, respectively. The sixth column records the word sense rank relative to the chosen concept (e.g., \texttt{1}). The seventh column records the ontological relation existing between the selected concept label and its parent (e.g., \texttt{IS\_A}). Finally, the eighth and the ninth columns provide the user reference information (e.g., \texttt{Mayukh Bagchi (mayukh.bagchi@unitn.it)}) and the version associated to each selected concept label (e.g., \texttt{USER: [DS-1-13012024]}). 

\section{Knowledge Representation}
Given an illustration of the DataScientia UKC namespace, let us now concentrate on the first version of the UKC-based knowledge representation developed for the DataScientia initiative following the formalism and process proposed in the thesis. First, illustrative examples of portions of the first version of the DataScientia language teleontology are discussed. Notice that these representations follow the knowledge representation conventions discussed in chapter (4).

\begin{figure}[htp]
\includegraphics[width=16cm,height=10cm]{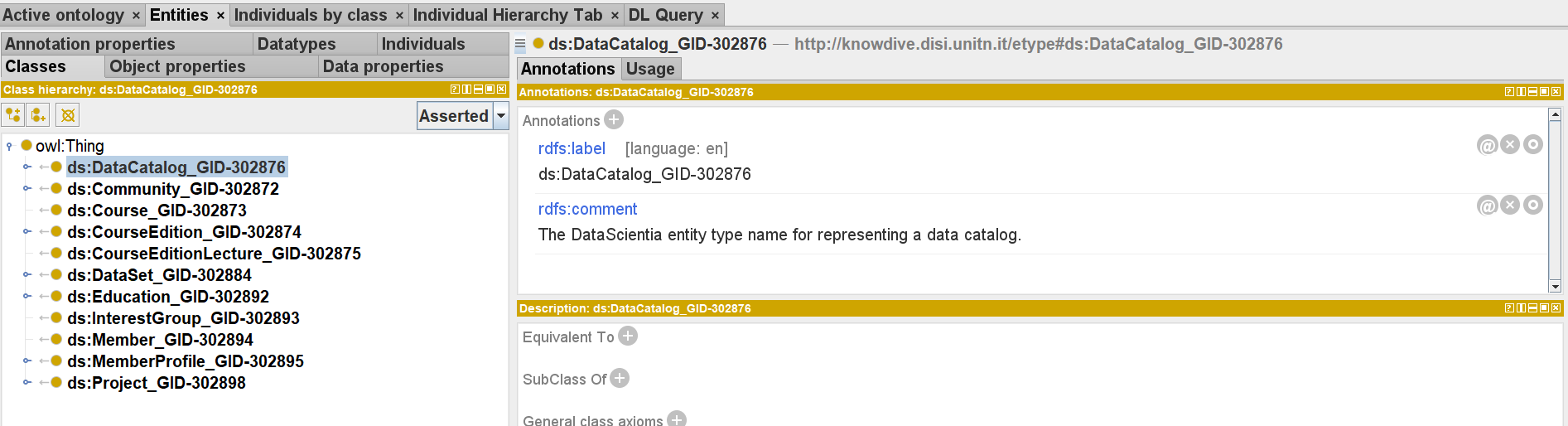}
\caption{A fragment of the top-level of the entity type name hierarchy of the DataScientia language teleontology.}
\centering
\label{dslt-ettlv}
\end{figure}

First, let us examine a fragment of the top-level of the DataScientia language teleontology entity type name hierarchy as visualized in the Figure \ref{dslt-ettlv}. The top-level entity type names are organized as a hierarchy as visualized from the left-half of the figure encoding concepts such as data catalog, community, course, course edition, dataset, ineterst groups, members, etc. Further, the meaning of each entity type name (e.g., \texttt{ds:DataCatalog\_GID-302876}) is uniquely identified and disambiguated by a UKC GID (e.g., \texttt{302876}) appended to the label of the name (e.g., \texttt{ds:DataCatalog\_GID-302876}). The label of the entity type names also display the prefix (e.g., \texttt{ds}) of the DataScientia UKC namespace from which it has been reused. The entity type name is annotated with information such as an \texttt{rdfs:comment} on the right-half of the figure modelled as an annotation property. Further, notice the entity type name is not defined, i.e., there are no associated axiomatic assertions and/or property constraint, e.g., disjointness assertions, class expressions via asserted property constraints, etc., associated to the selected entity type name \texttt{ds:DataCatalog\_GID-302876}. 

\begin{figure}[htp]
\includegraphics[width=16cm,height=10cm]{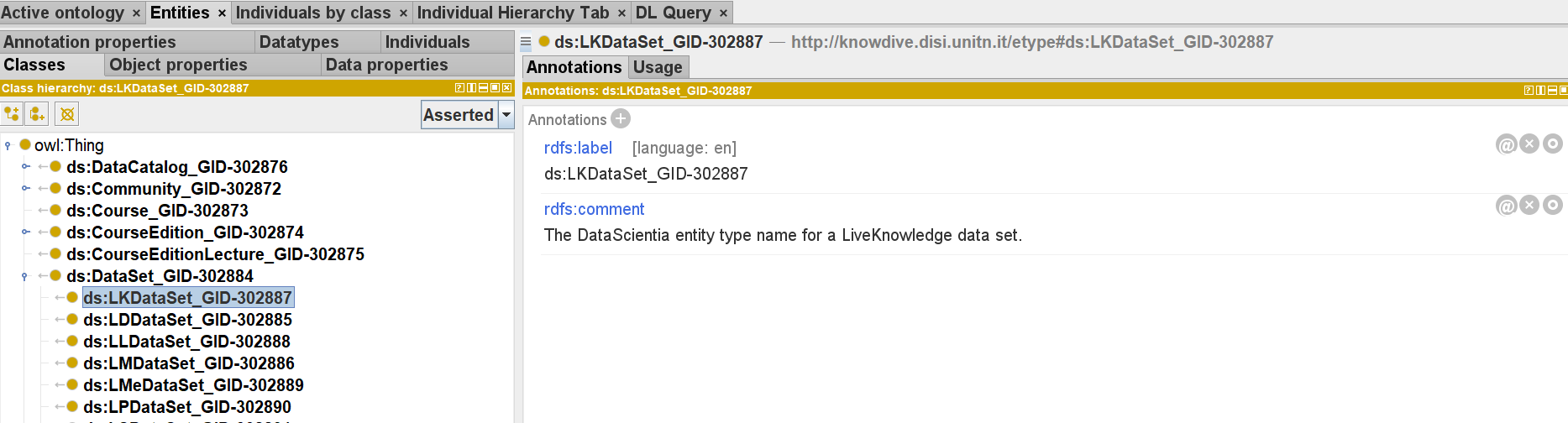}
\caption{A fragment of an expanded view of the  entity type name hierarchy of the DataScientia language teleontology.}
\centering
\label{dslt-etexp}
\end{figure}

Second, let us exemplify a fragment of an expanded view of the top-level of the DataScientia language teleontology entity type name hierarchy as depicted in the Figure \ref{dslt-etexp}. A portion of the top-level entity type names as described before, i.e., the sub-hierarchy of \texttt{ds:DataSet\_GID-302884}, have been expanded as visualized from the left-half of the figure encoding concepts such as LL Dataset, LK Dataset and LP Dataset. The meaning of each entity type name in the expanded hierarchy (e.g., \texttt{ds:LKDataSet\_GID-302887}) is uniquely identified and disambiguated by a UKC GID (e.g., \texttt{302887}) appended to the label of the name (e.g., \texttt{ds:LKDataSet\_GID-302887}). As from before, the label of the entity type names also display the prefix (e.g., \texttt{ds}) of the DataScientia UKC namespace from which it has been reused. The entity type name is annotated with information such as an \texttt{rdfs:comment} on the right-half of the figure modelled as an annotation property. Finally, similar to the explanation before, the entity type name is not defined, i.e., there are no associated axiomatic assertions and/or property constraint, e.g., disjointness assertions, class expressions via asserted property constraints, etc., associated to the selected entity type name \texttt{ds:LKDataSet\_GID-302887}.

\begin{figure}[htp]
\includegraphics[width=16cm,height=10cm]{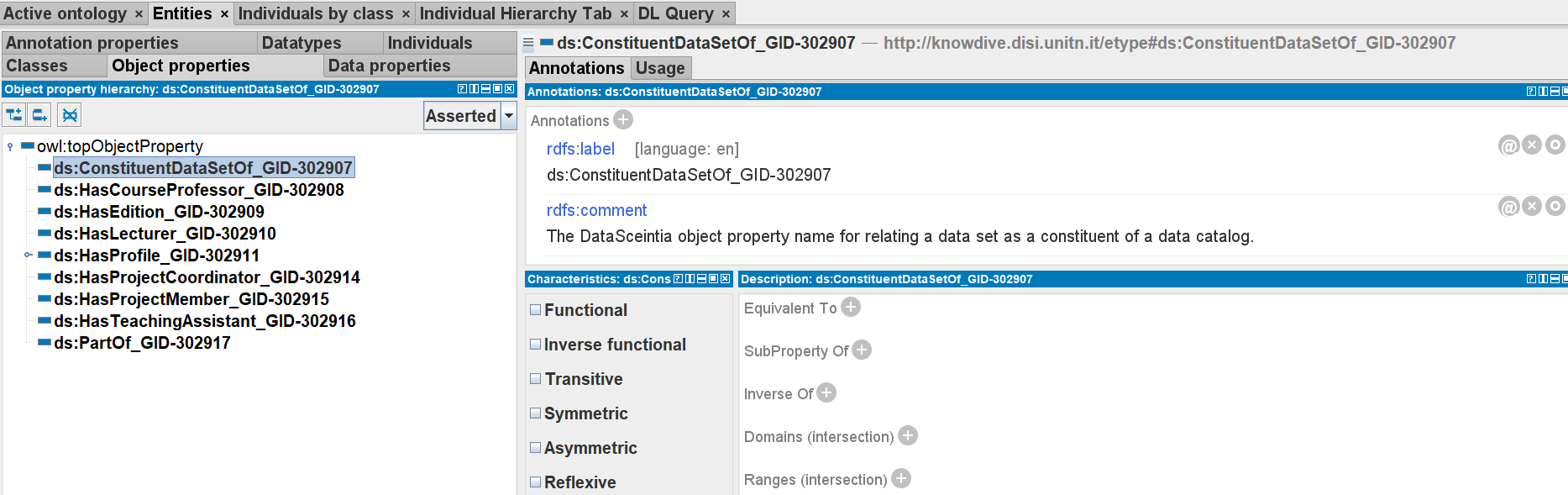}
\caption{A fragment of the  object property name hierarchy of the DataScientia language teleontology.}
\centering
\label{dslt-oph}
\end{figure}

Third, let us examine a fragment of the object property names of the DataScientia language teleontology visualized in the Figure \ref{dslt-oph}. The object property names are organized as a hierarchy as visualized from the left-half of the figure. Further, the meaning of each object property name (e.g., \texttt{ds:ConstituentDataSetOf\_GID-302907}) is uniquely identified and disambiguated by a UKC GID (e.g., \texttt{302907}) appended to the label of the name (e.g., \texttt{ds:ConstituentDataSetOf\_GID-302907}). Like before, the label of the object property names also display the prefix (e.g., \texttt{ds}) of the DataScientia UKC namespace from which it has been reused. The object property name is also annotated with information such as an \texttt{rdfs:comment} on the right-half of the figure modelled as an annotation property. Finally, it is interesting to note that the object property name is not associated to axiomatic assertions, e.g., characteristic assertions, disjointness assertions, domain and/or range assertions, etc, associated to the selected entity type name \texttt{ds:ConstituentDataSetOf\_GID-302907}.

\begin{figure}[htp]
\includegraphics[width=16cm,height=10cm]{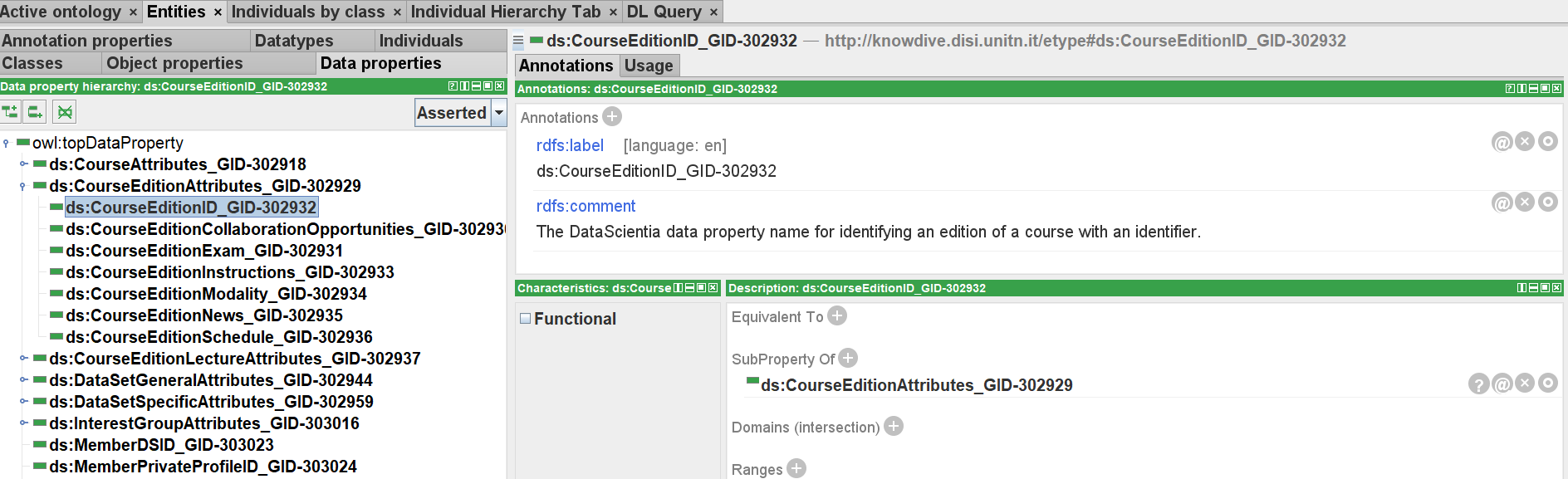}
\caption{A fragment of the  data property name hierarchy of the DataScientia language teleontology.}
\centering
\label{dslt-dph}
\end{figure}

Finally, let us examine a fragment of the data property names of the DataScientia language teleontology visualized in the Figure \ref{dslt-dph}. The data property names are organized as a hierarchy as visualized from the left-half of the figure. Further, the meaning of each data property name (e.g., \texttt{ds:CourseEditionID\_GID-302932}) is uniquely identified and disambiguated by a UKC GID (e.g., \texttt{302932}) appended to the label of the name (e.g., \texttt{ds:CourseEditionID\_GID-302932}). The label of the data property names also display the prefix (e.g., \texttt{ds}) of the DataScientia UKC namespace from which it has been reused. The object property name is also annotated with information such as an \texttt{rdfs:comment} on the right-half of the figure modelled as an annotation property. Lastly, it is interesting to note that the object property name is not associated to axiomatic assertions, e.g., characteristic assertions, disjointness assertions, domain and/or range assertions, data type assertions, etc, associated to the selected entity type name \texttt{ds:CourseEditionID\_GID-302932}. 

Let us now concentrate on illustrations-cum-elucidation of the first version of the DataScientia knowledge teleontology.

\begin{figure}[htp]
\includegraphics[width=16cm,height=8cm]{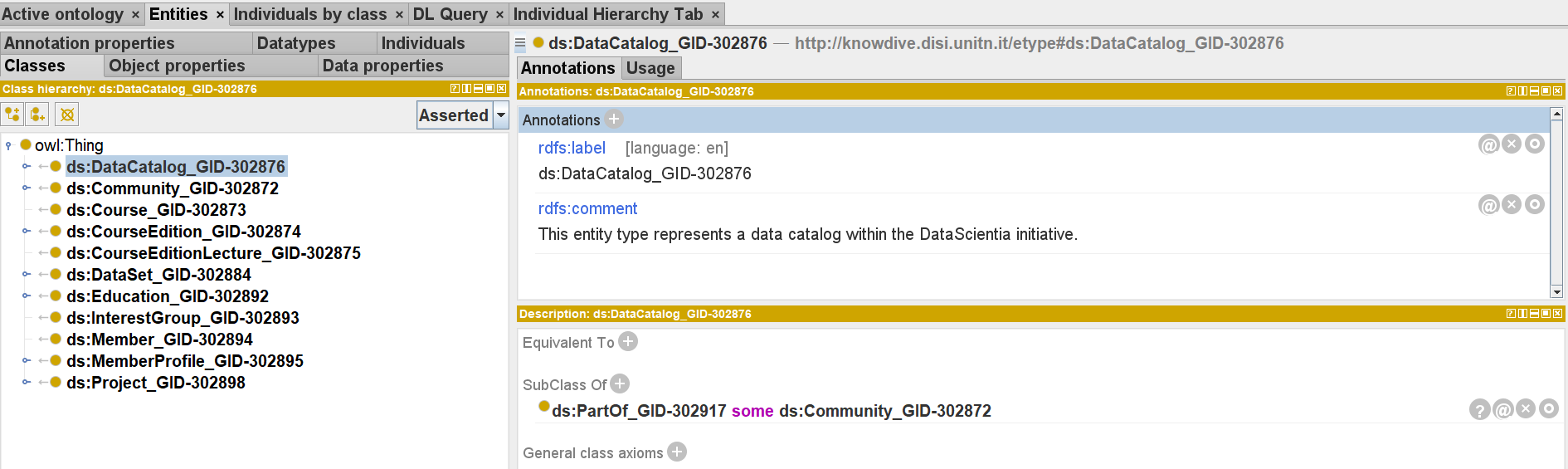}
\caption{A fragment of the top-level of the  entity type hierarchy of the DataScientia knowledge teleontology.}
\centering
\label{dskt-ettlv}
\end{figure}

First, let us focus on a fragment of the top-level entity type view of the DataScientia knowledge teleontology as visualized in the Figure \ref{dskt-ettlv}. The top-level of the entity types are organized as a hierarchy as visualized from the left-half of the figure encoding entity types such as data catalog, community, course, course edition, dataset, ineterst groups, members, etc. Further, the meaning of the name of each entity type (e.g., \texttt{ds:DataCatalog\_GID-302876}) is uniquely identified and disambiguated by a UKC GID (e.g., \texttt{302876}) appended to the label of the name (e.g., \texttt{ds:DataCatalog\_GID-302876}). Further, in sync with the language teleontology, the label of the entity types also display the prefix (e.g., \texttt{ds}) of the DataScientia UKC namespace from which it has been reused. The entity types are annotated with information such as an \texttt{rdfs:comment} on the right-half of the figure modelled as an annotation property. Further, notice the entity types are defined, i.e., there are axiomatic assertions and/or property constraint associated to the entity type using the object and data properties defined. For example, for the entity type \texttt{ds:DataCatalog\_GID-302876}, the axiomatic descriptions include, e.g., object restriction constraint (e.g., \texttt{ds:PartOf\_GID-302917 \textbf{some} ds:Community\_GID-302872}).

\begin{figure}[htp]
\includegraphics[width=16cm,height=8cm]{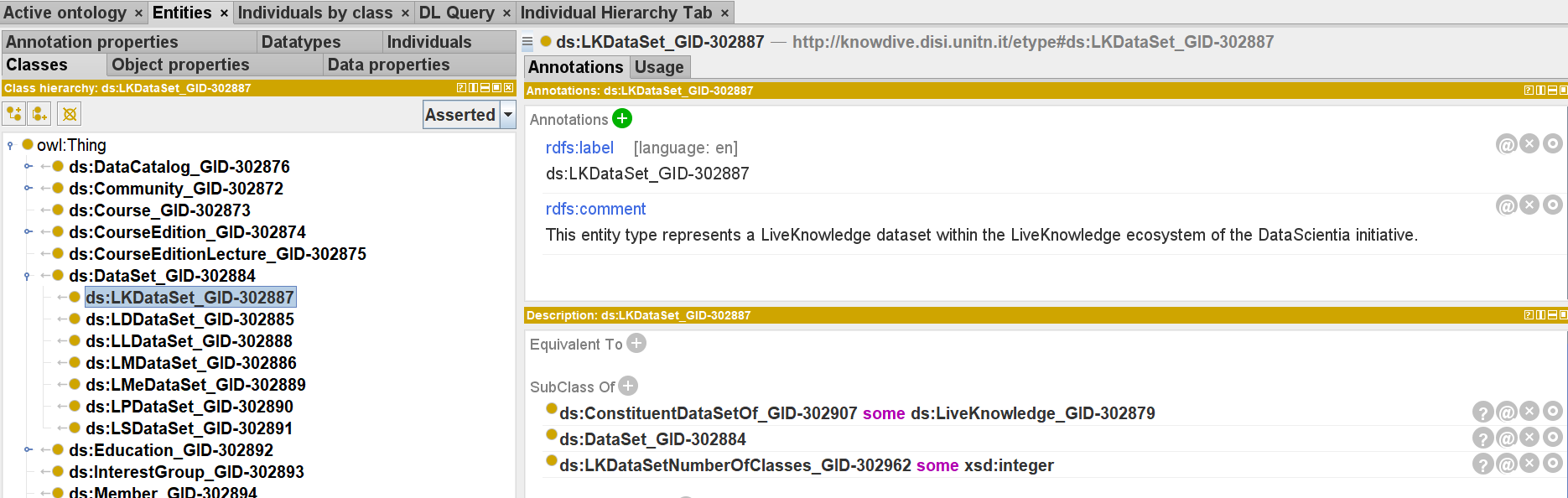}
\caption{A fragment of an expanded view of the  entity type hierarchy of the DataScientia knowledge teleontology.}
\centering
\label{dskt-etexp}
\end{figure}

Second, let us examine a fragment of an expanded view of the top-level of the DataScientia knowledge teleontology visualized in the Figure \ref{dskt-etexp}. The entity types are organized as a hierarchy as visualized from the left-half of the figure with the sub-hierarchy of \texttt{ds:DataSet\_GID-302884} expanded and shown in the figure encoding entity types such as LL Dataset, LK Dataset and LP Dataset. Further, the meaning of the name of each entity type (e.g., \texttt{ds:LKDataSet\_GID-302887}) is uniquely identified and disambiguated by a UKC GID (e.g., \texttt{300887}) appended to the label of the name (e.g., \texttt{ds:LKDataSet\_GID-302887}). The entity types are annotated with information such as an \texttt{rdfs:comment} on the right-half of the figure modelled as an annotation property. Finally, the entity types are defined, i.e., there are axiomatic assertions and/or property constraint associated to the entity type using the object and data properties defined. For example, for the entity type \texttt{ds:LKDataSet\_GID-302887}, the axiomatic descriptions include, e.g., data restriction constraint (e.g., \texttt{ds:LKDataSetNumberOfClasses\_GID-302962 \textbf{some} xsd:integer}), object restriction constraint (e.g., \\ \texttt{ds:ConstituentDataSetOf\_GID-302907 \textbf{some} ds:LiveKnowledge\_GID-302879}).

\begin{figure}[htp]
\includegraphics[width=16cm,height=8cm]{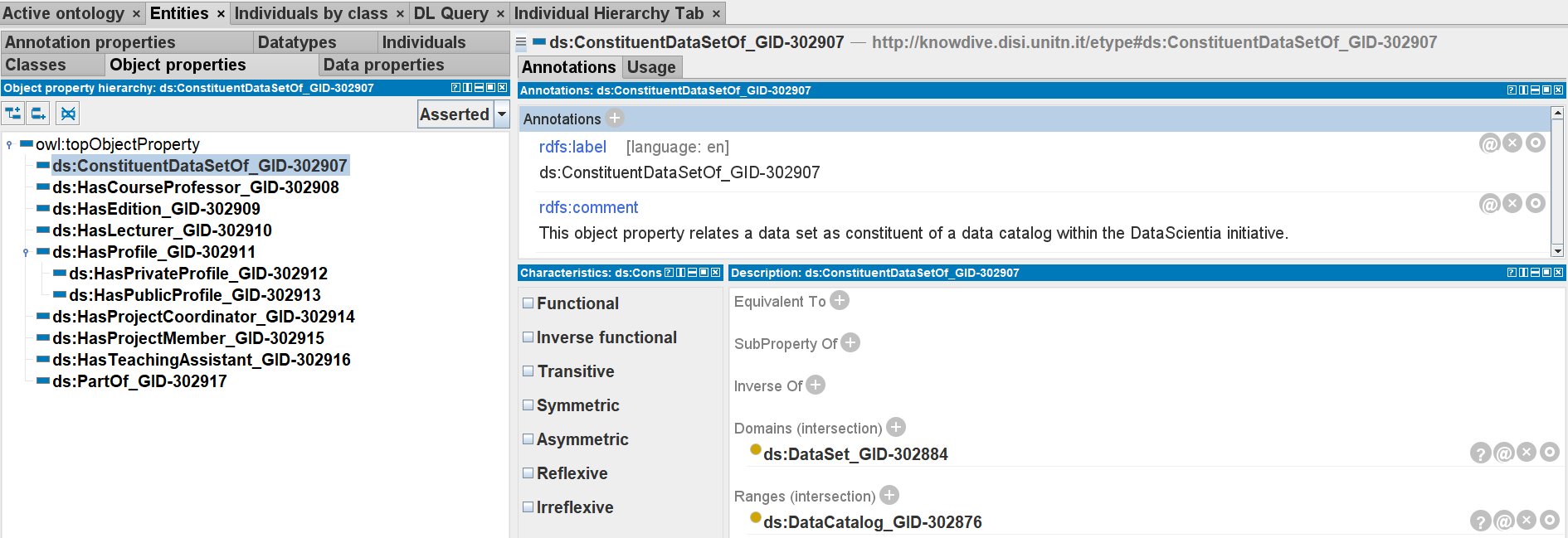}
\caption{A fragment of the  object property hierarchy of the DataScientia knowledge teleontology.}
\centering
\label{dskt-oph}
\end{figure}

Third, let us examine a fragment of the object property hierarchy of the DataScientia knowledge teleontology as visualized in the Figure \ref{dskt-oph}. The object properties are organized as a hierarchy as visualized from the left-half of the figure. Further, the meaning of each object property (e.g., \texttt{ds:ConstituentDataSetOf\_GID-302907}) is uniquely identified and disambiguated by a UKC GID (e.g., \texttt{302907}) appended to the label of the name of the object property (e.g., \texttt{ds:ConstituentDataSetOf\_GID-302907}). The object properties are also annotated with information such as an \texttt{rdfs:comment} on the right-half of the figure modelled as an annotation property. Last but not the least, it is interesting to note that the object property is well-defined using axiomatic assertions, e.g., domain assertion (\texttt{ds:DataSet\_GID-302884}) and range assertion (\texttt{ds:DataCatalog\_GID-302876}), associated to the selected object property \\ \texttt{ds:ConstituentDataSetOf\_GID-302907}. 

\begin{figure}[htp]
\includegraphics[width=16cm,height=8cm]{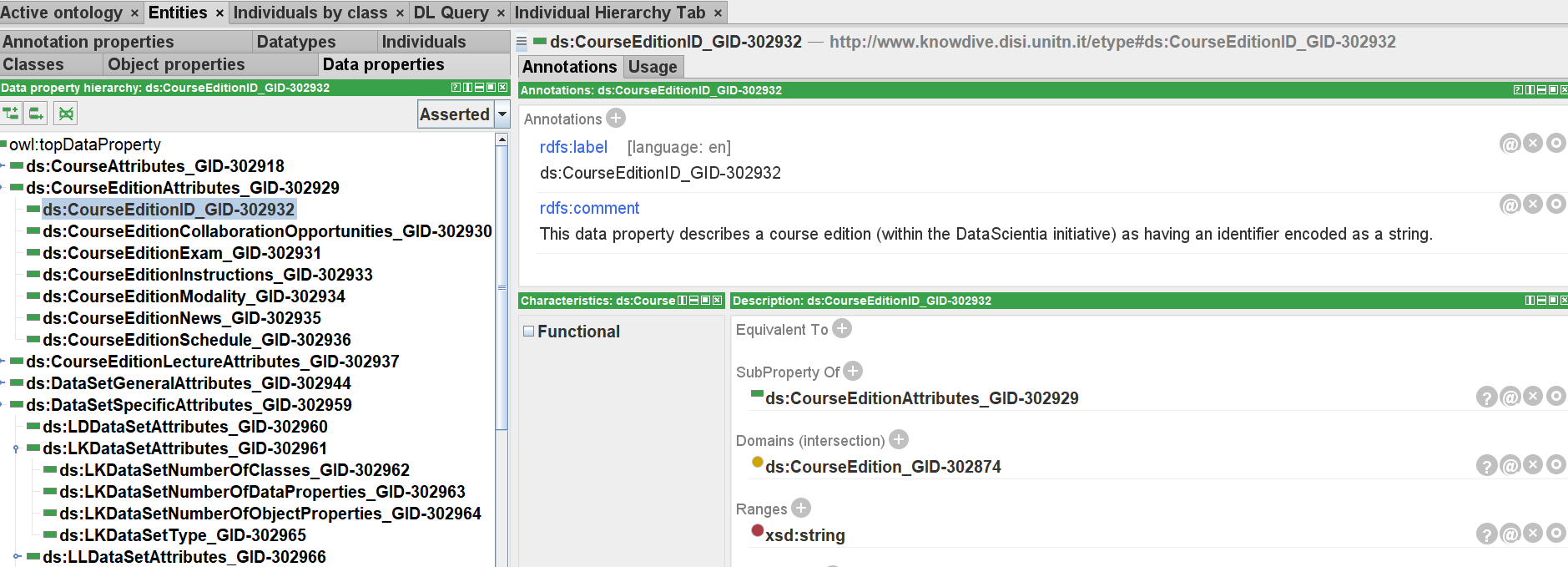}
\caption{A fragment of the  data property hierarchy of the DataScientia knowledge teleontology.}
\centering
\label{dskt-dph}
\end{figure}

Finally, let us concentrate on a fragment of the data property hierarchy of the DataScientia knowledge teleontology as visualized in the Figure \ref{dskt-dph}. The data properties are organized as a hierarchy as visualized from the left-half of the figure. Further, the meaning of each data property (e.g., \texttt{ds:CourseEditionID\_GID-302932}) is uniquely identified and disambiguated by a UKC GID (e.g., \texttt{302932}) appended to the label of the name of the data property (e.g., \texttt{ds:CourseEditionID\_GID-302932}). The data properties are also annotated with information such as an \texttt{rdfs:comment} on the right-half of the figure encoded as an annotation property. Lastly, notice that the data property is well-defined using axiomatic assertions, e.g., domain assertion (\texttt{ds:CourseEdition\_GID-302874}) and range assertion (\texttt{xsd:string}) with string as the chosen data type, associated to the selected data property \texttt{ds:CourseEditionID\_GID-302932}. 

Notice that an exhaustive explanation of the content of the DataScientia UKC namespace, language teleontology and knowledge teleontology is out of scope of the thesis as the thesis is primarily focused on using examples of DataScientia as proof-of-concept to exemplify the stratified language and knowledge representation formalisms advanced in the thesis.

Last but not the least, notice that the language and knowledge representations developed as part of the DataScientia initiative can be qualitatively evaluated in terms of whether the representations accommodate different levels of representation heterogeneity in terms of conceptual unity/diversity, language unity/diversity and knowledge unity/diversity. To that end, for example, in the figure \ref{dsns}, conceptual unity is established by modelling the concept \texttt{302876}. On the other hand, given the conceptual unity, we can establish and accommodate conceptual diversity by modelling different specializations of the concept \texttt{302876}, e.g., \texttt{302877}, \texttt{302879}, etc.. Similarly, as from the figure \ref{dsns}, we can establish language unity by modelling \texttt{ds:DataCatalog} (identified via the concept \texttt{302876}) in the English DataScientia domain language. On the other hand, given the language unity, we can establish language diversity by modelling different specializations of data catalogs, e.g., \texttt{ds:LiveData} (identified via the concept \texttt{302877}) and \texttt{ds:LiveKnowledge} (identified via the concept \texttt{302879}). Finally, as evidenced from the figure \ref{dskt-etexp} (partially illustrated), we can establish knowledge unity by modelling the etype \texttt{ds:LKDataSet\_GID-302887} in terms of a set of basic data properties. On the other hand, given the knowledge unity, we can establish knowledge diversity by modelling \texttt{ds:LKDataSet\_GID-302887} with the same basic set of data properties extended by different sets of (more specific) data properties, as per the level of abstraction required for a specific knowledge modelling scenario as and when it arises in the context of the DataScientia initiative. In the context of the above qualitative evaluation, it is important to reinforce the importance of the essentiality (as well as, the consequences in case of its non-proposal) of the stratified representation formalism-based language and knowledge representation proposed in this thesis towards accommodating the stratified character of representation heterogeneity.

\section{Summary}
To summarize, this chapter elucidated and exemplified a proof-of-concept for the UKC-based language and knowledge representations proposed in this thesis in the context of the DataScientia initiative. The next chapter presents highlights of a similar proof-of-concept for the UKC-based language and knowledge representations in the context of the JIDEP research and innovation project.

    \chapter{CASE STUDY ON MATERIALS MODELLING}

\section{Introduction}
As the second proof-of-concept, the thesis briefly illustrates language and knowledge representation of composite materials produced in the JIDEP\footnote{https://www.jidep.eu/} Horizon Europe research and innovation project following the UKC-based language and knowledge representation formalism and process advanced in the thesis. The JIDEP project’s objective was to create an Entity Graph-driven material passports of products and components \cite{MPO}, whereby, for example, raw materials such as a certain quantity of carbon fibre extracted from an automotive monocoque are considered as products. Considering this objective, the language and knowledge resources to describe and integrate representationally heterogeneous data about products, components, composite materials, composite manufacturing processes, etc., to ultimately enable the development and publishing of material passports were created by the University of Trento in collaboration with the University of Cambridge and other industrial partners within the project. To that end, first, the section provides a brief informal background on selected concepts relevant to model the language and knowledge representations in accordance with the objectives of the JIDEP project. Second, the section illustrates and elucidates snapshots of a first version of the UKC-based language and knowledge representation of composite materials developed as a part of the JIDEP project following the representation formalism and process proposed in the thesis. Finally, the section concludes with a brief discussion on how the language and knowledge representations produced within the project can be qualitatively evaluated in the context of the overarching problem of representation heterogeneity as proposed and considered in this thesis.

\section{Background}
First, a state-of-the-art review of the knowledge representations of materials modelling existing in the applied ontology community was conducted as (input) background knowledge by the University of Cambridge in collaboration with the University of Trento towards bootstrapping the development of the UKC-based composite materials language and knowledge representation later illustrated in this section. Some of the ontology-based knowledge representations reviewed are listed as follows.
\begin{itemize}
    \item Elementary Multiperspective Material Ontology (EMMO)\footnote{https://emmo-repo.github.io/}, an ontology focused on the materials modelling domain created by the European Materials Modelling Council (EMMC)\footnote{https://emmc.eu/}.
    \item CHAMEO Ontology \cite{del2022chameo}, a reference ontology for the description of materials characterization procedures, providing definitions at the methodological level.
    \item Mechanical Testing Ontology \cite{morgado2020mechanical}, an application of the EMMO to the field of mechanical testing.
\end{itemize}

Given the above review, the JIDEP partners decided to model the knowledge of composite materials by defining a central concept termed composite material (also called composite). The concept of composite was specialized into composite based on matrix material, composite based on reinforcement type, composite based on the manufacturing process, composite based on the orientation of reinforcement, composite based on the functional requirement, composite based on resin type, composite based on the type of fibre and composite based on structure. The concept - composite based on matrix material - has multiple sub-concepts, including polymer matrix composite (PMC), metal matrix composite (MMC) and ceramic matrix composite (CMC). The concept - composite based on reinforcement type - has further sub-concepts such as fibre-reinforced composite (FRC) and particulate-reinforced composite. Similar to the cases above, the concept of composite based on manufacturing process has many sub-concepts, including hand lay-up, pultrusion, filament winding, injection moulding, resin transfer moulding (RTM), compression moulding, vacuum infusion and autoclave. Composite based on the orientation of reinforcement has several sub-concepts, such as
longitudinal orientation, transverse orientation, random orientation and uni-directional
orientation. Composite based on functional requirement is classified into structural, barrier, thermal, and electromagnetic. Composite based on resin type is classified into the thermosetting,
thermoplastic, and hybrid resin. Composite based on the type of fibre is classified into glass fibre, carbon fibre, aramid fibre, silicon fibre, boron fibre and inorganic fibre. Composite based on the structure has several subclasses, including woven, non-woven, knitted, and braided structures. The modelling of composite materials also included the modelling of manufacturing aspects. Composite manufacturing processes include the wet lay-up process, resin transfer moulding process, pultrusion process, filament winding process, compression moulding process, injection moulding process, vacuum infusion process and autoclave process. The wet lay-up process has two sub-concepts: hand lay-up and spray lay-up. The resin transfer moulding process is specialized into closed mould RTM and open mould RTM. The filament winding method subsumes the following concepts: axial filament winding, helical filament winding and radial filament winding. The compression moulding process has two sub-concepts: hot press and vacuum-assisted compression moulding. The injection moulding process has the following sub-concepts: reaction injection moulding (RIM),
structural reaction injection moulding (SRIM) and resin transfer injection moulding (RTIM).
The autoclave process is classified into high-pressure autoclave moulding and high-temperature autoclave moulding.

Notice that an exhaustive explanation of both the reviewed knowledge representations as well as the detailed content of the informal specification of the JIDEP composite materials knowledge can be found in the dedicated JIDEP public deliverable \cite{jidep} and is extrinsic to the core focus of this thesis.

\section{Language Representation}
\begin{figure}[htp]
\includegraphics[width=16cm,height=6cm]{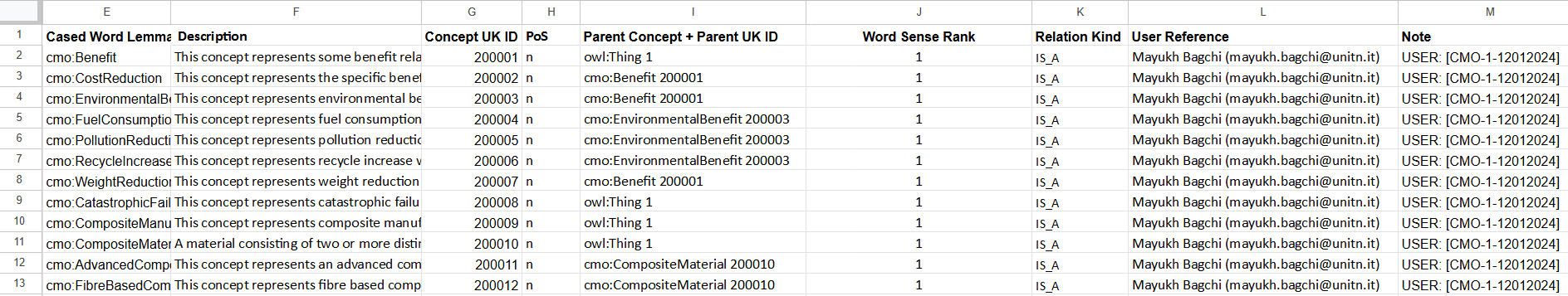}
\caption{A fragment of the JIDEP composite materials UKC namespace.}
\centering
\label{cmns}
\end{figure}

Given an overview of the JIDEP composite materials knowledge, let us now concentrate on the first version of the UKC-based language representation developed for the JIDEP project following the formalism and process proposed in the thesis. An illustrative example of a portion of the fully annotated spreadsheet of the JIDEP composite materials UKC namespace is provided in Figure \ref{cmns} which is generated following an end-to-end execution of the UKC annotation process described in section (3.4) of chapter (3). Note that the namespace prefix \texttt{cmo} refers to terms from the JIDEP UKC namespace encoding concepts of knowledge of composite materials (a selection of which is exemplified in the preceding sub-section). Further, notice also the fact that since the UKC annotation process has been executed, the terms in this version of the JIDEP composite materials UKC namespace are integrated into the lexical-semantic hierarchy of the UKC as a domain language.

\noindent The first column \texttt{Cased Word Lemma} lists sequentially (the hierarchy of) concept label(s) which underwent UKC annotation (e.g., \texttt{cmo:Benefit}). The second and the third columns encodes a gloss of the relative term (e.g., \texttt{\textit{This concept represents some benefit related to composite materials}}) and the concept GID (e.g., \texttt{200001}) following the UKC annotation process. The fourth and the fifth columns record the part of speech category (e.g., \texttt{n (noun)}) and the parent concept as well as the parent concept GID (e.g., \texttt{owl:Thing 1}) of the selected concept/term, respectively. The sixth column records the word sense rank relative to the chosen concept (e.g., \texttt{1}). The seventh column records the ontological relation existing between the selected concept label and its parent (e.g., \texttt{IS\_A}). Finally, the eighth and the ninth columns provide the user reference information (e.g., \texttt{Mayukh Bagchi (mayukh.bagchi@unitn.it)}) and the version associated to each selected concept label (e.g., \texttt{USER: [CMO-1-12012024]}).  

\section{Knowledge Representation}
Given an illustration of the JIDEP composite materials UKC namespace, let us now concentrate on the first version of the UKC-based knowledge representation developed for JIDEP composite materials knowledge following the formalism and process proposed in the thesis. First, illustrative examples of portions of the first version of the JIDEP (composite materials) language teleontology are discussed. Notice that these representations follow the knowledge representation formalism discussed in chapter (4). Further, notice that the modelling of data properties were not considered for the version of the JIDEP language and knowledge representations report in this section.

\begin{figure}[htp]
\includegraphics[width=16cm,height=10cm]{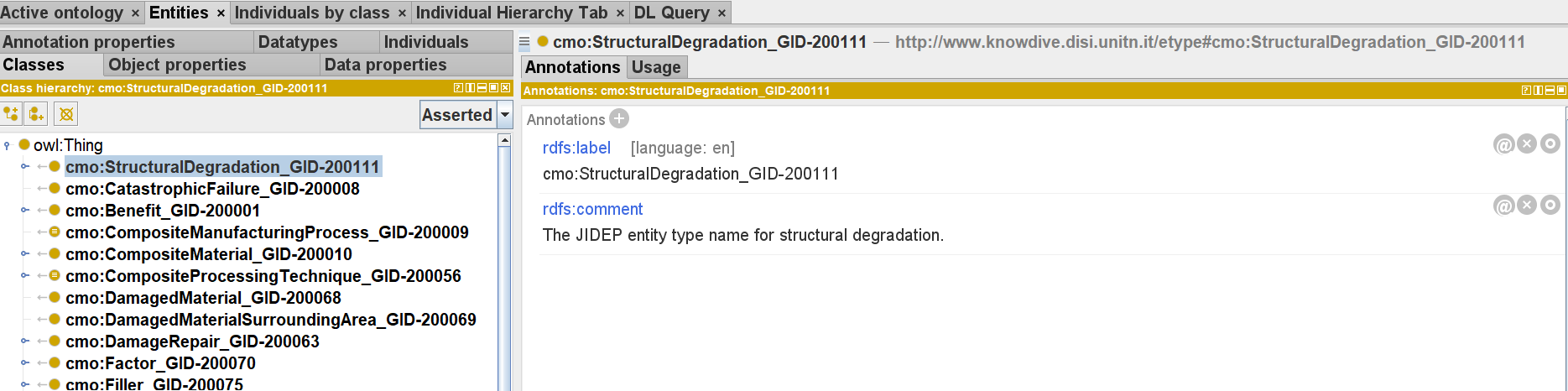}
\caption{A fragment of the entity type name hierarchy of the JIDEP language teleontology.}
\centering
\label{cslt-etexp}
\end{figure}

First, let us examine a fragment of the JIDEP language teleontology entity type name hierarchy as visualized in the Figure \ref{cslt-etexp}. The entity type names are organized as a hierarchy as visualized from the left-half of the figure encoding concepts such as structural degradation, composite materials, composite manufacturing processes, etc. Further, the meaning of each entity type name (e.g., \texttt{cmo:StructuralDegradation\_GID-200111}) is uniquely identified and disambiguated by a UKC GID (e.g., \texttt{200111}) appended to the label of the name (e.g., \texttt{cmo:StructuralDegradation\_GID-200111}). The label of the entity type names also display the prefix (e.g., \texttt{cmo}) of the JIDEP UKC namespace from which it has been reused. The entity type name is annotated with information such as an \texttt{rdfs:comment} on the right-half of the figure modelled as an annotation property. Further, notice the entity type name is not defined, i.e., there are no associated axiomatic assertions and/or property constraint, e.g., disjointness assertions, class expressions via asserted property constraints, etc., associated to the selected entity type name \texttt{cmo:StructuralDegradation\_GID-200111}. 

\begin{figure}[htp]
\includegraphics[width=16cm,height=8cm]{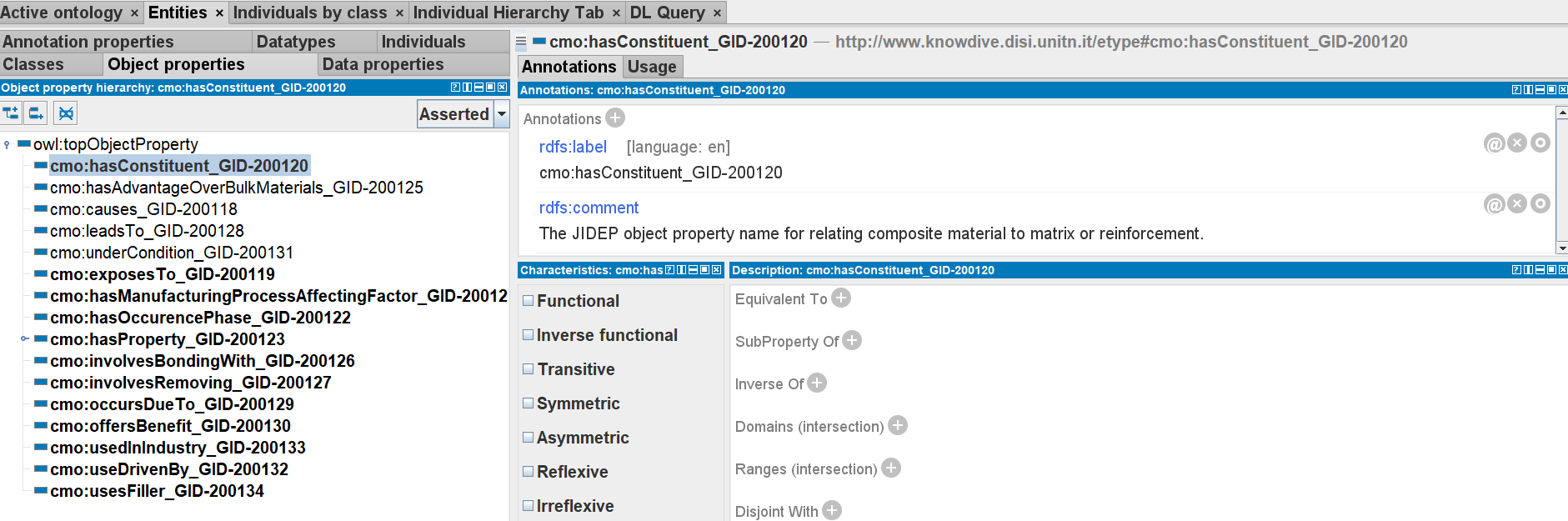}
\caption{A fragment of the object property name hierarchy of the JIDEP language teleontology.}
\centering
\label{cslt-oph}
\end{figure}

Second, let us examine a fragment of the object property names of the JIDEP language teleontology visualized in the Figure \ref{cslt-oph}. The object property names are organized as a hierarchy as visualized from the left-half of the figure. Further, the meaning of each object property name (e.g., \texttt{cmo:hasConstituent\_GID-200120}) is uniquely identified and disambiguated by a UKC GID (e.g., \texttt{200120}) appended to the label of the name (e.g., \texttt{cmo:hasConstituent\_GID-200120}). Like before, the label of the object property names also display the prefix (e.g., \texttt{cmo}) of the JIDEP UKC namespace from which it has been reused. The object property name is also annotated with information such as an \texttt{rdfs:comment} on the right-half of the figure modelled as an annotation property. Finally, it is interesting to note that the object property name is not associated to axiomatic assertions, e.g., characteristic assertions, disjointness assertions, domain and/or range assertions, etc, associated to the selected entity type name \texttt{cmo:hasConstituent\_GID-200120}.   

Let us now concentrate on illustrations-cum-elucidation of the first version of the JIDEP (composite materials) knowledge teleontology.

\begin{figure}[htp]
\includegraphics[width=16cm,height=8cm]{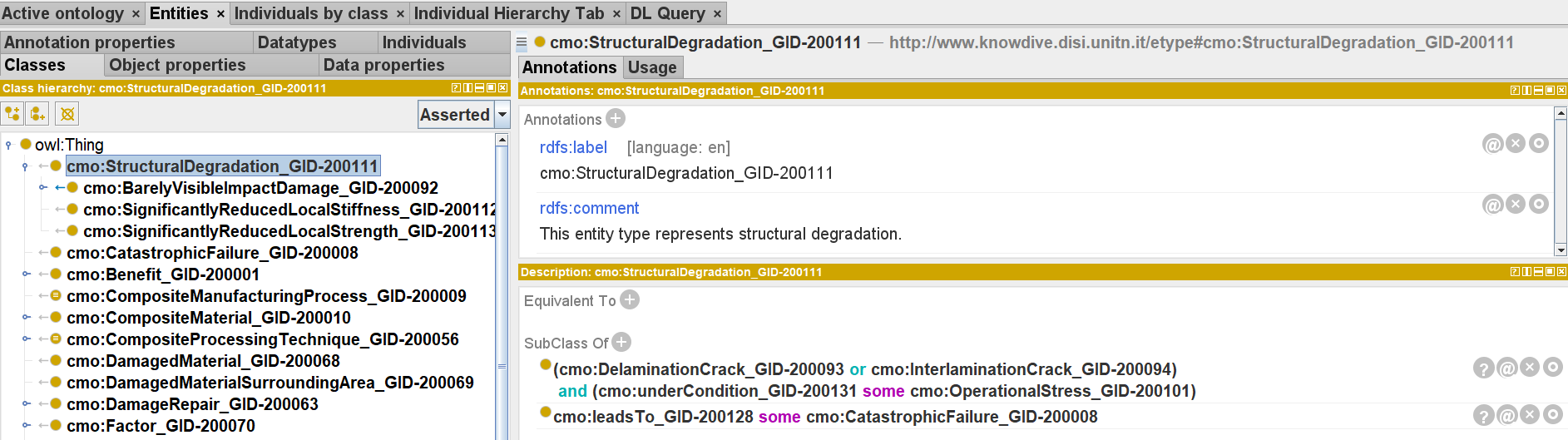}
\caption{A fragment of the entity type hierarchy of the JIDEP knowledge teleontology.}
\centering
\label{cskt-etexp}
\end{figure}

Let us first focus on a fragment of the JIDEP knowledge teleontology as visualized in the Figure \ref{cskt-etexp}. The entity types are organized as a hierarchy as visualized from the left-half of the figure encoding entity types such as structural degradation, composite materials, composite manufacturing processes, etc. Further, the meaning of the name of each entity type (e.g., \texttt{cmo:StructuralDegradation\_GID-200111}) is uniquely identified and disambiguated by a UKC GID (e.g., \texttt{200111}) appended to the label of the name (e.g., \texttt{cmo:StructuralDegradation\_GID-200111}). Finally, notice the entity types are defined, i.e., there are axiomatic assertions associated to the entity type using the object properties defined. For example, for the entity type \texttt{cmo:StructuralDegradation\_GID-200111}, the axiomatic descriptions include, e.g., object restriction constraint (e.g., \texttt{cmo:leadsTo\_GID-200128 \textbf{some} \\ cmo:CatastrophicFailure\_GID-200008}), etc.

\begin{figure}[htp]
\includegraphics[width=16cm,height=8cm]{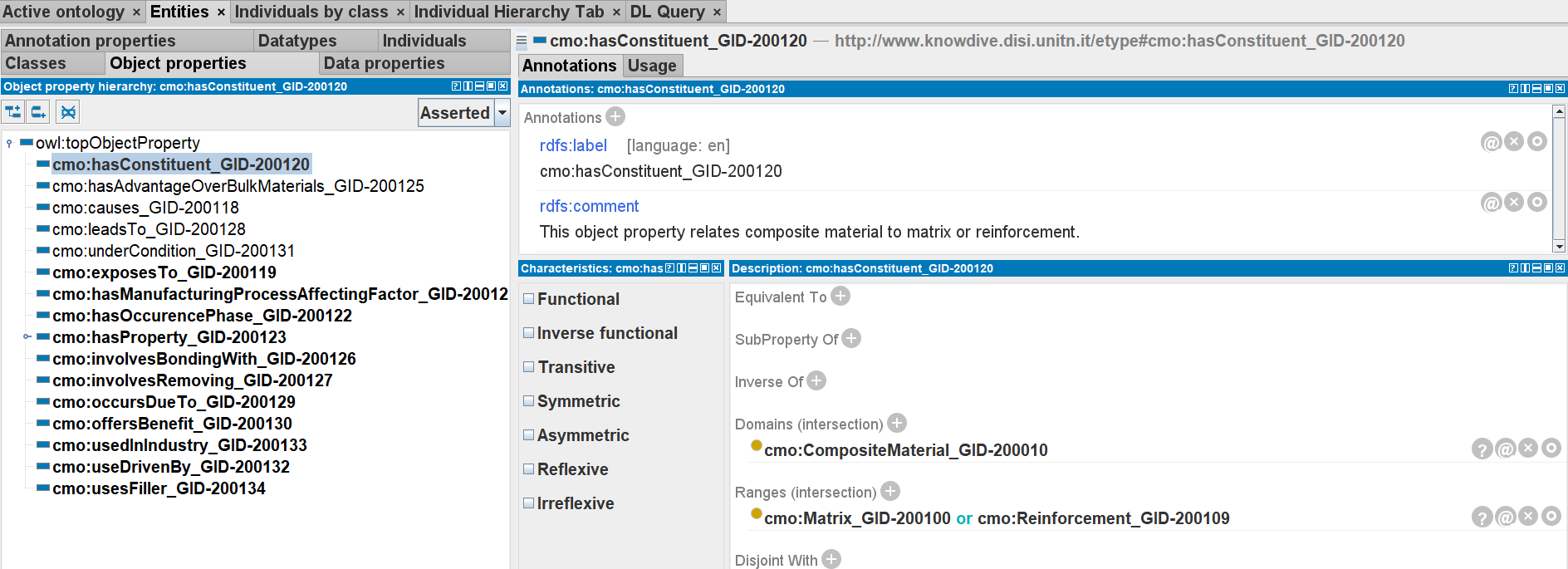}
\caption{A fragment of the object property hierarchy of the JIDEP knowledge teleontology.}
\centering
\label{cskt-oph}
\end{figure}

Finally, let us examine a fragment of the object property hierarchy of the JIDEP knowledge teleontology as visualized in the Figure \ref{cskt-oph}. The object properties are organized as a hierarchy as visualized from the left-half of the figure. Further, the meaning of each object property (e.g., \texttt{cmo:hasConstituent\_GID-200120}) is uniquely identified and disambiguated by a UKC GID (e.g., \texttt{200120}) appended to the label of the name of the object property. Last but not the least, it is interesting to note that the object property is well-defined using axiomatic assertions, e.g., domain assertion \texttt{cmo:CompositeMaterial\_GID-200010} and range assertion \texttt{cmo:Matrix\_GID-200100 \textbf{or} cmo:Reinforcement\_GID-200109}, associated to the selected object property \\ \texttt{cmo:hasConstituent\_GID-200120}.

Notice that an exhaustive explanation of the content of the JIDEP composite materials UKC namespace, language teleontology and knowledge teleontology is out of scope of the thesis as the thesis is primarily focused on using examples of the JIDEP project as proof-of-concept to exemplify the stratified language and knowledge representation formalisms advanced in the thesis.

Finally, notice that the language and knowledge representations developed as part of the JIDEP project can be qualitatively evaluated in terms of whether the representations accommodate different levels of representation heterogeneity in terms of conceptual unity/diversity, language unity/diversity and knowledge unity/diversity. To that end, for example, in the figure \ref{cmns}, conceptual unity is established by modelling the concept \texttt{200003}. On the other hand, given the conceptual unity, we can establish and accommodate conceptual diversity by modelling different specializations of the concept \texttt{200003}, e.g., \texttt{200004}, \texttt{200005}, etc.. Similarly, as from the figure \ref{cmns}, we can establish language unity by modelling \texttt{cmo:EnvironmentalBenefit} (identified via the concept \texttt{200003}) in the English JIDEP domain language. On the other hand, given the language unity, we can establish language diversity by modelling different specializations of environmental benefits, e.g., \texttt{ds:FuelConsumptionReduction} (identified via the concept \texttt{200004}) and \texttt{ds:PollutionReduction} (identified via the concept \texttt{200005}). Finally, as evidenced from the figure \ref{cskt-etexp} (partially illustrated), we can establish knowledge unity by modelling the etype \texttt{cmo:StructuralDegradation\_GID-200111} in terms of a set of basic data properties in the context of JIDEP composite materials modelling. On the other hand, given the knowledge unity, we can establish knowledge diversity by modelling \texttt{cmo:StructuralDegradation\_GID-200111} with the same basic set of data properties extended by different sets of (more specific) data properties, as per the level of abstraction required for a specific scenario as and when it arises in the context of materials knowledge modelling. In the context of the above qualitative evaluation, it is also important to stress the centrality (as well as, the consequences in case of its non-proposal) of the stratified representation formalism-based language and knowledge representation proposed in this thesis towards accommodating the stratified character of representation heterogeneity.

\section{Summary}
To summarize, this chapter elucidated and exemplified a proof-of-concept for the UKC-based language and knowledge representations proposed in this thesis in the context of the JIDEP research and innovation project. The next chapter summarizes the highlights of the thesis and concludes with potential future lines of research work based on the thesis.

    \chapter{CONCLUSION}

\section{Takeaways}
The thesis proposed a refined characterization of the problem of semantic heterogeneity as representation heterogeneity in the context of motivating applications like KG based multilingual data integration. To that end, the thesis advanced a top-down solution approach to the problem of representation heterogeneity in terms of several characteristically independent but functionally interrelated solution components which are localized to tackle specific aspects of representation heterogeneity but, in unison, are also general enough to tackle representation heterogeneity in its entirety. First, the thesis advanced a representation formalism which is stratified into concept level, language level, knowledge level and data level to accommodate the unity and diversity existing within each level. Second, given the stratified representation formalism, the thesis advocated for a top-down language representation using Universal Knowledge Core (UKC), UKC namespaces and domain languages to tackle conceptual and language level unity and diversity. Third, given the stratification of representation and language representation, the thesis proposed a top-down knowledge representation using the notions of language teleontology and knowledge teleontology to tackle knowledge level unity and diversity. Fourth, the thesis described the use and development of the existing LiveKnowledge Catalog for enabling iterative reuse and sharing of language and knowledge resources with an aim to increasingly minimize representation heterogeneity. Finally, the \textit{kTelos} methodology is elucdiated in terms of how it evolved and how it integrated and unified the solution components above into a single process to iteratively generate the language and knowledge representations addressing representation heterogeneity. Last but not the least, the thesis also included poof-of-concepts relevant to DataScientia (data catalogs) and JIDEP (materials modelling) in terms of the language and knowledge representations adopted for the projects. 

\section{Limitations}
There are several limitations towards a complete realization of the ideas advanced by this thesis. First, the UKC-based language and knowledge representation ecosystem is yet to be completely developed in terms of its constituent UKC namespaces, domain languages, language teleontologies and knowledge teleontologies verticalized into different topics. Second, the ideas advanced by the thesis are also limited in terms of their practical applicability can enhance the development of language and knowledge-enriched Large Language Models (LLMs). Third, the content and services of the LiveKnowledge catalog is yet to be completely developed with a focus towards increased sharing and reuse of language and knowledge representations.

\section{Next Steps}
The limitations stated above are planned to be developed and addressed as future lines of research. First, the UKC-based language and knowledge representation ecosystem has to be developed further in terms of its constituent UKC namespaces, domain languages, language teleontologies and knowledge teleontologies verticalized into topics such as society and territory and data catalogs (in much finer detail than what is illustrated as proof-of-concept in this thesis). Second, it would also be interesting to study how the above representations can be tuned and instantiated to enable the development of Large Language Models (LLMs) in collaboration with researchers who are experts in LLMs. Third, the LiveKnowledge catalog has to be continuously updated, enriched and maintained to facilitate increased sharing and reuse of the above language and knowledge representations. Finally, it is also key to develop end-to-end implementations of \textit{kTelos} so as to develop and refine the guiding principles of modelling in order to generate reusable language and knowledge representations of increased quality and value.

\bibliographystyle{plain}
\bibliography{source/biblio.bib}

\appendix
    \chapter{GLOSSARY}
This appendix collects and enlists succinct description of some important keywords introduced and used throughout the thesis.
\begin{itemize}    
    \item[] \textbf{UNIVERSAL KNOWLEDGE CORE}: The Universal Knowledge Core (UKC) is a multilingual lexical-semantic resource which forms the foundation of language representation adopted in this thesis. The UKC represents what exists in the world (via concepts) and what we can name and define (via words) within the scope of a specific natural language. It was originally conceived at the University of Trento, Italy and has since been continuously developed and enriched in terms of the number of natural languages aligned to the UKC and the varied services offered by the UKC. The UKC is composed of two characteristically distinct but functionally interrelated components, namely, the Concept Core and the Language Core. In the UKC, the meaning of words are represented not only with sets of synonymous words, i.e., synsets (via the Language Core), but also using uniquely identifiable supralingual concepts (via the Concept Core) clustering together the synsets which, in different natural languages, codify the same meaning.

    \item[] \textbf{TOPIC}: A Topic is defined as an informal label under which informal controlled vocabularies belonging to a number of cognate universes of discourse may be (continuously) aggregated and collected. Therefore, a topic is composed of:
    \begin{itemize}
    \item one or more W3C namespaces, and/or,
    \item one or more classification schemes, and/or,
    \item one or more (terminological) standards.
    \end{itemize}

    \item[] \textbf{(W3C) NAMESPACE}: A (W3C) namespace can be defined as an informal language resource, identified by an internationalized resource identifier (IRI) reference, encoding any hierarchical collection of names which can be from amongst the options below:
    \begin{itemize}
    \item a collection of element names, and/or,
    \item a collection of attribute names, and/or,
    \item a collection of properties (FOAF), and/or,
    \item a collection of concepts (e.g., as in a WordNet), and many other uses are likely to arise.
    \end{itemize}

    \item[] \textbf{CLASSIFICATION}: A classification scheme is an informal language resource encoding a hierarchy of terms representing concepts in the context of a specific subject or a specific universe of discourse or a specific specialized purpose. Such a scheme often includes synonyms, definitions, explanations and example of each term in its hierarchy so as to refer to the semantics of the concept with least possible ambiguity.

    \item[] \textbf{STANDARD}: A terminology standard can be defined as an informal language resource encoding a hierarchy of terms which addresses a key requirement for unambiguous data and information interoperability, i.e., the ability to represent concepts in an unambiguous manner between a sender and receiver of information. A standard is always developed and continuously maintained by a leading organization. They are widely employed, e.g., in a domain, for data and information interoperability and they are often \textit{adapted} to suit local requirements, e.g., in terms of language translations, adding or deprecating terms.  

    \item[] \textbf{UKC NAMESPACE}: A UKC namespace is a formal language representation encoding a hierarchically related set of uniquely identifiable concept names, via unique UKC GIDs, in which each concept name has a single unambiguous meaning and represents potential entity types and properties in one or multiple allied topics. A consequence of the above definition is the fact that a UKC namespace is fully aligned to the lexical-semantic hierarchy of the UKC and conforms to the language representation formalism of the UKC.

    \item[] \textbf{DOMAIN LANGUAGE}: A domain language is defined as a controlled vocabulary representing hierarchically related set of uniquely identifiable concept names, via unique UKC GIDs, in which each concept name has a single unambiguous meaning and represents potential entity types and properties across one or multiple natural languages and/or UKC namespaces relevant to a specific and/or allied topic(s). It is a sub-hierarchy of the UKC lexical-semantic hierarchy, i.e., it represents what exists and what can be named in one or more topic(s) out of the totality of the UKC (depending on the requirements of an application scenario).
    
    \item[] \textbf{LANGUAGE TELEONTOLOGY}: A Language Teleontology is defined as a (intermediate) knowledge representation formally representing, reusing the semantically disambiguated words of domain language(s) modelled in the Language Representation phase, the uniquely identifiable concept names relevant to model entity types and properties relevant to the scope of a knowledge graph engineering project. A language teleontology uses three primary representational constructs: 
    \begin{itemize}
    \item entity type names for modelling the concept names which will, at a later stage, be defined as entity types,
    \item object property names for modelling the concept names which will, at a later stage, be defined as entity types, and
    \item data property names for modelling the concept names which will, at a later stage, be defined as entity types.
    \end{itemize}

    \item[] \textbf{KNOWLEDGE TELEONTOLOGY}: A Knowledge Teleontology is defined as a knowledge representation which, reusing the concept names formalized in a language teleontology, defines entity types, object properties modelling the interrelationships between entity types and and data properties encoding descriptions of an entity type, within the scope of a topic which can be exploited in potential knowledge graph engineering projects relevant to that topic or allied topics. To that end, a knowledge teleontology is defined using four primary representational constructs: 
    \begin{itemize}
    \item entity types which define the various classes of entities that share common attributes,
    \item object properties which define the interrelationships amongst different entity types,
    \item data properties which describe the attributes of different entity types, and,
    \item data types which asserts the acceptable range of data properties within a knowledge teleontology.
    \end{itemize} 
\end{itemize}

\end{document}